\RequirePackage{fix-cm} 
\documentclass[a4paper, twoside, reqno, dvips, 12pt]{amsart}
\usepackage{fixltx2e}   

\usepackage{etex}

\usepackage[latin1]{inputenc}
\usepackage[T1]{fontenc}

\usepackage{esint}
\usepackage{dsfont}
\usepackage{xspace}
\usepackage{amsgen}
\usepackage{amsthm}
\usepackage{amssymb}
\usepackage{amsmath}
\usepackage{wasysym}
\usepackage{upgreek}
\usepackage{amsfonts}
\usepackage{scalerel}
\usepackage{stmaryrd}
\usepackage{mathtools}
\usepackage{stackengine}

\usepackage{relsize}
\usepackage[mathcal, mathscr]{euscript}

\usepackage{mathrsfs}
\DeclareMathAlphabet{\mathscrbf}{OMS}{mdugm}{b}{n}

\usepackage{a4wide}

\usepackage[dvipsnames, table]{xcolor}
\definecolor{bckg}{RGB}{20.8, 20.8, 20.8}
\definecolor{oneblue}{rgb}{0.0, 0.0, 0.85}
\definecolor{Lightblue}{RGB}{214, 214, 214}
\definecolor{bluepigment}{rgb}{0.2, 0.2, 0.6}
\definecolor{charcoal}{rgb}{0.21, 0.27, 0.31}
\definecolor{denimblue}{rgb}{0.08, 0.38, 0.74}
\definecolor{Lightgray}{rgb}{0.89, 0.89, 0.89}
\definecolor{darkgrey}{rgb}{0.273, 0.281, 0.30}
\definecolor{darkelectricblue}{rgb}{0.33, 0.41, 0.47}

\usepackage[sort&compress, comma, square, numbers]{natbib}

\usepackage{psfrag}
\usepackage{graphicx}
\usepackage{subfigure}
\usepackage{morefloats}
\usepackage{indentfirst}

\usepackage{acronym}
\usepackage{microtype}
\usepackage[labelsep=period,%
            labelfont={bf,sf,color=bluepigment},%
            justification=raggedright]{caption}

\usepackage[perpage, symbol]{footmisc}

\usepackage[usenames, dvipsnames, pdf]{pstricks}
\usepackage{epsfig}
\usepackage{pst-grad} 
\usepackage{pst-plot} 

\usepackage[colorlinks,
           urlcolor=oneblue,
           linkcolor=denimblue,
           citecolor=NavyBlue,
           bookmarksopen=false,
           pdfpagemode=UseNone,
           pagebackref]{hyperref}

\usepackage[explicit]{titlesec}

\titleformat{\section}[block]
  {\color{NavyBlue}\Large\sffamily\bfseries}
  {}
  {0.0em}
  {\colorbox{bckg!5}{\strut\parbox{\dimexpr\linewidth-2\fboxsep\relax}{\thesection. #1}}}
  [\vspace*{0.33em}]

\titleformat{name=\section,numberless}[block]
  {\color{NavyBlue}\Large\sffamily\bfseries}
  {}
  {0.0em}
  {\colorbox{bckg!5}{\strut\parbox{\dimexpr\linewidth-2\fboxsep\relax}{#1}}}
  [\vspace*{0.33em}]

\titleformat{\subsection}
  {\color{NavyBlue}\large\sffamily\bfseries}
  {}
  {0.0em}
  {\colorbox{bckg!5}{\parbox{\dimexpr\linewidth-2\fboxsep\relax}{\thesubsection. #1}}}
  [\vspace*{0.33em}]

\titleformat{name=\subsection,numberless}
  {\color{NavyBlue}\Large\sffamily\bfseries}
  {}
  {0em}
  {\colorbox{bckg!5}{\parbox{\dimexpr\linewidth-2\fboxsep\relax}{#1}}}
  [\vspace*{0.33em}]

\titleformat{\subsubsection}
  {\color{bluepigment}\sffamily\normalsize\bfseries}
  {\thesubsubsection}
  {0.5em}
  {#1}
  [\vspace*{0.33em}]

\titleformat{\paragraph}[runin]
  {\color{bluepigment}\sffamily\small\bfseries}
  {}
  {0em}
  {#1}

\titlespacing{\section}{0.0em}{1.5em plus 2pt minus 2pt}%
{1.0em plus 2pt minus 2pt}[0em]
\titlespacing{\subsection}{0.5em}{1.5em plus 2pt minus 2pt}%
{1.0em}[0em]
\titlespacing{\subsubsection}{0.5em}{1.5em plus 2pt minus 2pt}%
{1.0em plus 2pt minus 2pt}[0em]

\usepackage{titletoc}

\contentsmargin{0.5em}
\newlength{\tocsep} 
\titlecontents{section}[\tocsep]
  {\addvspace{10pt}\bfseries\sffamily}
  {\contentslabel[\thecontentslabel]{\tocsep}}
  {}
  {\ \titlerule*[0.75pc]{.}\ \thecontentspage}
  []
\titlecontents{subsection}[\tocsep]
  {\addvspace{8pt}\sffamily}
  {\contentslabel[\thecontentslabel]{\tocsep}}
  {}
  {\ \titlerule*[0.5pc]{.}\ \thecontentspage}
  []
\titlecontents*{subsubsection}[\tocsep]
  {\addvspace{2pt}\footnotesize\sffamily}
  {}
  {}
  {\ \titlerule*[0.35pc]{.}\ \thecontentspage}
  [\\*]

\makeatletter
\def\@setauthors{%
  \begingroup
  \def\thanks{\protect\thanks@warning}%
  \trivlist
  \centering\footnotesize \@topsep30\p@\relax
  \advance\@topsep by -\baselineskip
  \item\relax
  \author@andify\authors
  \def\\{\protect\linebreak}%
  \textsc{\normalsize\textcolor{darkelectricblue}{\authors}}%
  \ifx\@empty\contribs
  \else
    ,\penalty-3 \space \@setcontribs
    \@closetoccontribs
  \fi
  \endtrivlist
  \endgroup
}
\def\@settitle{\begin{center}%
  \baselineskip14\p@\relax
    \bfseries
    \textsc{\Large\textcolor{charcoal}{\@title}}
  \end{center}%
}
\makeatother

\usepackage{enumitem}
\setlist[description]{%
  topsep=30pt,               
  itemsep=5pt,               
  font={\bfseries\sffamily\color{NavyBlue}}, 
}

\usepackage{fancyhdr}
\usepackage{lastpage}

\newcommand*\Title{\textcolor{bluepigment}{Dispersive shallow water wave modelling. Part IV}}
\newcommand*\Authors{\textcolor{bluepigment}{G.~Khakimzyanov, D.~Dutykh \& O.~Gusev}}
\newcommand*{\plogo}{\textcolor{gray}{{\texttt{arXiv.org} / \textsc{hal}}}} 

\numberwithin{equation}{section}

\newtheorem{remark}{Remark}

\newcommand{\up}[1]{$^{\mathrm{\small\textsf{#1}}}$} 

\newcommand{\nuc}{\upnu}

\renewcommand{\chi}{\upchi}

\newcommand{\pc}{\check{p}}

\renewcommand{\tau}{\uptau}
\newcommand{\ud}{\mathrm{d}}
\newcommand{\ue}{\mathrm{e}}
\newcommand{\D}{\mathscr{D}}
\renewcommand{\phi}{\varphi}
\newcommand{\J}{\mathcal{J}}
\newcommand{\M}{\mathcal{M}}

\newcommand{\W}{\mathcal{W}}
\newcommand{\Gm}{\mathbb{G}}
\newcommand{\Dd}{\mathcal{D}}
\newcommand{\Ff}{\mathscr{F}}

\newcommand{\Gg}{\mathscr{G}}
\newcommand{\Pp}{\mathscr{P}}
\newcommand{\Rr}{\mathscr{R}}

\newcommand{\Qq}{\mathscr{Q}}
\newcommand{\ups}{\upupsilon}
\renewcommand{\beta}{\upbeta}

\renewcommand{\O}{\mathcal{O}}

\renewcommand{\H}{\mathcal{H}}
\renewcommand{\leq}{\leqslant}

\renewcommand{\S}{\mathcal{S}}
\newcommand{\Sr}{\mathfrak{S}}

\renewcommand{\alpha}{\upalpha}

\newcommand{\g}{\boldsymbol{g}}
\newcommand{\n}{\boldsymbol{n}}
\newcommand{\x}{\boldsymbol{x}}
\renewcommand{\omega}{\upomega}

\renewcommand{\Pi}{\mathfrak{P}}

\renewcommand{\u}{\boldsymbol{u}}
\renewcommand{\v}{\boldsymbol{v}}
\renewcommand{\lambda}{\uplambda}

\newcommand{\hw}{h_{\,\mathrm{w}}}

\newcommand{\const}{\mathrm{const}}

\newcommand{\DDelta}{\boldsymbol{\Delta}}

\newcommand{\Ll}{\mathlarger{\mathlarger{\mathcal{L}}}}

\newcommand{\Pnh}{\power}
\newcommand{\pb}{\varrho}
\renewcommand{\k}{\upkappa}
\renewcommand{\r}{\Upsilon}
\newcommand{\La}{\Uplambda}
\newcommand{\h}{\mathsf{h}}
\newcommand{\m}{\mathsf{m}}
\newcommand{\s}{\mathsf{s}}
\newcommand{\ld}{\ell_{\,d}}
\newcommand{\B}{\mathscr{B}}
\newcommand{\F}{\mathcal{F}}
\newcommand{\K}{\mathcal{K}}
\newcommand{\km}{\mathsf{km}}
\newcommand{\Ka}{\mathcal{K}a}

\newcommand{\Fd}{\mathring{\F}}

\newcommand{\vO}{\boldsymbol{0}}
\newcommand{\rro}{\mathring{\Upsilon}}

\newcommand{\Ko}{\mathring{\mathcal{K}}}
\newcommand{\Lab}{\boldsymbol{\Uplambda}}

\newcommand{\Fdb}{\mathring{\boldsymbol{\F}}}
\newcommand\openbigstar[1][0.7]{%
  \scalerel*{%
    \stackinset{c}{-.125pt}{c}{}{\scalebox{#1}{\color{white}{$\bigstar$}}}{%
      $\bigstar$}%
  }{\bigstar}
}
\newcommand{\power}{\raisebox{.15\baselineskip}{\Large\ensuremath{\wp}}}

\newcommand{\ie}{\emph{i.e.}\/ }
\newcommand{\eg}{\emph{e.g.}\/ }

\renewcommand{\div}{\grad\scal}

\newcommand{\divf}{\bar{\grad}\scal}
\newcommand{\scal}{\boldsymbol{\cdot}}
\newcommand{\grad}{\boldsymbol{\nabla}}

\newcommand{\abs}[1]{\lvert\, #1\, \rvert}

\newcommand{\pd}[2]{\frac{\partial\/ #1}{\partial\/ #2}}

\newcommand{\eqdef}{\mathop{\stackrel{\,\mathrm{def}}{:=}\,}}
\newcommand{\defeq}{\mathop{\stackrel{\,\mathrm{def}}{=:}\,}}

\DeclareMathOperator*{\dprime}{\prime\prime}

\usepackage{acronym}

\begin{document}

\title[\Title]{Dispersive shallow water wave modelling. Part IV: Numerical simulation on a globally spherical geometry}

\author[G.~Khakimzyanov]{Gayaz Khakimzyanov}
\address{\textbf{G.~Khakimzyanov:} Institute of Computational Technologies, Siberian Branch of the Russian Academy of Sciences, Novosibirsk 630090, Russia}
\email{Khak@ict.nsc.ru}

\author[D.~Dutykh]{Denys Dutykh$^*$}
\address{\textbf{D.~Dutykh:} LAMA, UMR 5127 CNRS, Universit\'e Savoie Mont Blanc, Campus Scientifique, F-73376 Le Bourget-du-Lac Cedex, France}
\email{Denys.Dutykh@univ-savoie.fr}
\urladdr{http://www.denys-dutykh.com/}
\thanks{$^*$ Corresponding author}

\author[O.~Gusev]{Oleg Gusev}
\address{\textbf{O.~Gusev:} Institute of Computational Technologies, Siberian Branch of the Russian Academy of Sciences, Novosibirsk 630090, Russia}
\email{gusev\_oleg\_igor@mail.ru}



\begin{titlepage}
\thispagestyle{empty} 
\noindent
{\Large Gayaz \textsc{Khakimzyanov}}\\
{\it\textcolor{gray}{Institute of Computational Technologies, Novosibirsk, Russia}}
\\[0.02\textheight]
{\Large Denys \textsc{Dutykh}}\\
{\it\textcolor{gray}{CNRS--LAMA, Universit\'e Savoie Mont Blanc, France}}
\\[0.02\textheight]
{\Large Oleg \textsc{Gusev}}\\
{\it\textcolor{gray}{Institute of Computational Technologies, Novosibirsk, Russia}}
\\[0.08\textheight]

\vspace*{1cm}

\colorbox{Lightblue}{
  \parbox[t]{1.0\textwidth}{
    \centering\huge\sc
    \vspace*{0.7cm}
    
    \textcolor{bluepigment}{Dispersive shallow water wave modelling. Part IV: Numerical simulation on a globally spherical geometry}
    
    \vspace*{0.7cm}
  }
}

\vfill 

\raggedleft     
{\large \plogo} 
\end{titlepage}


\newpage
\thispagestyle{empty} 
\par\vspace*{\fill}   
\begin{flushright} 
{\textcolor{denimblue}{\textsc{Last modified:}} \today}
\end{flushright}


\newpage
\maketitle
\thispagestyle{empty}


\begin{abstract}

In the present manuscript we consider the problem of dispersive wave simulation on a rotating globally spherical geometry. In this Part IV we focus on numerical aspects while the model derivation was described in Part III. The algorithm we propose is based on the splitting approach. Namely, equations are decomposed on a uniform elliptic equation for the dispersive pressure component and a hyperbolic part of shallow water equations (on a sphere) with source terms. This algorithm is implemented as a two-step predictor-corrector scheme. On every step we solve separately elliptic and hyperbolic problems. Then, the performance of this algorithm is illustrated on model idealized situations with even bottom, where we estimate the influence of sphericity and rotation effects on dispersive wave propagation. The dispersive effects are quantified depending on the propagation distance over the sphere and on the linear extent of generation region. Finally, the numerical method is applied to a couple of real-world events. Namely, we undertake simulations of the \textsc{Bulgarian} 2007 and \textsc{Chilean} 2010 tsunamis. Whenever the data is available, our computational results are confronted with real measurements.


\bigskip
\noindent \textbf{\keywordsname:} finite volumes; splitting method; nonlinear dispersive waves; spherical geometry; rotating sphere; Coriolis force \\


\end{abstract}


\newpage
\tableofcontents
\thispagestyle{empty}


\newpage
\section{Introduction}

Until recently, the modelling of long wave propagation on large scales has been performed in the framework of Nonlinear Shallow Water Equations (NSWE) implemented under various software packages \cite{Kolar1994}. This model is hydrostatic and non-dispersive \cite{Stoker1957}. Among popular packages we can mention, for example, the \textsc{TUNAMI} code \cite{Imamura1996a} based on a conservative finite difference leap-frog scheme on real bathymetries. This code has been extensively used for tsunami wave modeling by various groups (see \eg \cite{Zaitsev2005}). The code \textsc{MOST} uses the directional splitting approach \cite{Titov1996, Titov1997} and is also widely used for the simulation of tsunami wave propagation and run-up \cite{Wei2008, Tang2012}. The \textsc{MGC} package \cite{Shokin2008} is based on a modified \textsc{MacCormack} finite difference scheme \cite{Fedotova2006}, which discretizes NSWE in spherical coordinates. Obviously, the \textsc{MGC} code can also work in \textsc{Cartesian} coordinates as well. This package was used to simulate the wave run-up on a real-world beach \cite{Gusyakov2008} and tsunami wave generation by underwater landslides \cite{Beisel2012}. Recently the \textsc{VOLNA} code was developed using modern second order finite volume schemes on unstructured grids \cite{Dutykh2009a}. Nowadays this code is essentially used for the quantification of uncertainties of the tsunami risk \cite{Beck2016}.

All numerical models described above assume the wave to be non-dispersive. However, in the presence of wave components with higher frequencies (or equivalently shorter wavelengths), the frequency dispersion effects may influence the wave propagation. Even in 1982 in \cite{Mirchina1982} it was pointed out:
\begin{quote}
  [\,\,\dots\,] \textit{the considerations and estimates for actual tsunamis indicate that nonlinearity and dispersion can appreciably affect the tsunami wave propagation at large distances}.
\end{quote}
Later this conclusion was reasserted in \cite{Pelinovsky1996a} as well. The catastrophic Sumatra event in 2004 \cite{Syno2006} along with subsequent events brought a lot of new data all around the globe and also from satellites \cite{Kulikov}. The detailed analysis of this data allowed to understand better which models and algorithms should be applied at various stages of a tsunami life cycle to achieve the desired accuracy \cite{DGK, Murty2006}. The main conclusion can be summarized as follows: for a complete and satisfactory description of a tsunami wave life cycle on global scales, one has to use a nonlinear dispersive wave model with moving (in the generation region \cite{Dutykh2007b}) realistic bathymetry. For trans-oceanic tsunami propagation one has to include also Earth's sphericity and rotation effects. A whole class of suitable mathematical models combining all these features was presented in the previous Part~III \cite{Khakimzyanov2016a} of the present series of papers.

At the present time we have a rather limited amount of published research literature devoted to numerical issues of long wave propagation in a spherical ocean. In many works (see \eg \cite{Glimsdal2006, Horrillo2006}) Earth's sphericity is not taken explicitly into account. Instead, the authors project Earth's surface (or at least a sub-region) on a tangent plane to Earth in some point and computations are then performed on a flat space using a \textsc{Boussinesq}-type (Weakly Nonlinear and Weakly Dispersive --- WNWD) model without taking into account the \textsc{Coriolis} force. We notice that some geometric defects are unavoidable in this approach. However, even in this simplified framework the importance of dispersive effects has been demonstrated by comparing the resulting wave field with hydrostatic (NSWE) computations.

In \cite{Lovholt2008} the authors studied the transoceanic propagation of a hypothetical tsunami generated by an eventual giant landslide which may take place at La Palma island, which the most north-westerly island of the Canary Islands, Spain. Similarly the authors employed a WNWD model, but this time written in spherical coordinates with Earth's rotation effects. However, the employed model could handle only static bottoms. As a result, the initial fields were generated using a different hydrodynamic model and, then, transferred into the WNWD model as the initial condition to compute the wave long time evolution. However, when waves approach the shore, another limitation of weakly nonlinear models becomes apparent --- in coastal regions nonlinear effects grow quickly and, thus, the computations should be stopped before the wave reaches the coast. Otherwise, the numerical results may loose their validity. In \cite{Lovholt2008} it was also shown that the wave dispersion may play a significant r\^ole on the resulting wave field. Namely, NSWE predict the first significant wave hitting the shore, while WNWD equations predict rather an undular bore in which the first wave amplitude is not necessarily the highest \cite{Lovholt2010}. Of course, these undular bores cannot be described in the framework of NSWE \cite{Peregrine1966, Grue2008}.

An even more detailed study of tsunami dispersion was undertaken recently \cite{Glimsdal2013}, where also a WNWD model was used, but the dispersion effect was estimated for several real-world events. The authors came to interesting `uncertain' conclusions: 
\begin{quote}
  [\,\dots\,] \textit{However, undular bores, which are not included in shallow-water theory, may evolve during shoaling. Even though such bores may double the wave height locally, their effect on inundation is more uncertain because the individual crests are short and may be strongly affected by dissipation due to wave breaking}.
\end{quote}
It was also noted that near coasts WNWD model provides unsatisfactory results, that is why fully nonlinear dispersive models should be employed to model all stages from tsunami generation to the inundation. The same year a fully nonlinear dispersive model on a sphere including \textsc{Coriolis} effect was derived in \cite{Kirby2013}. However, in contrast to another paper from the same group \cite{Grilli2012}, the horizontal velocity variable is taken as a trace of 3D fluid velocity on a certain surface laying between the bottom and free surface. The proposed model may have some drawbacks. First of all, the well-posed character of the \textsc{Cauchy} problem is not clear. A very similar (and much simpler) \textsc{Nwogu}'s model \cite{Nwogu1993} is known to possess instabilities for certain configurations of the bottom \cite{Lovholt2009}. We underline also that the authors of \cite{Kirby2013} did not present so far any numerical simulations with their fully nonlinear model. A study of dispersive and \textsc{Coriolis} effects were performed in the WNWD counterpart of their fully nonlinear equations. In order to solve numerically their spherical \textsc{Boussinesq}-type system a \textsc{Cartesian} TVD scheme previously described in \cite{Shi2012} was generalized to spherical coordinates. This numerical model was implemented as a part of well-known \textsc{FUNWAVE(-TVD)} code \cite{Shi2012a}.

A fully nonlinear weakly dispersive model on a sphere with the depth-averaged velocity variable was first derived in \cite{Fedotova2010}. Later it was shown in \cite{Fedotova2014a} that the same model can be derived without using the potential flow assumption. Moreover, this model admits an elegant conservative structure with the mass, momentum and energy conservations. In particular, the energy conservation allows to control the amount of numerical viscosity in simulations. The same conservative structure can be preserved while deriving judiciously weakly nonlinear models as well. Only the expressions of the kinetic energy and various fluxes vary from one model to another. In this way one may obtain the whole hierarchy of simplified shallow water models on a sphere enjoying the same formal conservative structure \cite{Shokin2015}. In particular, it allows to develop a unique numerical algorithm, which can be applied to all models in this hierarchy by changing only the fluxes and source terms in the numerical code.

In this study we develop a numerical algorithm to model shallow water wave propagation on a rotating sphere in the framework of a fully nonlinear weakly dispersive model, which will be described in the following Section. For numerical illustrations we consider first model problems on the perfect sphere (\ie the bottom is even). In this way we assess the influence of dispersive, sphericity and rotation effects depending on the propagation distance and on the size of the wave generation region. These methods are implemented in \textsc{NLDSW\_sphere} code which is used to produce numerical results reported below.

The present manuscript is organized as follows. The governing equations that we tackle in our study are set in Section~\ref{sec:problem}. The numerical algorithm is described in Section~\ref{sec:num}. Several numerical illustrations are described in Section~\ref{sec:simus}. Namely, we start with tests over a perfect rotating sphere in Section~\ref{sec:flat}. Then, as an illustration of medium scale wave propagation we study the \textsc{Bulgarian} 2007 tsunami in Section~\ref{sec:bulgary}. On large trans-oceanic scales we simulate the 2010 \textsc{Chilean} tsunami in Section~\ref{sec:chile}. Finally in Section~\ref{sec:disc} we outline the main conclusions and perspectives of our study. Some further details on the derivation of the non-hydrostatic pressure equation are provided in Appendix~\ref{app:der}.


\section{Problem formulation}
\label{sec:problem}

The detailed derivation of the fully nonlinear model considered in the present study can be found in the previous Part~III \cite{Khakimzyanov2016a}. Here we only repeat the governing equations:
\begin{equation}\label{eq:base1}
  (\H\,R\,\sin\theta)_{\,t}\ +\ \bigl[\,\H\,u\,\bigr]_{\,\lambda}\ +\ \bigl[\,\H\,v\,\sin\theta\,\bigr]_{\,\theta}\ =\ 0\,,
\end{equation}
\begin{multline}\label{eq:base2}
  (\H\,u\,R\,\sin\theta)_{\,t}\ +\ \Bigl[\,\H\,u^2\ +\ g\;\frac{\H^{\,2}}{2}\,\Bigr]_{\,\lambda}\ +\ \bigl[\,\H\,u\,v\,\sin\theta\,\bigr]_{\,\theta}\ =\ g\,\H\,h_{\,\lambda}\\
  -\H\,u\,v\,\cos\theta\ -\ \digamma\,\H\,v\,R\,\sin\theta\ +\ \Pnh_{\,\lambda}\ -\ \pb\,h_{\,\lambda}\,,
\end{multline}
\begin{multline}\label{eq:base3}
  (\H\,v\,R\,\sin\theta)_{\,t}\ +\ \bigl[\,\H\,u\,v\,\bigr]_{\,\lambda}\ +\ \Bigl[\,\bigl(\H\,v^2\ +\ g\;\frac{\H^{\,2}}{2}\bigr)\,\sin\theta\,\Bigr]_{\,\theta}\ =\ g\,\H\,h_{\,\theta}\,\sin\theta\\
  +\ g\;\frac{\H^{\,2}}{2}\;\cos\theta\ +\ \H\,u^2\,\cos\theta\ +\ \digamma\,\H\,u\,R\,\sin\theta\ +\ \bigl(\Pnh_{\,\theta}\ -\ \pb\,h_{\,\theta}\bigr)\,\sin\theta\,,
\end{multline}
where $R$ is the radius of a virtual sphere rotating with a constant angular velocity $\Omega$ around the axis $O\,z$ of a fixed \textsc{Cartesian} coordinate system $O\,x\,y\,z\,$. The origin $O$ of this coordinate system is chosen so that the plane $O\,x\,y$ coincides with sphere's equatorial plane.

In order to describe conveniently the fluid flow we choose also a spherical coordinate system $O\,\lambda\,\theta\,r$ whose origin is located at sphere's center and it rotates with the sphere. Here $\lambda$ is the longitude increasing in the rotation direction starting from a certain meridian ($0\ \leq\ \lambda\ <\ 2\,\pi$). The other angle $\theta\ \eqdef\ \dfrac{\pi}{2}\ -\ \phi$ is the \emph{complementary latitude} ($-\dfrac{\pi}{2}\ <\ \phi\ <\ \dfrac{\pi}{2}$). Finally, $r$ is the radial distance from sphere's center. The \textsc{Newtonian} gravity force\footnote{Here we understand the force per unit mass, \ie the acceleration.} acts on fluid particles and its vector $\g$ is directed towards virtual sphere's center. The total water depth $\H\ \eqdef\ \eta\ +\ h\ >\ 0$ is supposed to be small comparing to sphere's radius, \ie $\H\ \ll\ R\,$. That is why we can suppose that the gravity acceleration $g\ \eqdef\ \abs{\g}$ and fluid density $\rho$ are constants throughout the fluid layer. The functions $h\,(\lambda,\,\theta,\,t)$ (the bottom profile) and $\eta\,(\lambda,\,\theta,\,t)$ (the free surface excursion) are given as deviations from the still water level $\eta_{\,0\,0}\,(\theta)\,$, which is not spherical due to the rotation effect \cite{Khakimzyanov2016a}.

By $u$ and $v$ we denote the linear components of the velocity vector:
\begin{equation*}
  u\ \eqdef\ R\,u^{\,1}\,\sin\theta\,, \qquad
  v\ \eqdef\ R\,u^{\,2}\,,
\end{equation*}
where $u^{\,1}\ =\ \dot{\lambda}$ and $u^{\,2}\ =\ \dot{\theta}\,$. The \textsc{Coriolis} parameter $\digamma\ \eqdef\ 2\,\Omega\,\cos\theta$ is expressed through the complementary latitude $\theta$ and additionally we can assume that
\begin{equation}\label{eq:1.4}
  \theta_{\,0}\ \leq\ \theta\ \leq\ \pi\ -\ \theta_{\,0}\,,
\end{equation}
where $\theta_{\,0}\ =\ \const\ \ll\ 1$ is a small angle. In other words, the poles are excluded from our computations. In practice, it is not a serious limitations since on the Earth poles are covered with ice and no free surface flow takes place there. The quantities $\Pnh$ and $\pb$ are dispersive components of the depth-integrated pressure $\Pp$ and fluid pressure at the bottom $\pc$ respectively:
\begin{equation*}
  \Pp\ =\ \frac{g\,\H^{\,2}}{2}\ -\ \Pnh\,, \qquad
  \pc\ =\ g\,\H\ -\ \pb\,.
\end{equation*}
These dispersive components $\Pnh$ and $\pb$ can be computed according to the following formulas \cite{Khakimzyanov2016a}:
\begin{equation}\label{eq:Pdef}
  \Pnh\ =\ \frac{\H^{\,3}}{3}\;\Rr_{\,1}\ +\ \frac{\H^{\,2}}{2}\;\Rr_{\,2}\,, \qquad
  \pb\ =\ \frac{\H^{\,2}}{2}\;\Rr_{\,1}\ +\ \H\,\Rr_{\,2}\,,
\end{equation}
where
\begin{equation*}
  \Rr_{\,1}\ \eqdef\ \Dd\,(\div\u)\ -\ (\div\u)^2\,, \qquad
  \Rr_{\,2}\ \eqdef\ \Dd^{\,2}\,h\,, \qquad
  \u\ \eqdef\ \bigl(u^{\,1},\,u^{\,2}\bigr)\,.
\end{equation*}
To complete model presentation we remind also the definitions of various operators in spherical coordinates that we use:
\begin{equation*}
  \Dd\ \eqdef\ \partial_{\,t}\ +\ \u\scal\grad\,, \qquad
  \grad\ \eqdef\ \bigl(\partial_{\,\lambda},\,\partial_{\,\theta}\bigr)\,, \qquad
  \u\scal\grad\ \equiv\ u^{\,1}\,\partial_{\,\lambda}\ +\ u^{\,2}\,\partial_{\,\theta}\,,
\end{equation*}
\begin{equation*}
  \div\u\ \equiv\ u^{\,1}_{\,\lambda}\ +\ \frac{1}{\J}\;\bigl(\,\J\,u^{\,2}\bigr)_{\,\theta}\,, \qquad
  \J\ \eqdef\ -R^{\,2}\,\sin\theta\,.
\end{equation*}
In the most detailed form, functions $\Rr_{1,\,2}$ can be equivalently rewritten as
\begin{align*}
  \Rr_{1}\ &\equiv\ (\div\u)_{\,t}\ +\ \frac{1}{R\,\sin\theta}\;\Bigl[\,u\,(\div\u)_{\,\lambda}\ +\ v\,(\div\u)_{\,\theta}\,\sin\theta\,\Bigr]\ -\ (\div\u)^{\,2}\,, \\
  \Rr_{2}\ &\equiv\ (\Dd\,h)_{\,t}\ +\ \frac{1}{R\,\sin\theta}\;\Bigl[\,u\,(\Dd\,h)_{\,\lambda}\ +\ v\,(\Dd\,h)_{\,\theta}\,\sin\theta\,\Bigr]\,,
\end{align*}
with
\begin{align*}
  \div\u\ &\equiv\ \frac{1}{R\,\sin\theta}\;\Bigl[\,u_{\,\lambda}\ +\ (\,v\,\sin\theta)_{\,\theta}\,\Bigr]\,, \\
  \Dd\,h\ &\equiv\ h_{\,t}\ +\ \frac{1}{R\,\sin\theta}\;\Bigl[\,u\,h_{\,\lambda}\ +\ v\,h_{\,\theta}\,\sin\theta\,\Bigr]\,.
\end{align*}

The model \eqref{eq:base1} -- \eqref{eq:base3} is referred to as `fully nonlinear' one since it was derived without any simplifying assumptions on the wave amplitude \cite{Khakimzyanov2016a}. In other words, all nonlinear (but weakly dispersive) terms are kept in this model. So far we shall refer to this model as Fully Nonlinear Weakly Dispersive (FNWD) model. The FNWD model should be employed to simulate water wave propagation in coastal and even in slightly deeper regions over uneven bottoms. Weak dispersive effects in the FNWD model ensure that we shall obtain more accurate results than with simple NSWE. Moreover, the FNWD model contains the terms coming from moving bottom effects \cite{Kervella2007}. Consequently, we can model also the wave generation process by fast or slow bottom motions \cite{Dutykh2007b}, thus, allowing to model tsunami waves from their generation until the coasts \cite{Dutykh2009a}. In this way we extend the validity region of existing WNWD models \cite{Glimsdal2013, Lovholt2008,Lovholt2010} by including the wave generation regions along with the coasts where the nonlinearity becomes critical.

Concerning the \emph{linear} dispersive properties, it is generally believed that models with the depth-averaged velocity can be further improved in the sense of \textsc{Bona}--\textsc{Smith} \cite{BS} and \textsc{Nwogu} \cite{Nwogu1993}. However, nonlinear dispersive wave models tweaked in this way (see \eg \cite{Lynett2002}) have a clear advantage only in the linear one-dimensional (1D) situations. For nonlinear 3D computations (especially involving the moving bottom \cite{Dutykh2006}) the advantage of `improved' models becomes more obscure comparing to dispersive wave models with the depth-averaged velocity adopted in our study \cite{Chubarov2005, Shokin2007}. Moreover, the mathematical model after such transformations (or `improvements' as they are called in the literature) often looses the energy conservation and \textsc{Galilean}'s invariance properties. A successful attempt in this direction was achieved only recently \cite{Clamond2015c}.

Equations \eqref{eq:base1} -- \eqref{eq:base3} admit also a non-conservative form:
\begin{align}\label{eq:nc1}
  \H_{\,t}\ +\ \frac{1}{R\,\sin\theta}\;\Bigl[\,(\H\,u)_{\,\lambda}\ +\ (\H\,v\,\sin\theta)_{\,\theta}\,\Bigr]\ &=\ 0\,, \\
  u_{\,t}\ +\ \frac{1}{R\,\sin\theta}\;u\,u_{\,\lambda}\ +\ \frac{1}{R}\;v\,u_{\,\theta}\ +\ \frac{g}{R\,\sin\theta}\;\eta_{\,\lambda}\ &=\ \frac{1}{R\,\sin\theta}\;\frac{\Pnh_{\,\lambda}\ -\ \pb\,h_{\,\lambda}}{\H}\ -\ \frac{u\,v}{R}\;\cot\theta\ -\ \digamma\,v\,, \label{eq:nc2} \\
  v_{\,t}\ +\ \frac{1}{R\,\sin\theta}\;u\,v_{\,\lambda}\ +\ \frac{1}{R}\;v\,v_{\,\theta}\ +\ \frac{g}{R}\;\eta_{\,\theta}\ &=\ \frac{\Pnh_{\,\theta}\ -\ \pb\,h_{\,\theta}}{R\,\H}\ +\ \frac{u^{\,2}}{R}\,\cot\theta\ +\ \digamma\,u\,. \label{eq:nc3}
\end{align}
In the numerical algorithm presented below we use both conservative and non-conservative forms for our convenience.

If in conservative \eqref{eq:base1} -- \eqref{eq:base3} or non-conservative \eqref{eq:nc1} -- \eqref{eq:nc3} governing equations we neglect dispersive contributions, \ie $\Pnh\ \rightsquigarrow\ 0\,$, $\pb\ \rightsquigarrow\ 0\,$, then we recover NSWE on a rotating attracting sphere \cite{Cherevko2009a}:
\begin{align*}
  \H_{\,t}\ +\ \frac{1}{R\,\sin\theta}\;\Bigl[\,(\H\,u)_{\,\lambda}\ +\ (\H\,v\,\sin\theta)_{\,\theta}\,\Bigr]\ &=\ 0\,, \\
  u_{\,t}\ +\ \frac{1}{R\,\sin\theta}\;u\,u_{\,\lambda}\ +\ \frac{1}{R}\;v\,u_{\,\theta}\ +\ \frac{g}{R\,\sin\theta}\;\eta_{\,\lambda}\ &=\ -\ \frac{u\,v}{R}\;\cot\theta\ -\ \digamma\,v\,, \\ 
  v_{\,t}\ +\ \frac{1}{R\,\sin\theta}\;u\,v_{\,\lambda}\ +\ \frac{1}{R}\;v\,v_{\,\theta}\ +\ \frac{g}{R}\;\eta_{\,\theta}\ &=\ \frac{u^{\,2}}{R}\,\cot\theta\ +\ \digamma\,u\,. 
\end{align*}

In order to obtain a well-posed problem, we have to prescribe initially the free surface deviation from its equilibrium position along with the initial velocity field. Moreover, the appropriate boundary conditions have to be prescribed as well. The curvilinear coast-line is approximated by a family of closed polygons and on edges $e$ we prescribe the wall boundary condition:
\begin{equation}\label{eq:bc}
  \u\scal\n\,\bigr\rvert_{\,e}\ =\ 0\,,
\end{equation}
where $\n$ is an exterior normal to edge $e\,$. In this way, we replace the wave run-up problem by wave/wall interaction, where a solid wall is located along the shoreline on a certain prescribed depth $\hw\,$.


\section{Numerical algorithm}
\label{sec:num}

The system of equations \eqref{eq:base1} -- \eqref{eq:base3} is not of \textsc{Cauchy}--\textsc{Kovalevskaya}'s type, since momentum balance equations \eqref{eq:base2}, \eqref{eq:base3} involve mixed derivatives with respect to time and space of the velocity components $u\,$, $v\,$. We already encountered this difficulty in the globally flat case \cite{Khakimzyanov2016}. A direct (\eg finite difference) approximation of governing equations would lead to a complex system of fully coupled nonlinear algebraic equations in a very high dimensional space. In the globally flat case \cite{Khakimzyanov2016c} it was found out that it is more judicious to perform a preliminary decoupling\footnote{We underline that this decoupling procedure involves no approximation and the resulting system of equations is completely equivalent to the base model \eqref{eq:base1} -- \eqref{eq:base3}.} of the system into a scalar elliptic equation and a system of hyperbolic equations with source terms \cite{Khakimzyanov2016}. In the present work we realize the same idea for FNWD equations \eqref{eq:base1} -- \eqref{eq:base3} on a rotating attracting sphere. Similarly we shall rewrite the governing equations as a scalar elliptic equation to determine the dispersive component of the depth-integrated pressure $\Pnh$ and a hyperbolic system of shallow water type equations with some additional source terms. With this splitting we can apply the most appropriate numerical methods for elliptic and hyperbolic problems correspondingly.

The derivation of the elliptic equation for non-hydrostatic pressure component $\Pnh$ can be found in Appendix~\ref{app:der}. Here we provide only the final result:
\begin{equation}\label{eq:ell}
  \biggl[\,\frac{1}{\sin\theta}\;\Bigl\{\,\frac{\Pnh_{\,\lambda}}{\H}\ -\ \frac{\grad\Pnh\scal\grad h}{\H\,\r}\;h_{\,\lambda}\,\Bigr\}\,\biggr]_{\,\lambda}\ +\ \biggl[\,\Bigl\{\,\frac{\Pnh_{\,\theta}}{\H}\ -\ \frac{\grad\Pnh\scal\grad h}{\H\,\r}\;h_{\,\theta}\,\Bigr\}\,\sin\theta\,\biggr]_{\,\theta}\ -\ \Ko\,\Pnh\ =\ \F\,,
\end{equation}
where
\begin{equation*}
  \Ko\ \eqdef\ \K_{\,0\,0}\ +\ \pd{\K_{\,0\,1}}{\lambda}\ +\ \pd{\K_{\,0\,2}}{\theta}\,,
\end{equation*}
with
\begin{equation*}
  \K_{\,0\,0}\ \eqdef\ R^{\,2}\;\frac{12\,(\r\ -\ 3)}{\H^{\,3}\,\r}\;\sin\theta\,, \qquad 
  \K_{\,0\,1}\ \eqdef\ \frac{6\,h_{\,\lambda}}{\H^{\,2}\,\r\,\sin\theta}\,, \qquad
  \K_{\,0\,2}\ \eqdef\ \frac{6\,h_{\,\theta}}{\H^{\,2}\,\r}\;\sin\theta\,.
\end{equation*}
Here we introduced a new variable $\r$ defined as
\begin{equation*}
  \r\ \eqdef\ 4\ +\ \grad h\scal\grad h\ \equiv\ 4\ +\ \abs{\grad h}^{\,2}\,.
\end{equation*}
The scalar product $\grad\Pnh\scal\grad h$ can be easily expressed as
\begin{equation*}
  \grad\Pnh\scal\grad h\ \equiv\ \frac{1}{R^{\,2}}\;\biggl\{\,\frac{\Pnh_{\,\lambda}\,h_{\,\lambda}}{\sin^{\,2}\theta}\ +\ \Pnh_{\,\theta}\,h_{\,\theta}\,\biggr\}\,.
\end{equation*}
The right hand side $\F$ is defined as
\begin{multline*}
  \F\ \eqdef\ \biggl[\,\underbrace{\frac{1}{\sin\theta}\;\Bigl\{\,g\,\eta_{\,\lambda}\ +\ \frac{\Qq}{\r}\;h_{\,\lambda}\ -\ \La_{\,1}\,\Bigr\}}_{\displaystyle{\defeq\ \Fd_{\,1}}}\,\biggr]_{\,\lambda}\ +\ \biggl[\,\underbrace{\Bigl\{\,g\,\eta_{\,\theta}\ +\ \frac{\Qq}{\r}\;h_{\,\theta}\ -\ \La_{\,2}\,\Bigr\}\,\sin\theta}_{\displaystyle{\defeq\ \Fd_{\,2}}}\,\biggr]_{\,\theta}\\ 
  -\ R^{\,2}\;\frac{6\,\Qq}{\H\,\r}\;\sin\theta\ + \ \frac{2}{\sin\theta}\;\Bigl\{\,u_{\,\lambda}\ +\ (\,v\,\sin\theta)_{\,\theta}\,\Bigr\}^{\,2}\\ 
  -\ 2\,\bigl(u_{\,\lambda}\,v_{\,\theta}\ -\ v_{\,\lambda}\,u_{\,\theta}\bigr)\ -\ 2\,(u\,v)_{\,\lambda}\cot\theta\ -\ \bigl(\,v^{\,2}\,\cos\theta\bigr)_{\,\theta}\,,
\end{multline*}
where
\begin{multline*}
  \Qq\ \eqdef\ \bigl(\Lab\ -\ g\,\grad\eta\bigr)\scal\grad h\ +\ \frac{1}{R^{\,2}\,\sin\theta}\;\Bigl\{\,\frac{u^{\,2}}{\sin\theta}\;h_{\,\lambda\,\lambda}\ +\ 2\,u\,v\,h_{\,\lambda\,\theta}\ +\ v^{\,2}\,h_{\,\theta\,\theta}\,\sin\theta\,\Bigr\}\\
  +\ \underbrace{h_{\,t\,t}\ +\ 2\,\Bigl\{\,\frac{u}{R\,\sin\theta}\;h_{\,\lambda\,t}\ +\ \frac{v}{R}\;h_{\,\theta\,t}\,\Bigr\}}_{\displaystyle{\defeq\ \B}}\,,
\end{multline*}
with vector $\Lab\ \eqdef\ {}^{\top}\bigl(\La_{\,1},\,\La_{\,2}\bigr)$ whose components are
\begin{equation*}
  \La_{\,1}\ \eqdef\ -\bigl(\,2\,u\,v\,\cot\theta\ +\ \digamma\,v\,R\bigr)\,\sin\theta\,, \qquad
  \La_{\,2}\ \eqdef\ u^{\,2}\,\cot\theta\ +\ \digamma\,u\,R\,.
\end{equation*}
The term $\B$ contains all the terms coming from bottom motion effects. If the bottom is stationary, then $\B\ \equiv\ 0\,$.

The particularity of equation \eqref{eq:ell} is that it does not contain time derivatives of dynamic variables $\H\,$, $u$ and $v\,$. This equation is very similar to the elliptic equation derived in the globally flat case \cite{Khakimzyanov2016}. The differences consist only in terms coming from Earth's sphericity and rotation effects. It is not difficult to show that under total water depth positivity assumption $\H\ >\ 0$ and condition \eqref{eq:1.4}, equation \eqref{eq:ell} is uniformly elliptic. In order to have the uniqueness result of the \textsc{Dirichlet} problem for \eqref{eq:ell}, the coefficient $\Ko$ has to be positive defined \cite{Ladyzhenskaya1973}, \ie $\Ko\ >\ 0\,$. In the case of an even bottom (\ie $h\ \equiv\ h_{\,0}\ =\ \const$) the positivity condition takes the following form:
\begin{equation}\label{eq:well}
  \Ko\ =\ \frac{12\,R^{\,2}\,\sin\theta}{\H^{\,3}\,\rro}\;\biggl[\,\rro\ -\ 3\ -\ \frac{\Omega^{\,2}}{2\,g}\;\Bigl\{\,\H\,\sin^2\theta\ -\ \eta_{\,\theta}\sin(\,2\,\theta)\ +\ \frac{8\,\H}{\rro}\;\cos(\,2\,\theta)\,\Bigr\}\,\biggr]\ >\ 0\,,
\end{equation}
where this time the variable $\rro$ is defined simply as
\begin{equation*}
  \rro\ \eqdef\ 4\ +\ \frac{\Omega^{\,4}\,R^{\,2}}{4\,g^{\,2}}\;\sin^2(\,2\,\theta\,)\,.
\end{equation*}
Using the values of parameters for our planet, \ie
\begin{equation*}
  R\ =\ 6.38\times 10^{\,6}\;\m\,, \qquad
  \Omega\ =\ 7.29\times 10^{\,-5}\;\s^{\,-1}\,, \qquad
  g\ =\ 9.81\;\;\frac{\m}{\s^{\,2}}\,,
\end{equation*}
and assuming \eqref{eq:1.4} along with the fact that the gradient of the free surface elevation $\eta$ is bounded, one can show that $\Ko\ \gg\ 1\,$. Thus, the condition \eqref{eq:well} is trivially verified. In this case one can construct a finite difference operator with positive definite (grid-)operator. Theoretically, it is not excluded that for large bottom variations locally the coefficient $\Ko$ might become negative. In such cases, the conditioning of the discrete system is worsened and more iterations are needed to achieve the desired accuracy in solving equation \eqref{eq:ell}. In the globally flat case the analysis of these cases was performed in \cite{Khakimzyanov2016}. In practice, sometimes it is possible to avoid such complications by applying a prior smoothing operator to the bathymetry function $h\,(\lambda,\,\theta,\,t)\,$.


\subsection{Numerical scheme construction}

The finite difference counterpart of equation \eqref{eq:ell} can be obtained using the so-called integro-interpolation method \cite{Khakimzyanov2001}. In this Section we construct a second order approximation to be able to work on coarser grids for a fixed desired accuracy.

\begin{figure}
  \centering
  \subfigure[]{\includegraphics[width=0.32\textwidth]{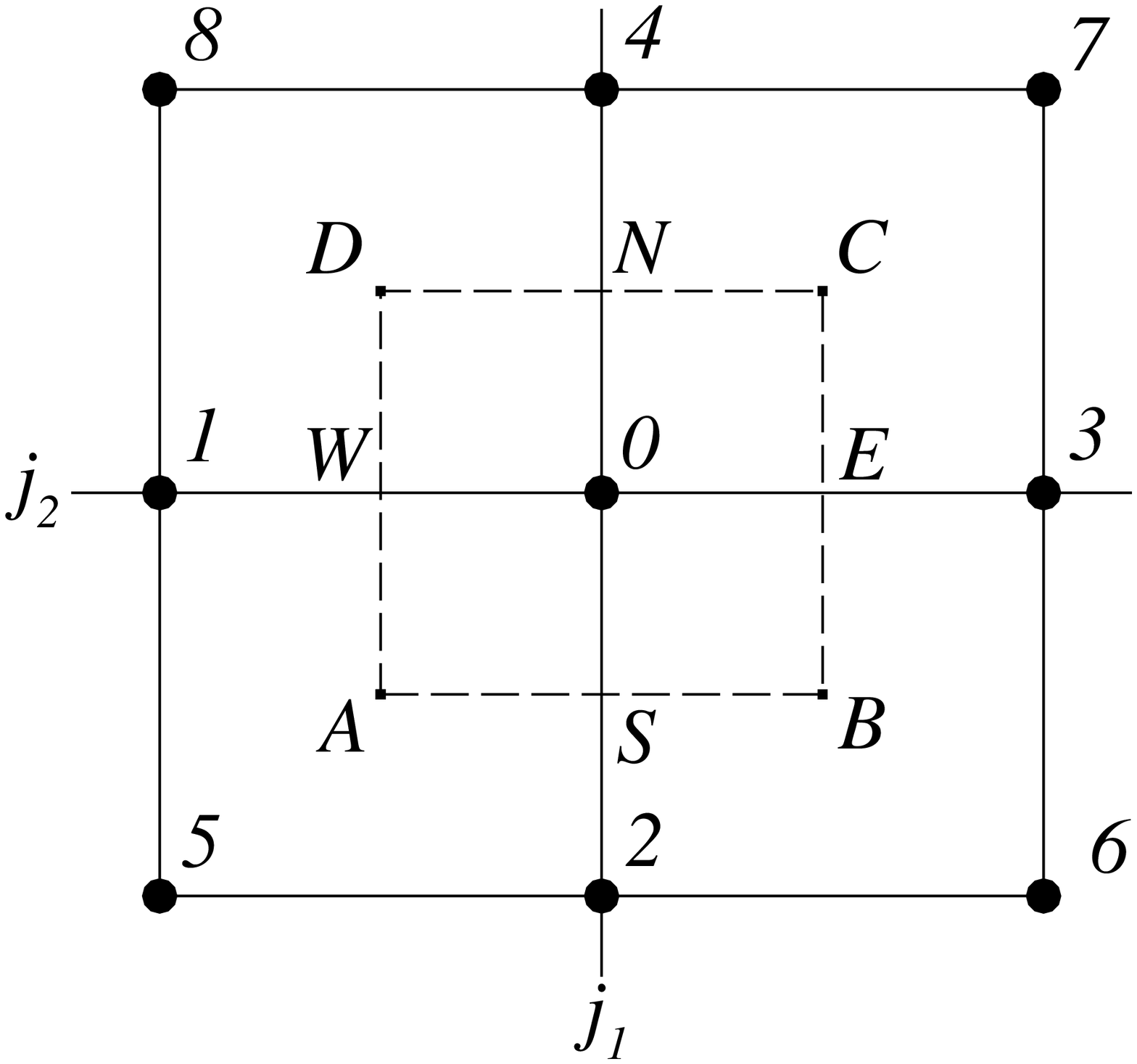}}
  \subfigure[]{\includegraphics[width=0.32\textwidth]{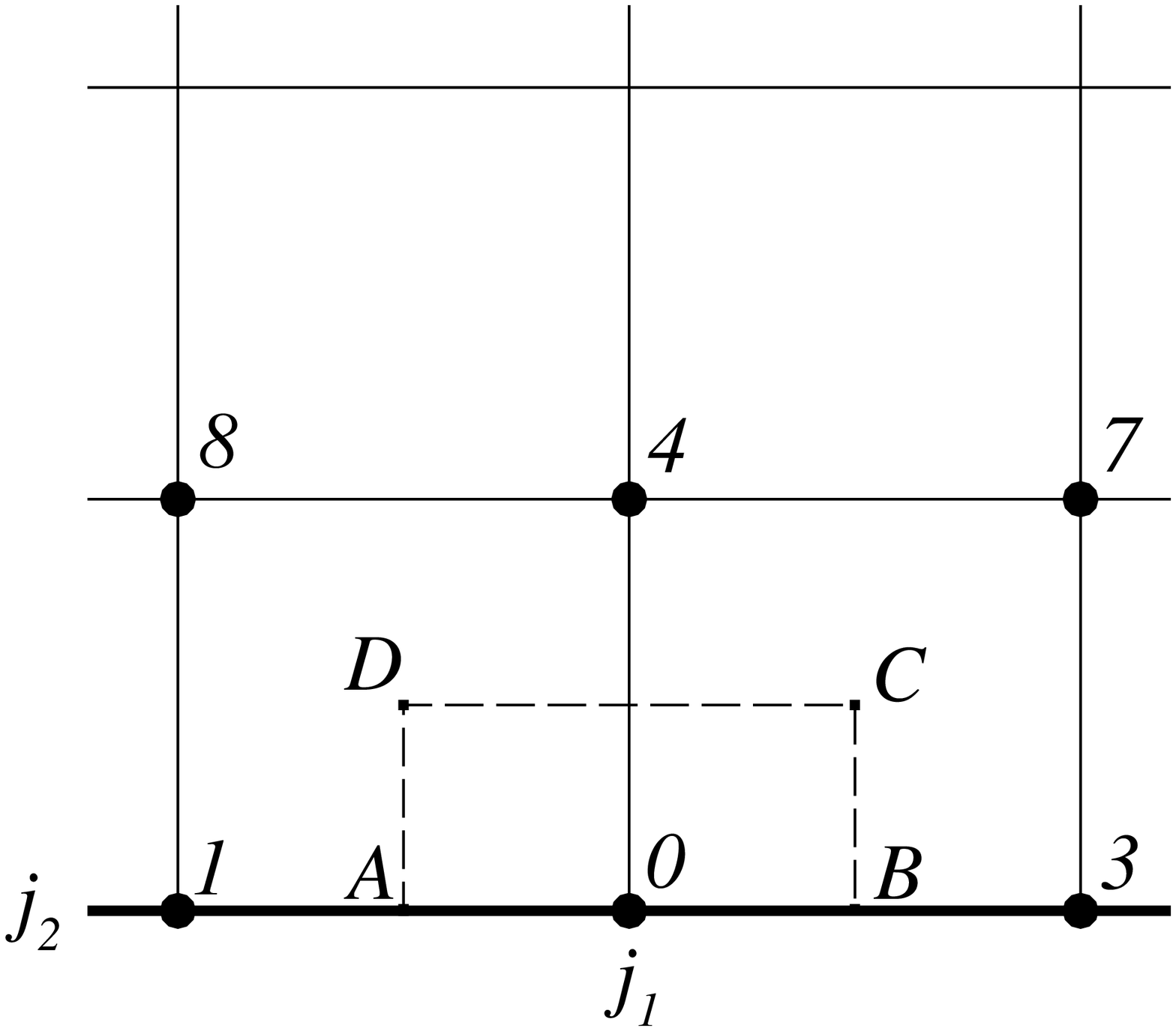}}
  \subfigure[]{\includegraphics[width=0.32\textwidth]{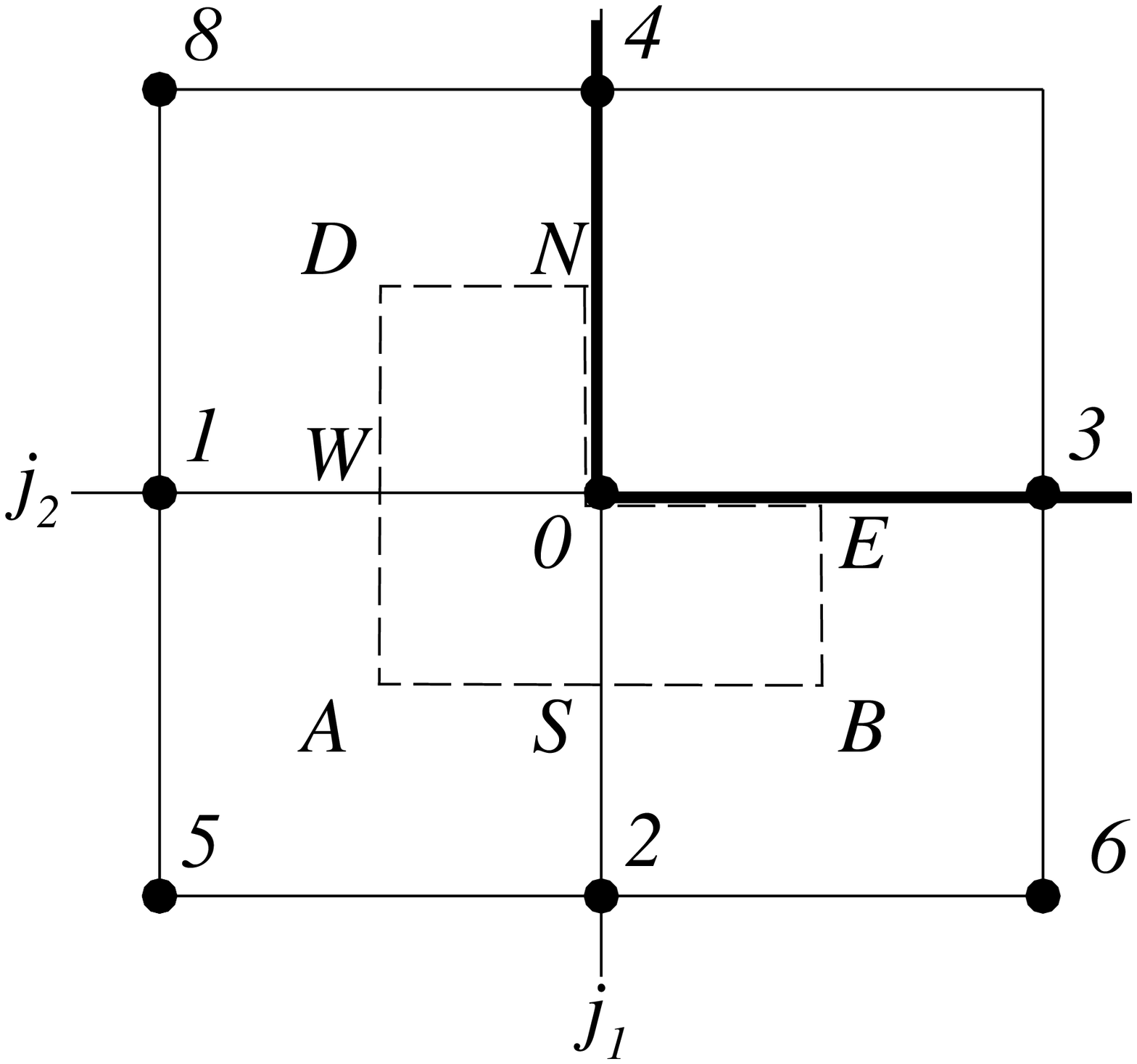}}
  \caption{\small\em Integration contours and finite difference stencils for equation \eqref{eq:ell} in an internal (a), boundary (b) and corner (c) nodes.}
  \label{fig:stencils}
\end{figure}

In spherical coordinates we consider a uniform grid with spacings $\Delta_{\,\lambda}$ and $\Delta_{\,\theta}$ along the axes $O\,\lambda$ and $O\,\theta$ correspondingly. In order to derive a numerical scheme, first we integrate equation \eqref{eq:ell} over a rectangle $A\,B\,C\,D$ depicted in Figure~\ref{fig:stencils}(\textit{a}), with vertices $A\,$, $B\,$, $C\,$ and $D$ being the geometrical centers of adjacent cells. After applying \textsc{Green}'s formula to the resulting integral, we obtain:
\begin{multline}\label{eq:2.7}
  \sqint_{\,BC}\,\Ff^{\,1}\,\ud\theta\ -\ \sqint_{\,AD}\,\Ff^{\,1}\,\ud\theta\ +\ \sqint_{\,DC}\,\Ff^{\,2}\,\ud\lambda\ -\ \sqint_{\,AB}\,\Ff^{\,2}\,\ud\lambda\\
  -\ \sqiint_{\,ABCD}\,\Ko\,\Pnh\;\ud\lambda\,\ud\theta\ =\ \sqiint_{\,ABCD}\,\F\;\ud\lambda\,\ud\theta\,.
\end{multline}
From the double integral on the right hand side we can extract a contour part as well:
\begin{equation*}
  \sqiint_{\,ABCD}\,\divf\Fdb\;\ud\lambda\,\ud\theta\ =\ \sqint_{\,BC}\,\Fd_{\,1}\,\ud\theta\ -\ \sqint_{\,AD}\,\Fd_{\,1}\,\ud\theta\ +\ \sqint_{\,DC}\,\Fd_{\,2}\,\ud\lambda\ -\ \sqint_{\,AB}\,\Fd_{\,2}\,\ud\lambda\,,
\end{equation*}
where $\divf\vec{(\,\cdot\,)}\ \eqdef\ (\,\cdot_{\,1})_{\,\lambda}\ +\ (\,\cdot_{\,2})_{\,\theta}$ is the `flat' divergence operator, vectors $\Fdb\ \eqdef\ \bigl(\,\Fd_{\,1},\,\Fd_{\,2}\bigr)\,$, $\bigl(\,\Ff^{\,1},\,\Ff^{\,2}\bigr)$ with components defined as
\begin{align*}
  \Ff^{\,1}\ &=\ \frac{1}{\sin\theta}\;\biggl\{\,\frac{\Pnh_{\,\lambda}}{\H}\ -\ \frac{\grad\Pnh\scal\grad h}{\H\,\r}\;h_{\,\lambda}\,\biggr\}\,, \qquad
  \Ff^{\,2}\ &=\ \biggl\{\,\frac{\Pnh_{\,\theta}}{\H}\ -\ \frac{\grad\Pnh\scal\grad h}{\H\,\r}\;h_{\,\theta}\,\biggr\}\,\sin\theta\,, \\
  \Fd_{\,1}\ &=\ \frac{1}{\sin\theta}\;\biggl\{\,g\,\eta_{\,\lambda}\ +\ \frac{\Qq}{\r}\;h_{\,\lambda}\ -\ \La_{\,1}\,\biggr\}\,, \qquad 
  \Fd_{\,2}\ &=\ \biggl\{\,g\,\eta_{\,\theta}\ +\ \frac{\Qq}{\r}\;h_{\,\theta}\ -\ \La_{\,2}\,\biggr\}\,\sin\theta\,.
\end{align*}
In order to compute approximatively the integrals, we use the trapezoidal numerical quadrature rule along with a second order finite difference approximation of the derivatives in cell centers \cite{Gusev2014}. As a result one can obtain a finite difference approximation for equation \eqref{eq:ell} with nine points stencil in every internal node of the grid.

In an analogous way one can construct finite difference approximations in boundary nodes adjacent to a non-permeable wall. Let us assume that the node $\x_{\,j_{\,1},\,j_{\,2}}\ =\ \bigl(\lambda_{\,j_{\,1}},\,\theta_{\,j_{\,2}}\bigr)$ belongs to the boundary $\Gamma$ lying along a parallel represented with a bold solid line in Figure~\ref{fig:stencils}(\textit{b}). The interior of the computational domain lies in North-ward direction to this parallel. If nodes $\x_{\,j_{\,1}-1,\,j_{\,2}}$ and $\x_{\,j_{\,1}+1,\,j_{\,2}}$ also belong to $\Gamma\,$, then, the integration contour $A\,B\,C\,D$ is chosen such that vertices $C$ and $D$ coincide with adjacent cell centers (sharing the common node $\x_{\,j_{\,1},\,j_{\,2}}$) and vertices $A\,$, $B$ coincide with centers of edges belonging to $\Gamma$ (see again Figure~\ref{fig:stencils}(\textit{b})). In this case the rectangle side $A\,B$ belongs completely to the boundary $\Gamma\,$. That is why boundary conditions for the function $\Pnh$ have to be used while computing the integrals in the integro-differential equation \eqref{eq:2.7}.


\subsubsection{Boundary conditions treatment}
\label{sec:bc}

In the present study we consider only the case of wall boundary conditions. In the situation depicted in Figure~\ref{fig:stencils}(\textit{b}) the boundary condition \eqref{eq:bc} becomes simply
\begin{equation*}
  v\,\bigr\rvert_{\,\Gamma}\ \equiv\ 0\,.
\end{equation*}
Thanks to equation \eqref{eq:nc3}, we obtain the following boundary condition for the variable $\Pnh\,$:
\begin{equation}\label{eq:bcv}
  \frac{1}{R}\;\biggl\{\,\frac{\Pnh_{\,\theta}\ -\ \pb\,h_{\,\theta}}{\H}\ -\ g\,\eta_{\,\theta}\,\biggr\}\ +\ \frac{u^{\,2}}{R}\;\cot\theta\ +\ \digamma\,u\ =\ 0\,, \qquad \x\ \in\ \Gamma\,.
\end{equation}
In Appendix~\ref{app:der} it is shown that the dispersive component of the fluid pressure at the bottom $\pb$ is related to other quantities $\H\,$, $u\,$, $v$ and $\Pnh$ as
\begin{equation}\label{eq:pbeq}
  \pb\ =\ \frac{1}{\r}\;\biggl\{\,\frac{6\,\Pnh}{\H}\ +\ \H\,\Qq\ +\ \grad\Pnh\scal\grad h\,\biggr\}\,.
\end{equation}
After substituting the last formula into equation \eqref{eq:bcv}, the required boundary condition takes the form:
\begin{equation*}
  \biggl\{\,\frac{1}{R}\;\Bigl(\,\frac{\Pnh_{\,\theta}}{\H}\ -\ \frac{\grad\Pnh\scal\grad h}{\H\,\r}\;h_{\,\theta}\Bigr)\ -\ \frac{6\,h_{\,\theta}}{R\,\H^{\,2}\,\r}\;\Pnh\,\biggr\}\,\biggr\rvert_{\,\Gamma}\ =\ \biggl\{\,\frac{1}{R}\;\Bigl(g\,\eta_{\,\theta}\ +\ \frac{\Qq}{\r}\;h_{\,\theta}\Bigr)\ -\ \frac{u^{\,2}}{R}\;\cot\theta\ -\ \digamma\,u\,\biggr\}\,\biggr\rvert_{\,\Gamma}\,,
\end{equation*}
or by using some notations introduced above we can simply write a more compact form:
\begin{equation}\label{eq:bc3}
  \bigl\{\,\Ff^{\,2}\ -\ \K_{\,0\,2}\,\Pnh\ -\ \Fd_{\,2}\,\bigr\}\,\bigr\rvert_{\,A\,B}\ =\ 0\,.
\end{equation}
The last boundary condition is used while approximating the integrals over the rectangle side $A\,B\,$. Consider the double integral in the left hand side of the integro-differential equation \eqref{eq:2.7}. It can be approximated as
\begin{multline*}
  \sqiint_{\,ABCD}\,\Ko\,\Pnh\;\ud\lambda\,\ud\theta\ \approx\ \sqiint_{\,ABCD}\,\K_{\,0\,0}\,\Pnh\;\ud\lambda\,\ud\theta\ +\\
  \biggl\{\,\sqint_{\,BC}\,\K_{\,0\,1}\,\ud\theta\ -\ \sqint_{\,AD}\,\K_{\,0\,1}\,\ud\theta\ +\ \sqint_{\,DC}\,\K_{\,0\,2}\,\ud\lambda\ -\ \sqint_{\,AB}\,\K_{\,0\,2}\,\ud\lambda\,\biggr\}\,\Pnh(\vO)\,,
\end{multline*}
and after introducing the following approximations:
\begin{equation*}
  \sqint_{\,AB}\,\Ff^{\,2}\,\ud\lambda\ \approx\ \Ff^{\,2}(\vO)\,\Delta_{\,\lambda}\,, \qquad
  \sqint_{\,AB}\,\K_{\,0\,2}\,\ud\lambda\ \approx\ \K_{\,0\,2}(\vO)\,\Delta_{\,\lambda}\,, \qquad
  \sqint_{\,AB}\,\Fd_{\,2}\,\ud\lambda\ \approx\ \Fd_{\,2}(\vO)\,\Delta_{\,\lambda}\,,
\end{equation*}
we come to an important conclusion: \emph{all the terms coming from integrals $\sqint_{\,AB}$ vanish thanks to the boundary condition \eqref{eq:bc3}.} Consequently, the implementation of boundary conditions turns out to be trivial in the integro-interpolation method employed in our study. For boundary cells of other types, we use the same method of integro-interpolating approximations to obtain difference equations. As an illustration, in Figure~\ref{fig:stencils}(\textit{c}) we depict another one of the eight possible configurations of angular nodes. Since we write a difference equation for every grid node (interior and boundary), the total number of equations coincides with the total number of discretization points.


\subsubsection{Computational miscellanea}

In the most general case a realistic computational domain for the wave propagation has a complex shape. Generally it is not convex and due to the existence of islands it might be multiply connected. It has some implications for linear solvers that we can use to solve the difference equations described above. First of all, due to the large scale nature of problems considered in this study, we privilege iterative schemes for the sake of computational efficiency. Then, due to geometrical and topological reasons described above, we adopt a simple but efficient method of Successive Over--Relaxation (SOR) \cite{Young1950}. This method contains a free parameter $\varpi$, which can be used to accelerate the convergence. The optimal value of $\varpi^{\,\star}\,$, which ensures the fastest convergence is in general unknown. This question was studied theoretically for the \textsc{Poisson} equation with \textsc{Dirichlet} boundary conditions in a rectangle \cite{Samarskii2001}. So, in this case the optimal value of $\varpi^{\,\star}$ was shown to belong to the interval $(1,\,2)\,$. For example, if the mesh is taken uniform in each side $\ell$ of a square with the spacing $h\ =\ \frac{\ell}{N}\,$, then
\begin{equation*}
  \varpi^{\,\star}\ =\ \frac{2}{1\ +\ \sin\;\frac{\pi\,h}{\ell}}\,.
\end{equation*}
If we take $\ell\ =\ 1$ and $N\ =\ 30$ we obtain the optimal value $\varpi^{\,\star}\ \approx\ 1.81\,$. Another observation is that in the limit $N\ \to\ +\infty$ the optimal value $\varpi^{\,\star}\ \to\ 2\,$. In our numerical experiments we observed the same tendency: with the mesh refinement the optimal relaxation parameter $\varpi^{\,\star}$ for the discretized equation \eqref{eq:ell} approaches $2$ as well. In practice, we took the values $\varpi^{\,\star}\ \in\ [\,1.85,\,1.95\,]$ depending on the degree of refinement.

For given functions $\Pnh$ and $\pb$ the system of conservative equations \eqref{eq:base1} -- \eqref{eq:base3} is of hyperbolic type under the conditions \eqref{eq:1.4} and water depth positivity $\H\,(\x,\,t)\ >\ 0\,$, $\forall t\ >\ 0\,$. Thanks to this property we have in our disposal the whole arsenal of numerical tools that have been developed for hyperbolic systems of equations \cite{Godunov1987, Godlewski1990}. In the one-dimensional case we opted for predictor--corrector schemes \cite{Khakimzyanov2016d, Khakimzyanov2016} with a free scheme parameter $\theta_{\,j_{\,1},\,j_{\,2}}^{\,n}\,$. A judicious choice of this parameter ensures the TVD property and the monotonicity of solutions at least for scalar equations \cite{Khakimzyanov2015a}. Some aspects of predictor--corrector schemes in two spatial dimensions are described in \cite{Shokina2012}. In the present study we employ the predictor--corrector scheme with $\theta_{\,j_{\,1},\,j_{\,2}}^{\,n}\ \equiv\ 0\,$, which minimizes the numerical dissipation (and makes the scheme somehow more fragile). This choice is not probably the best for hyperbolic NSWE, but for non-hydrostatic FNWD models it works very well due to the inherent dispersive regularization property of solutions. Moreover, on the predictor stage we compute directly the quantities $\H\,$, $u$ and $v$ (instead of computing the fluxes) since they are needed to compute the coefficients along with the right hand side $\F$ of equation \eqref{eq:ell}.

Let us describe briefly the numerical algorithm we use to solve the extended system of equations \eqref{eq:base1} -- \eqref{eq:base3}, \eqref{eq:ell} (for more details see also \cite{Gusev2014, Khakimzyanov2016}). At the initial moment of time we are given by the free surface elevation $\eta\,(\x,\,0)$ and velocity vector $\u\,(\x,\,0)\,$. Moreover, if the bottom is not static, additionally we have to know also the quantities $h_{\,t}\,(\x,\,0)$ and $h_{\,t\,t}\,(\x,\,0)\,$. These data suffice to determine the initial distribution of the depth-integrated pressure $\Pnh$ by solving numerically the elliptic equation \eqref{eq:ell}. Finally, the dispersive pressure component on the bottom $\pb$ is computed by a finite difference analogue of equation \eqref{eq:pbeq}. By recurrence, let us assume to we know the same data on the $n$\up{th} time layer $t\ =\ t^{\,n}\,$: $\H^{\,n}\,$, $\u^{\,n}\,$, $\Pnh^{\,n}$ and $\pb^{\,n}\,$. Then, we employ the predictor--corrector scheme, each time step of this scheme consists of two stages. On the predictor stage we compute the quantities $\H^{\,n+\frac{1}{2}}\,$, $\u^{\,n+\frac{1}{2}}$ in cell centers as a solution of explicit discrete counterparts of equations \eqref{eq:nc1} -- \eqref{eq:nc3} (with right hand sides taken from the time layer $t^{\,n}$). Then, one solves a difference equation to determine $\Pnh^{\,n+\frac{1}{2}}\,$. The coefficients and the right hand side are evaluated using new values $\H^{\,n+\frac{1}{2}}$ and $\u^{\,n+\frac{1}{2}}\,$. From formula \eqref{eq:pbeq} one infers the value of $\pb^{\,n+\frac{1}{2}}\,$. All the values computed at the predictor stage $\H^{\,n+\frac{1}{2}}\,$, $\u^{\,n+\frac{1}{2}}\,$, $\Pnh^{\,n+\frac{1}{2}}$ and $\pb^{\,n+\frac{1}{2}}$ are then used at the corrector stage to determine the new values $\H^{\,n+1}$ and $\u^{\,n+1}\,$. At the corrector stage we employ the conservative form of equations \eqref{eq:base1} -- \eqref{eq:base3}. In the very last step we compute also the values of $\Pnh^{\,n+1}$ and $\pb^{\,n+1}\,$. The algorithm described becomes a spherical analogue of the well-known \textsc{Lax}--\textsc{Wendroff} scheme if one neglects the dispersive terms.

An important property of the proposed numerical algorithm is its well-balanced character if the bottom is steady (\ie $\B\ \equiv\ 0$) and the sphere is not too deformed (\ie $\Ko\ >\ 0$). In other words, it preserves exactly the so-called `lake-at-rest' states where the fluid is at rest $\u^{\,n}\ \equiv\ \vO$ and the free surface is unperturbed $\eta^{\,n}\ \equiv\ 0\,$. Then, it can be rigorously shown that this particular state will be preserved in the following layer $t\ =\ t^{\,n+1}$ as well. It is achieved by balanced discretizations of left and right hand sides in the momentum equations \eqref{eq:base2}, \eqref{eq:base3}. This task is not a priori trivial since the equilibrium free surface shape is not spherical due to Earth's rotation effects \cite{Khakimzyanov2016a}.


\section{Numerical illustrations}
\label{sec:simus}

Currently, there is a well-established set of test problems \cite{Synolakis2008} which are routinely used to validate numerical codes for tsunami propagation and run-up. These tests can be used also for inter-comparison of various algorithms in 1D and 2D \cite{Horrillo2015}. However, currently, there do not exist such (generally admitted) tests for nonlinear dispersive wave models on a rotating sphere. The material presented below can be considered as a further effort to constitute such a database.


\subsection{Wave propagation over a flat rotating sphere}
\label{sec:flat}

\begin{figure}
  \centering
  \subfigure[]{\includegraphics[width=0.48\textwidth]{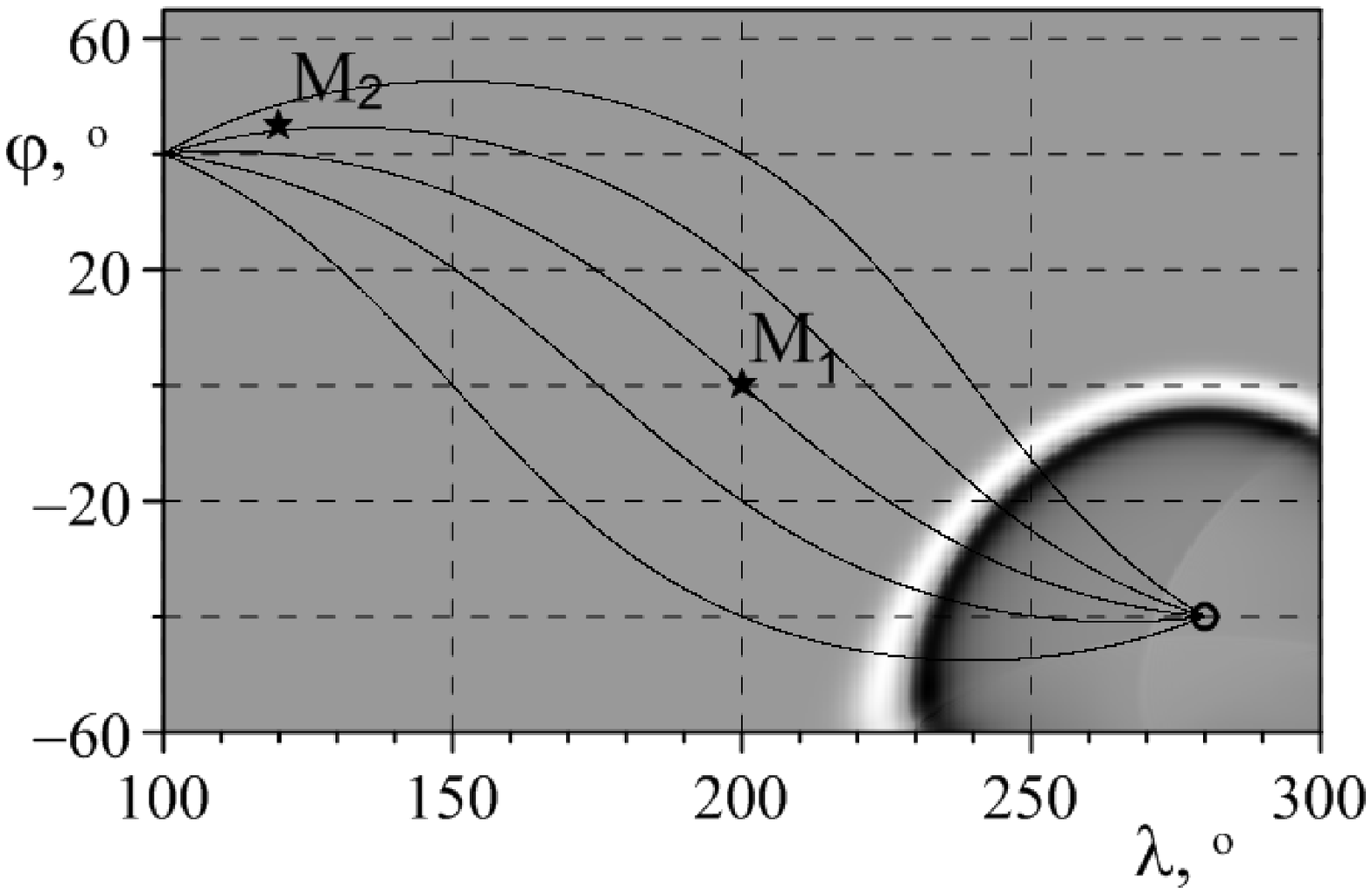}}
  \subfigure[]{\includegraphics[width=0.48\textwidth]{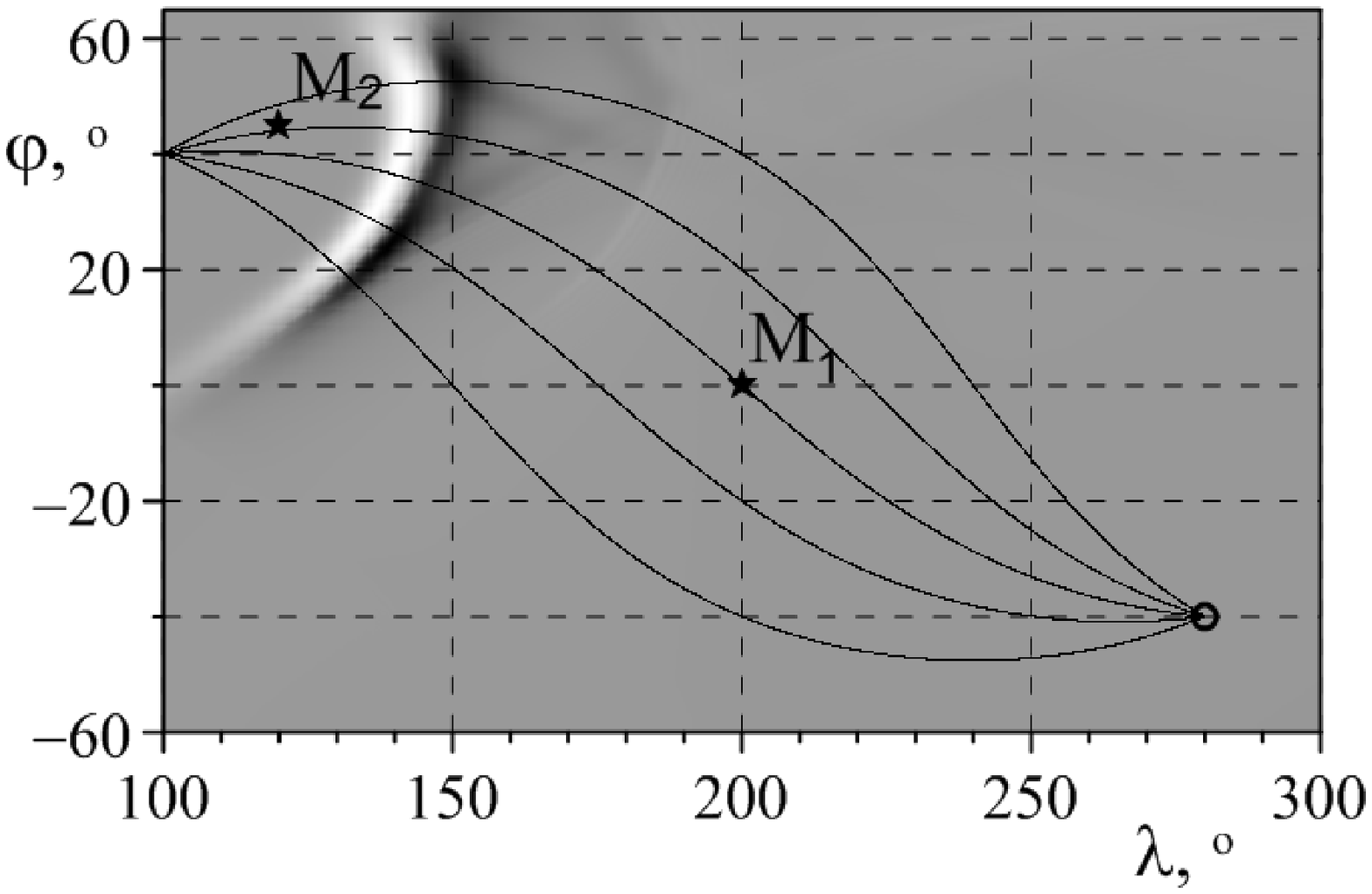}}
  \caption{\small\em The computational domain and free surface elevations computed with the FNWD model for the source $\W_{\,3}$ at $t\ =\ 6\;\h$ (a) and $t\ =\ 23\;\h$ (b). The locations of two synthetic wave gauges are depicted with the symbol $\star\,$. The center of the source region is shown with symbol $\circ\,$.}
  \label{fig:2}
\end{figure}

Consider a simple bounded spherical domain which occupies the region from $100^{\circ}$ to $300^{\circ}$ from the West to the East and from $-60^{\circ}$ to $65^{\circ}$ from the South to the North. From now on we use for simplicity the geographical latitude $\phi\ \equiv\ \dfrac{\pi}{2}\ -\ \theta$ instead of the variable $\theta\,$. The considered domain is depicted in Figure~\ref{fig:2} and contains a large portion of the \textsc{Pacific Ocean}, excluding, of course, the poles (see restriction \eqref{eq:1.4}). The idealization consists in the fact that we assume the (undisturbed) water depth is constant, \ie $h\ \equiv\ 4\;\km\,$. The initial condition consists of a \textsc{Gau\ss{}ian}-shaped bump put on the free surface
\begin{equation}\label{eq:ic0}
  \eta\,(\lambda,\,\phi,\,0)\ =\ \alpha_{\,0}\,\ue^{-\varpi\,\rho^{\,2}(\lambda,\,\phi)}\,,
\end{equation}
with zero velocity field in the fluid bulk. Function $\rho\,(\lambda,\,\phi)$ is a great-circle distance between the points $(\lambda,\,\phi)$ and $(\lambda_{\,0},\,\phi_{\,0})\,$, \ie
\begin{equation}\label{eq:ic}
  \rho\,(\lambda,\,\phi)\ \eqdef\ R\cdot\arccos\,\bigl\{\,\cos\phi\,\cos\phi_{\,0}\,\cos(\,\lambda\ -\ \lambda_{\,0})\ +\ \sin\phi\,\sin\phi_{\,0})\,\bigr\}\,.
\end{equation}
In our numerical simulations we take the initial amplitude $\alpha_{\,0}\ =\ 5\;\m\,$, the \textsc{Gau\ss{}ian} center is located at $\bigl(\lambda_{\,0},\,\phi_{\,0}\bigr)\ =\ \bigl(280^{\circ},\,-40^{\circ}\bigr)\,$. The parameter $\varpi$ is chosen from three values $8\times 10^{-10}\,$, $8\times 10^{-11}$ and $8\times 10^{-12}\;\m^{\,-2}$. It corresponds to effective linear source sizes equal approximatively to $\W_{\,1}\ \approx\ 107.3\;\km\,$, $\W_{\,2}\ \approx\ 339\;\km$ and $\W_{\,3}\ \approx\ 1073\;\km$ respectively. The effective source size is defined as the diameter of the circle $\S_{\,10}$ serving as the level-set $\dfrac{\alpha_{\,0}}{10}$ of the initial free surface elevation $\eta\,(\lambda,\,\phi,\,0)\,$, \ie
\begin{equation*}
  \S_{\,10}\ \eqdef\ \Bigl\{\,(\lambda,\,\phi)\;\vert\;\eta\,(\lambda,\,\phi,\,0)\ =\ \frac{\alpha_{\,0}}{10}\,\Bigr\}\,.
\end{equation*}
On the boundary of the computational domain we prescribed \textsc{Sommerfeld}-type non-radiation boundary conditions \cite{Gusev2014}. In our opinion, this initial condition has the advantage of being symmetric, comparing to the asymmetric source proposed in \cite{Kirby2013}. Indeed, if one neglects the \textsc{Earth} rotation effect (\ie $\Omega\ \equiv\ 0$), our initial condition will generate symmetric solutions in the form of concentric circles drawn on sphere's surface. If one does not observe them numerically, it should be the first red flag. In the presence of \textsc{Earth}'s rotation (\ie $\Omega\ >\ 0$), the deviation of wave fronts from concentric circles for $t\ >\ 0$ characterizes \textsc{Coriolis}'s force effects (see Section~\ref{sec:coriolis}).

Oscillations of the free surface were recorded in our simulations by two synthetic wave gauges located in points $\M_{\,1}\ =\ \bigl(200^{\circ},\, 0^{\circ}\bigr)$ and $\M_{\,2}\ =\ \bigl(120^{\circ},\, 45^{\circ}\bigr)$ (see Figure~\ref{fig:2}). All simulations were run with the resolution $6001 \times 3751$ of nodes (unless explicitly stated to the contrary). The physical simulation time was set to $T\ =\ 30\;\h\,$. In our numerical experiments we observed that the CPU time for NSWE runs is about five times less than FNWD computations. Sequential FNWD runs took about $6$ days. This gives the first idea of the `price' we pay to have non-hydrostatic effects.


\subsubsection{Sphericity effects}
\label{sec:sphere}

The effects of \textsc{Earth}'s sphericity are studied by performing direct comparisons between our FNWD spherical model \eqref{eq:base1} -- \eqref{eq:base3} and the same FNWD on the plane (see Part~I \cite{Khakimzyanov2016c} for the derivation and Part~II \cite{Khakimzyanov2016} for the numerics). In the plane case the initial condition was constructed in order to have the same linear sizes $\W_{\,1,\,2,\,3}$ as in the spherical case. Namely, it is given by formula \eqref{eq:ic0} with function $\rho(x,\,y)$ replaced by the \textsc{Euclidean} distance to the center $\bigl(\,x_{\,0},\,y_{\,0}\,\bigr)\,$:
\begin{equation*}
  \rho\,(\,x,\,y\,)\ =\ \sqrt{(x\ -\ x_{\,0})^{\,2}\ +\ (y\ -\ y_{\,0})^{\,2}}\,.
\end{equation*}
The computational domain was a plane rectangle with sides lengths approximatively equal to those of the spherical rectangle depicted in Figure~\ref{fig:2}. The source was located in the point $\bigl(\,x_{\,0},\,y_{\,0}\,\bigr)$ such that the distance to the South--West corner is the same as in the spherical configuration. Synthetic wave gauges $\M_{\,1,\,2}$ were located on the plane in order to preserve the distance from the source. Moreover, in order to isolate sphericity effects, we turn off \textsc{Earth}'s rotation in computations presented in this Section, \ie $\Omega\ \equiv\ 0\,$.

\begin{figure}
  \centering
  \subfigure[]{\includegraphics[width=0.32\textwidth]{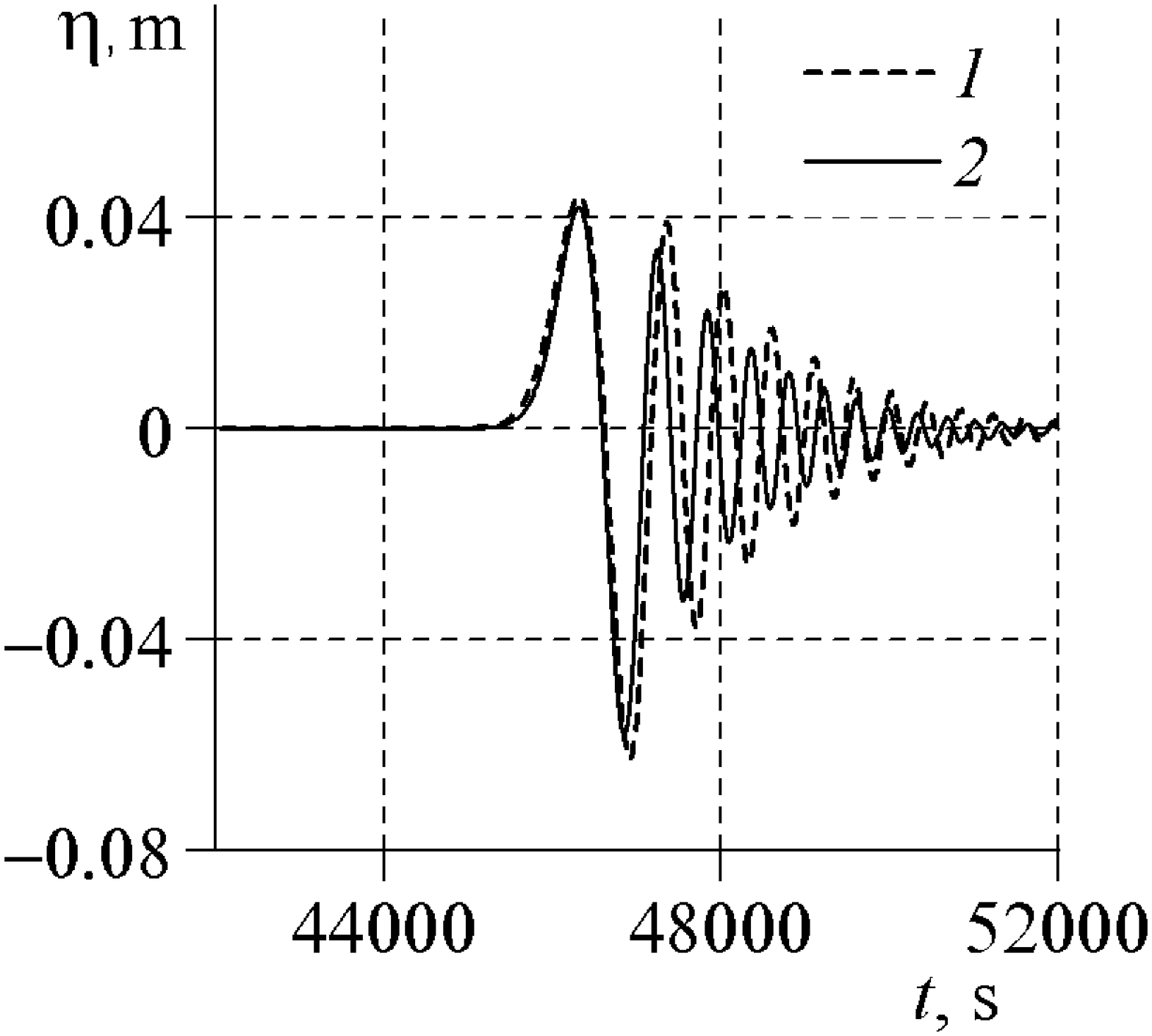}}
  \subfigure[]{\includegraphics[width=0.32\textwidth]{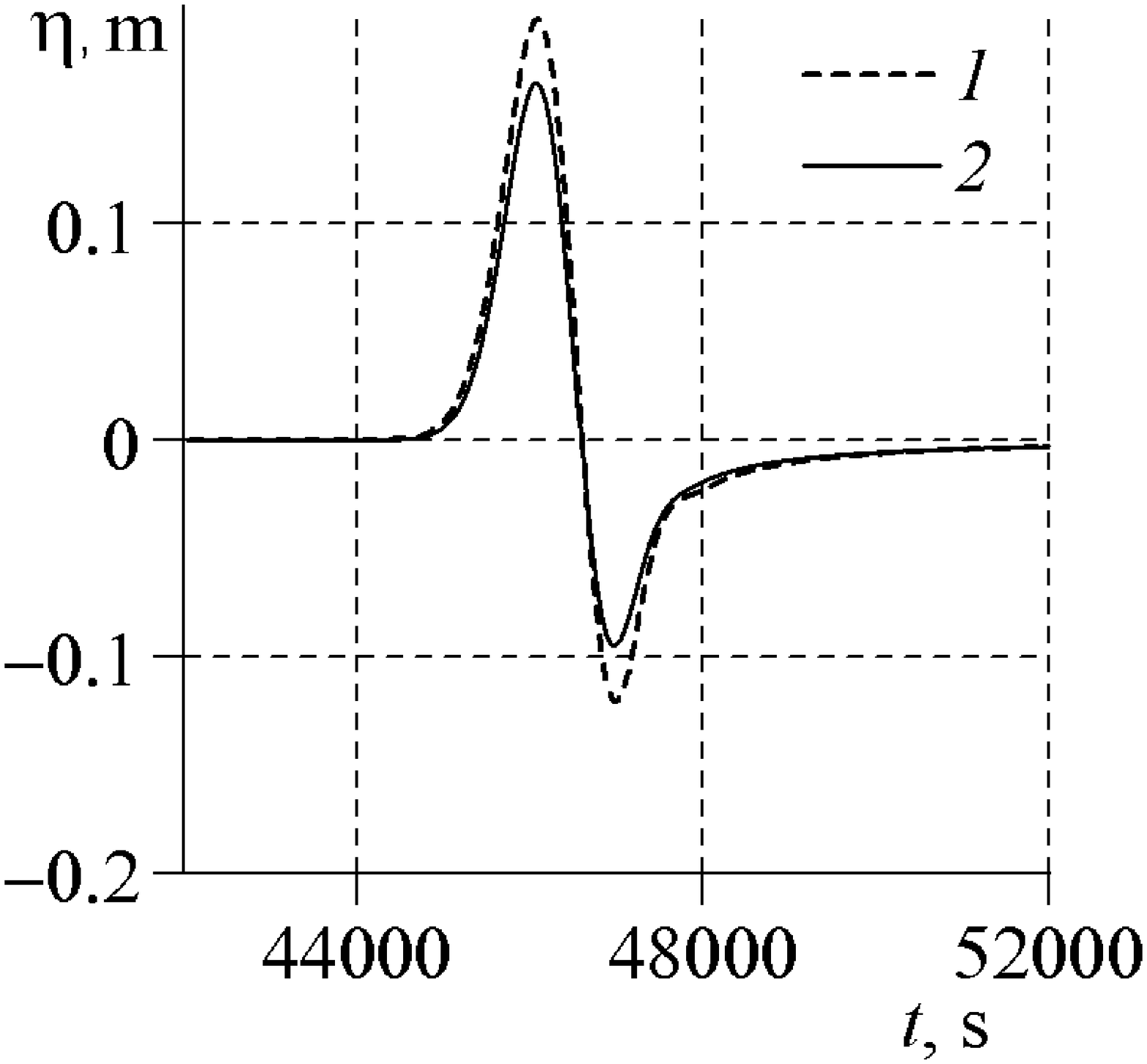}}
  \subfigure[]{\includegraphics[width=0.32\textwidth]{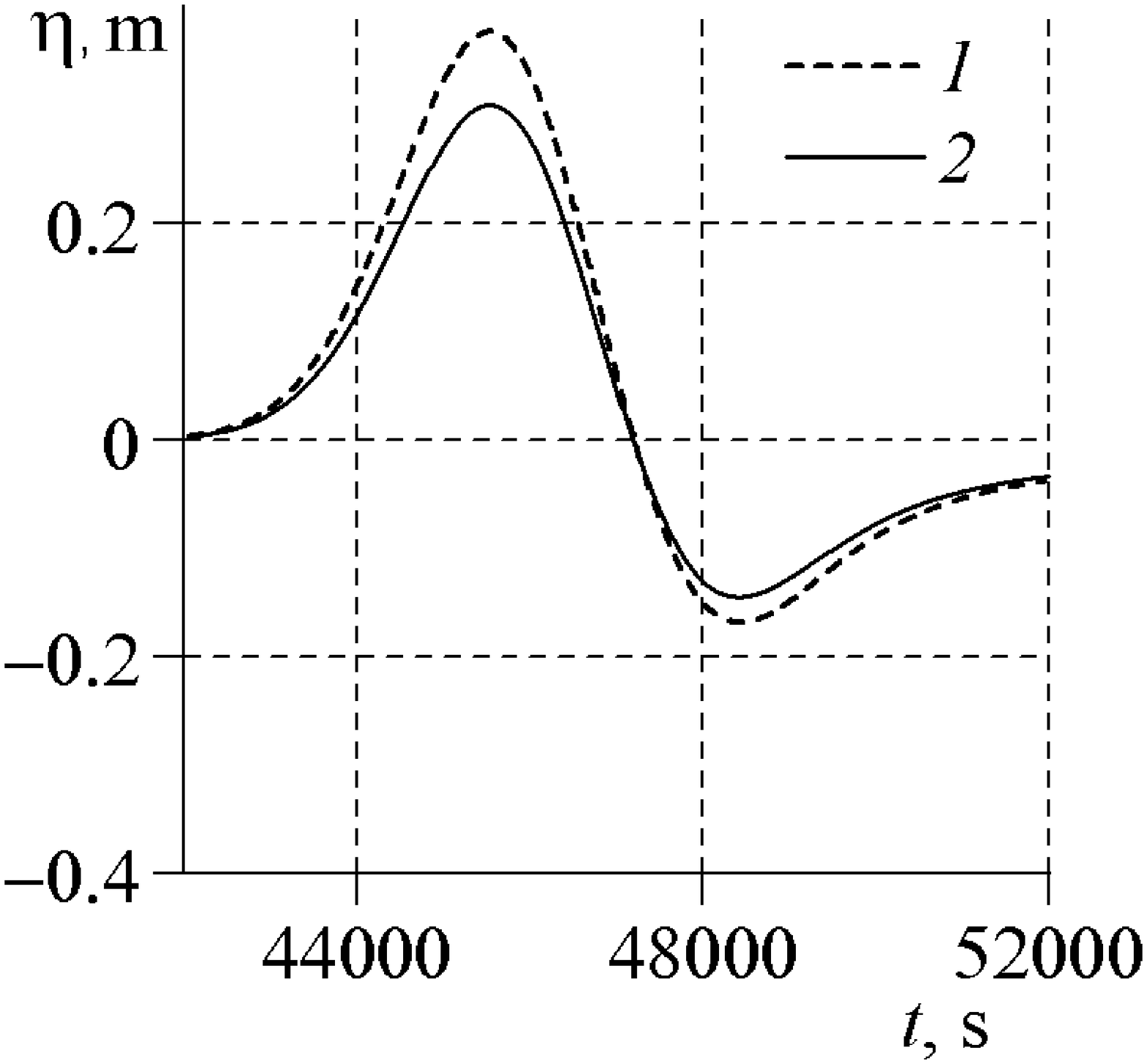}}
  \subfigure[]{\includegraphics[width=0.32\textwidth]{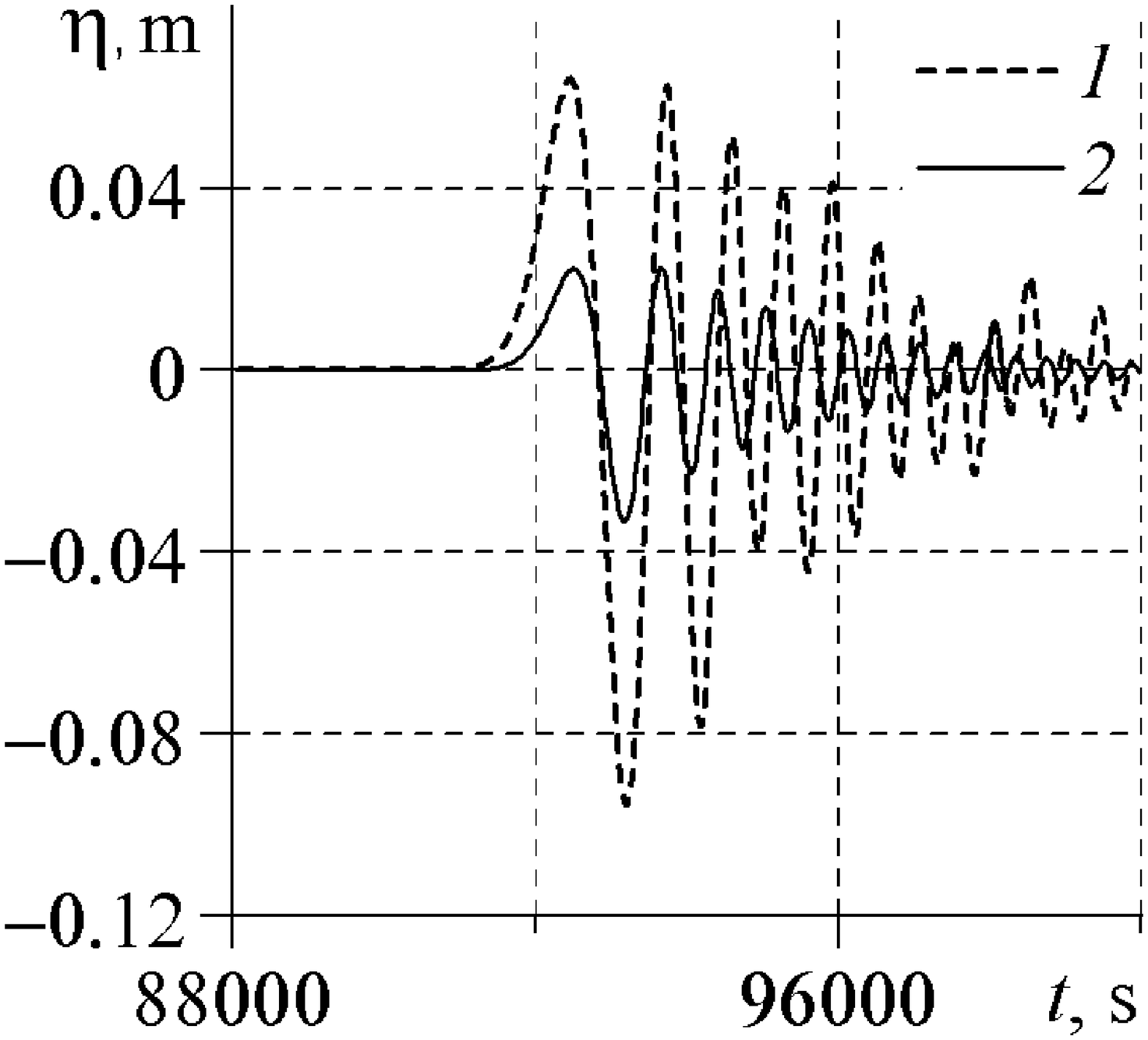}}
  \subfigure[]{\includegraphics[width=0.32\textwidth]{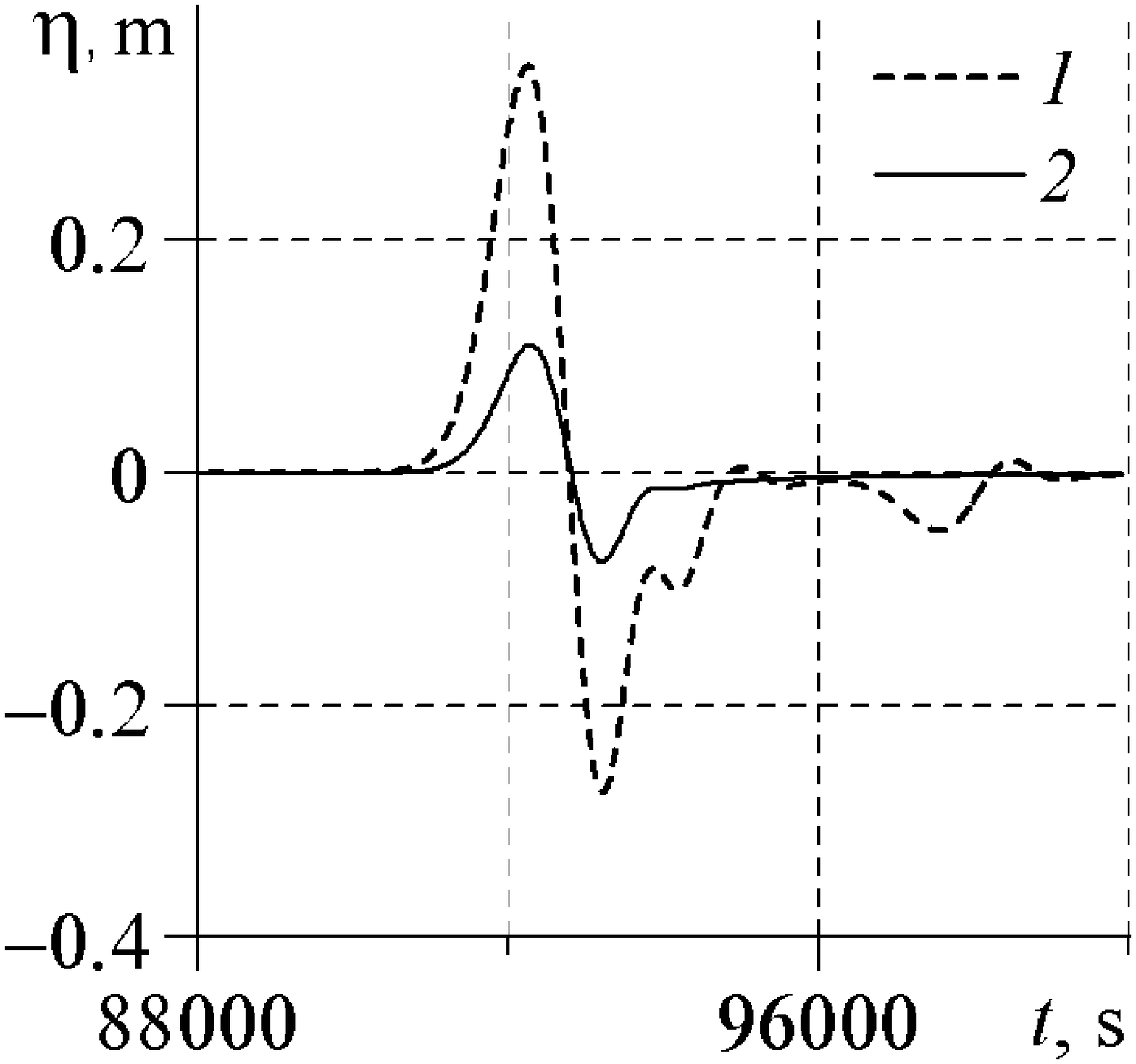}}
  \subfigure[]{\includegraphics[width=0.32\textwidth]{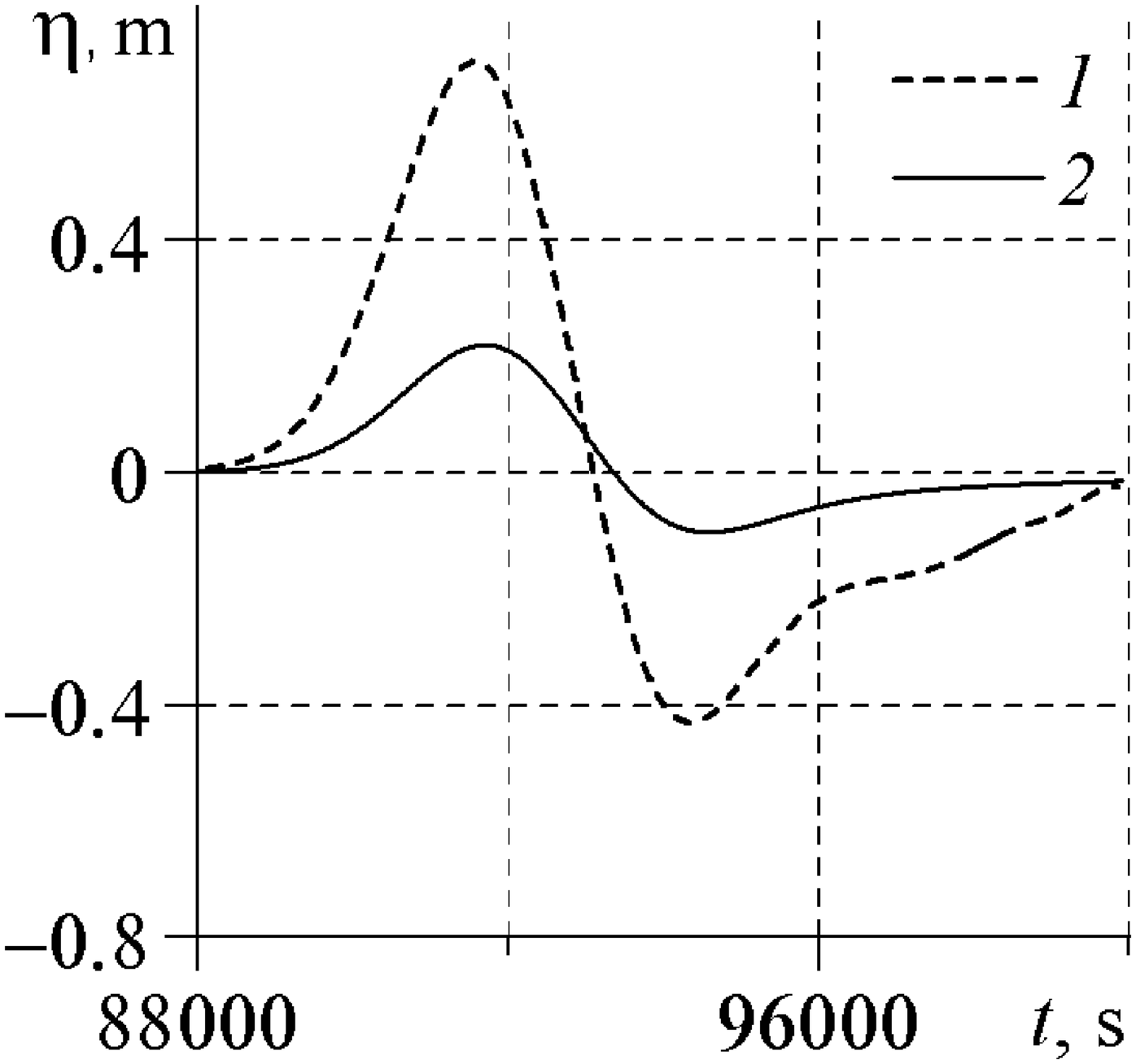}}
  \caption{\small\em Synthetic wave gauge records registered in points $\M_{\,1}\,$(a -- c) and $\M_{\,2}\,$(d -- f). FNWD predictions on a sphere are given with dashed lines (1) and the flat model is depicted with the solid line (2). The effective linear source sizes are $\W_{\,1}\,$(a, d), $\W_{\,2}\,$(b, e) and $\W_{\,3}\,$(c, f).}
  \label{fig:3}
\end{figure}

In Figure~\ref{fig:3} one can see the synthetic gauge records computed with the spherical and `flat' FNWD models. It can be clearly seen that sphericity effects become more and more important when the source size $\W$ increases. We notice also that in all cases the wave amplitudes predicted with the spherical FNWD model are higher than in `flat' FNWD computations. In general, the differences recorded at wave gauge $\M_{\,1}$ are rather moderate. However, when we look farther from the source, \eg at location $\M_{\,2}$ the discrepancies become flagrant --- the difference in wave amplitudes may reach easily several times (as always the `flat' model underestimates the wave). The explanation of this phenomenon is purely \emph{geometrical}. In the plane case the geometrical spreading along the rays starting at the origin is monotonically decreasing (the wave amplitude decreases like $\O\bigl(r^{\,-\frac{1}{2}}\bigr)\,$, where $r$ is the distance from the source). On the other hand, in the spherical geometry the rays starting from the origin $\bigl(\lambda_{\,0},\,\phi_{\,0}\bigr)$ go along initially divergent great circles that intersect in a diametrically opposite point to $\bigl(\lambda_{\,0},\,\phi_{\,0}\bigr)$ (this point has coordinates $\bigl(80^{\circ},\,40^{\circ}\bigr)$ in our case). As a result, the amplitude first decreases in geometrically divergent areas, but then there is a wave focusing phenomenon when the rays convergent in one point. A few such (divergent/convergent) rays are depicted in Figure~\ref{fig:2} with solid lines emanating from the point $\bigl(\lambda_{\,0},\,\phi_{\,0}\bigr)\,$. We intentionally put a wave gauge into the point $\M_{\,2}$ close to the focusing area in order to illustrate this phenomenon. Otherwise, the amplification could be made even larger.

Such amplification phenomena should take place, in principle, on the whole sphere. It did not happen in our numerical simulation since we employ \textsc{Sommerfeld}-type non-radiation boundary conditions. It creates sub-regions which are not attainable by the waves. It happens since the rays emanating from the origin $\bigl(\lambda_{\,0},\,\phi_{\,0}\bigr)$ cross the boundary of the computational domain (and, thus, the waves propagating along these rays are lost). To give an example of such `shaded' regions we can mention a neighbourhood of the point $\bigl(100^{\circ},\,-60^{\circ}\bigr)\,$. That is why the wave field looses its symmetry as it can be seen in Figure~\ref{fig:2}(\textit{b}): the edges of the wave front get smeared due to the escape of information through the boundaries. On the other hand, in Figure~\ref{fig:2}(\textit{a}) we show the free surface profile at earlier times where boundary effects did not affect yet the wave front which conserved a circle-like shape.

We should mention an earlier work \cite{Tkalich2007} where the importance of sphericity effects was also underlined. In that work the authors employed a WNWD model to simulate the Indian Ocean tsunami of the $25$\up{th} December 2004 \cite{Syno2006}. The computational domain was significantly smaller than in our computations presented above (no more than $40^{\circ}$ in each direction). That is why the sphericity effects were less pronounced than in Figure~\ref{fig:3}. More precisely, \emph{only} $30$\% discrepancy was reported in \cite{Tkalich2007} comparing to the `flat' computation.


\subsubsection{Coriolis effects}
\label{sec:coriolis}

\begin{figure}
  \centering
  \subfigure[]{\includegraphics[width=0.32\textwidth]{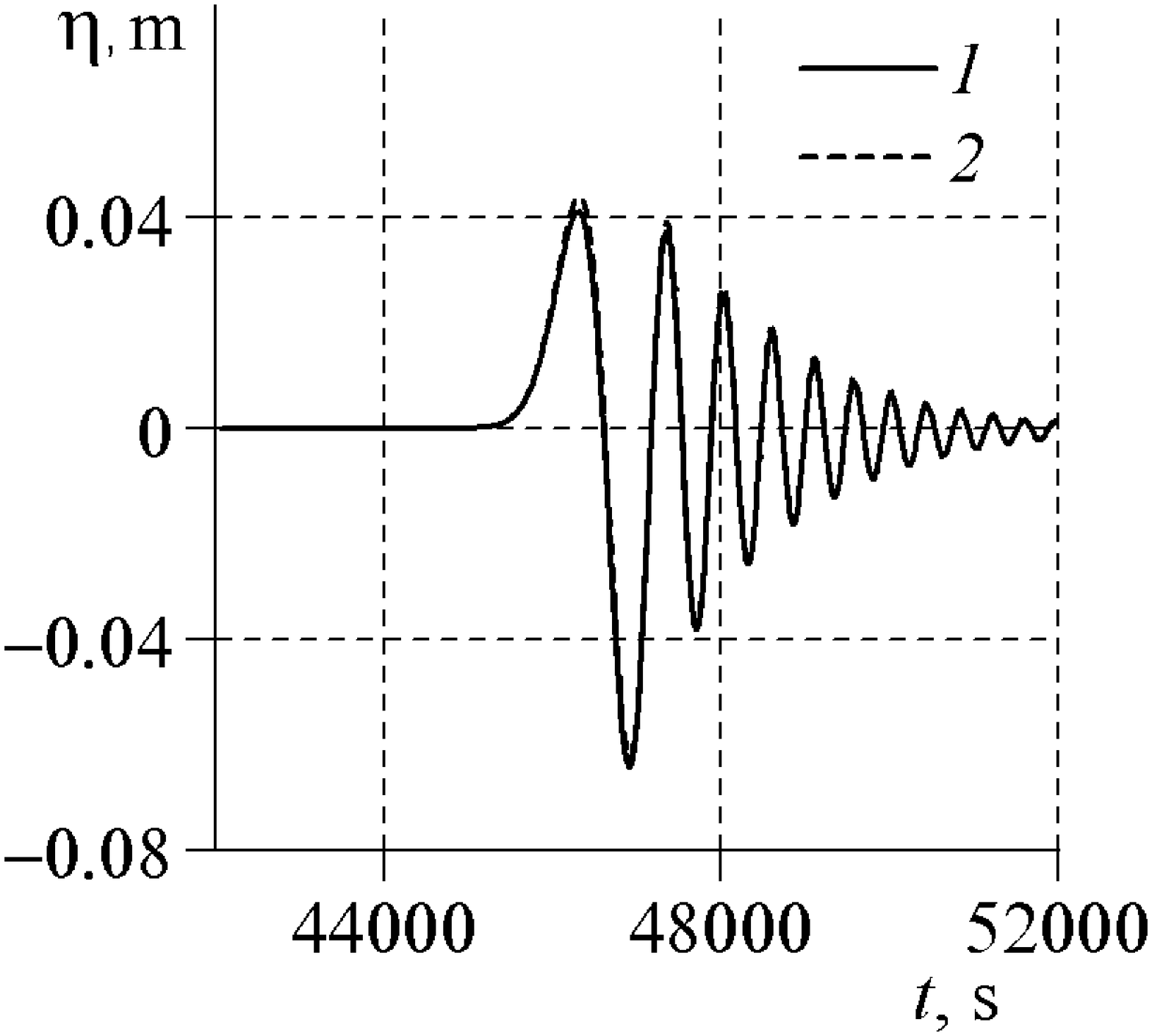}}
  \subfigure[]{\includegraphics[width=0.32\textwidth]{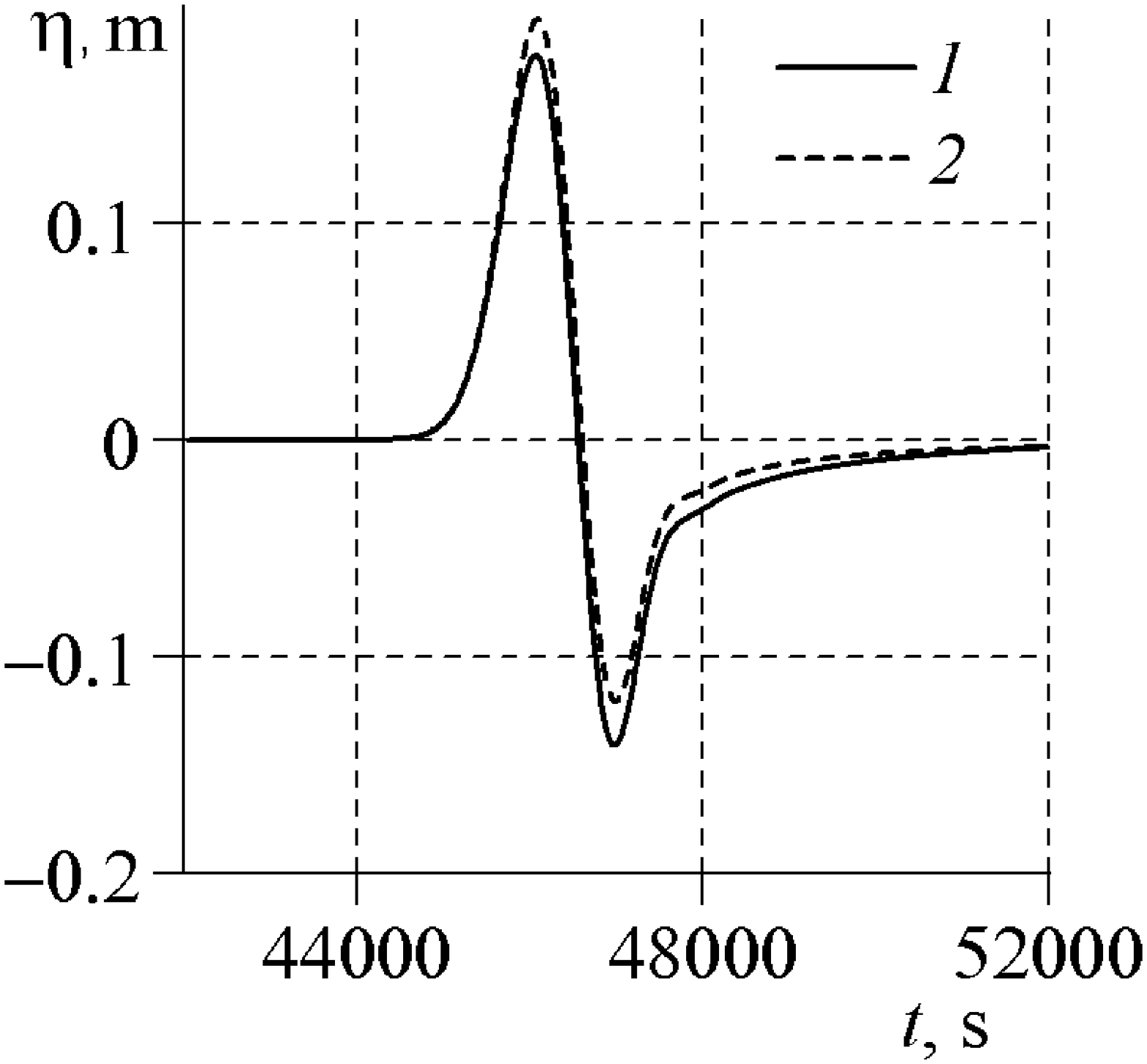}}
  \subfigure[]{\includegraphics[width=0.32\textwidth]{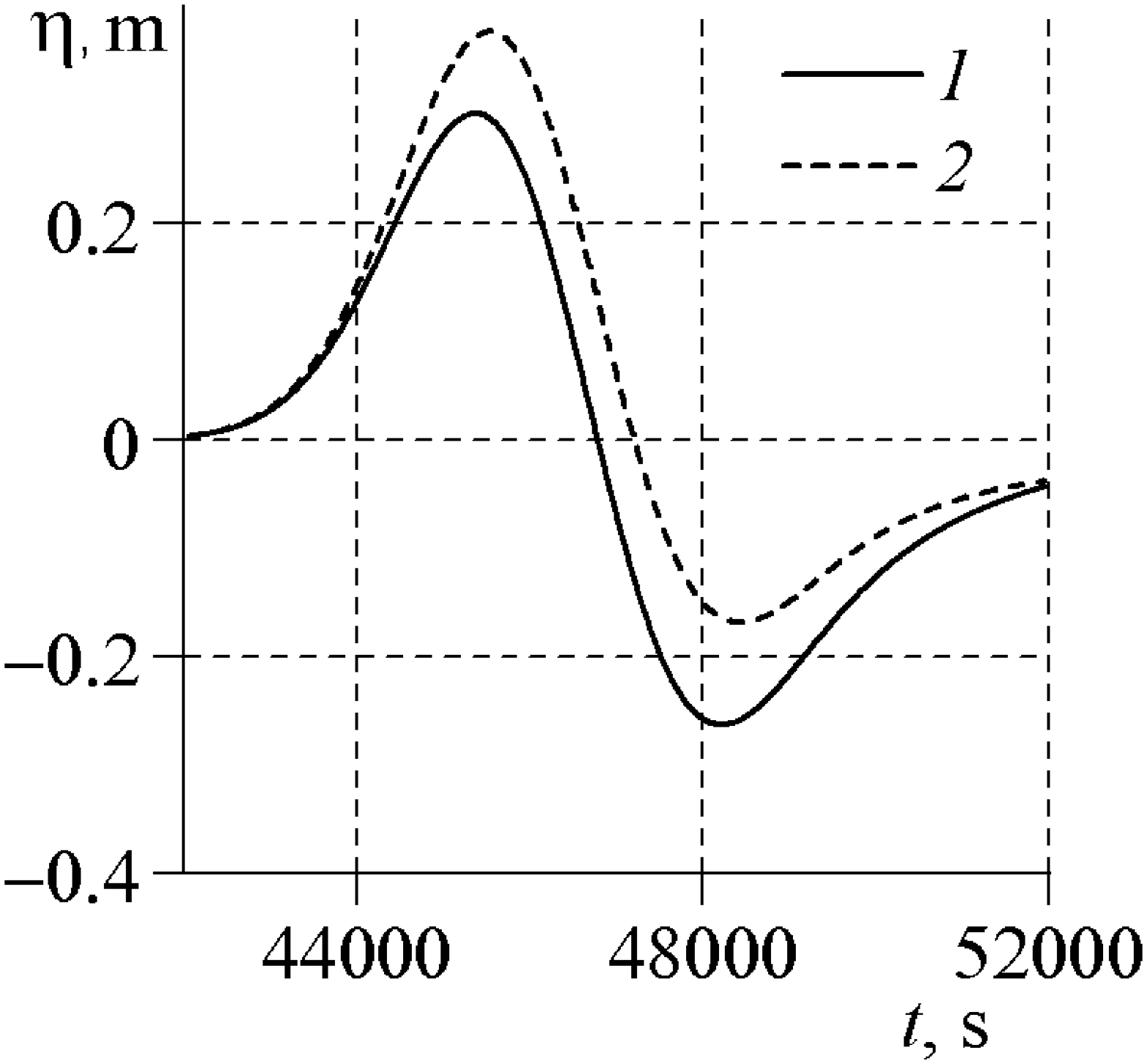}}
  \subfigure[]{\includegraphics[width=0.32\textwidth]{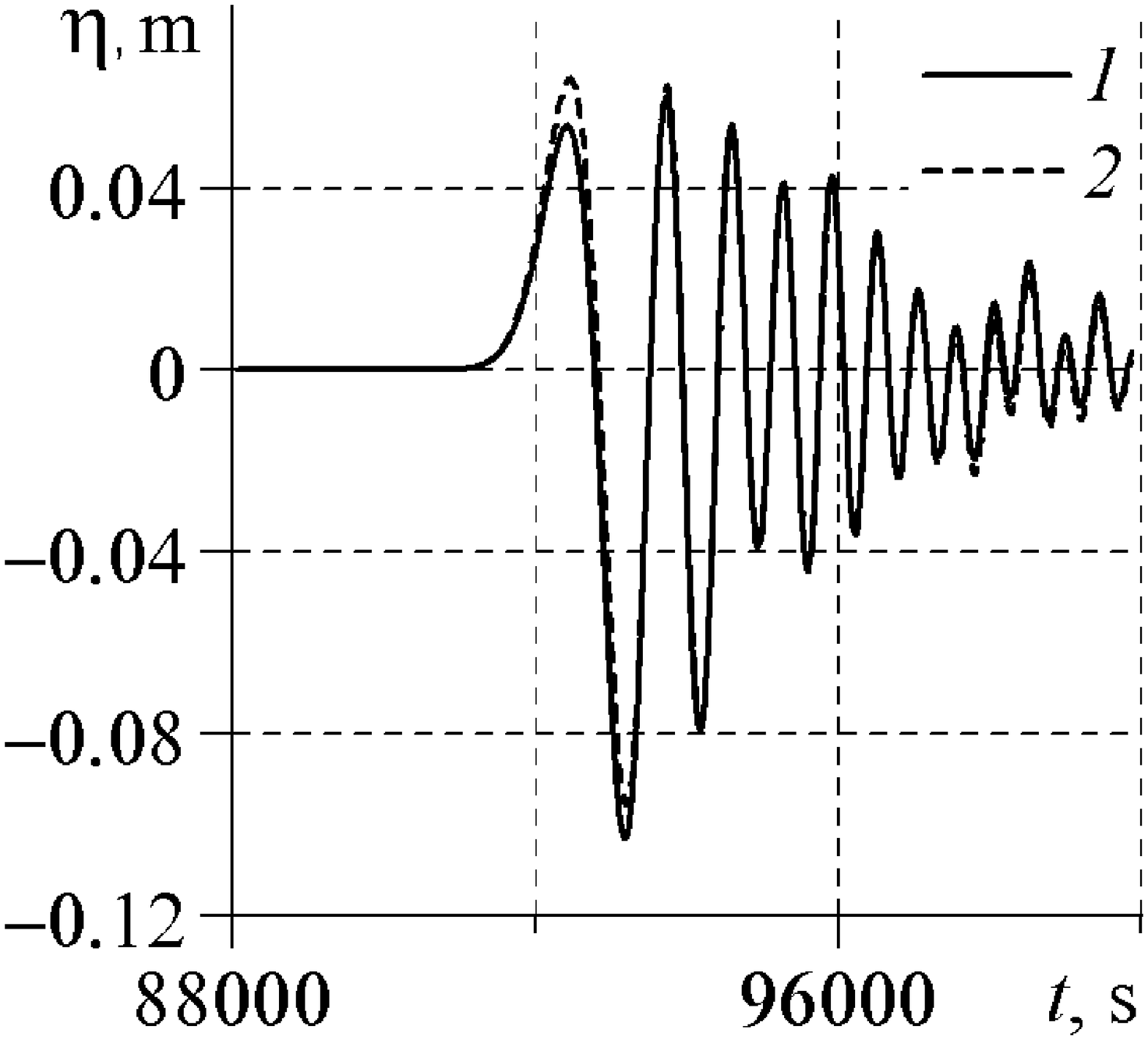}}
  \subfigure[]{\includegraphics[width=0.32\textwidth]{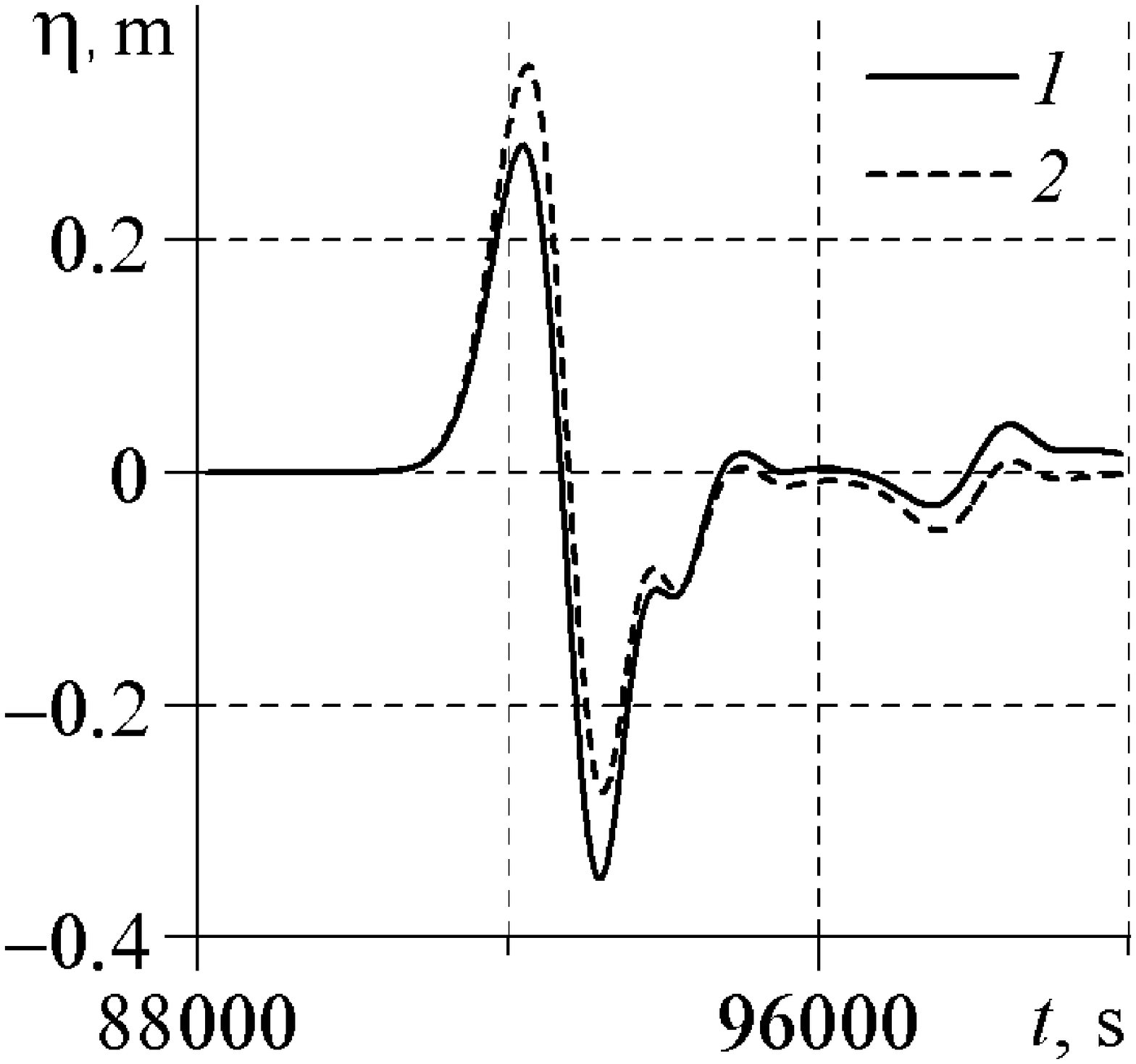}}
  \subfigure[]{\includegraphics[width=0.32\textwidth]{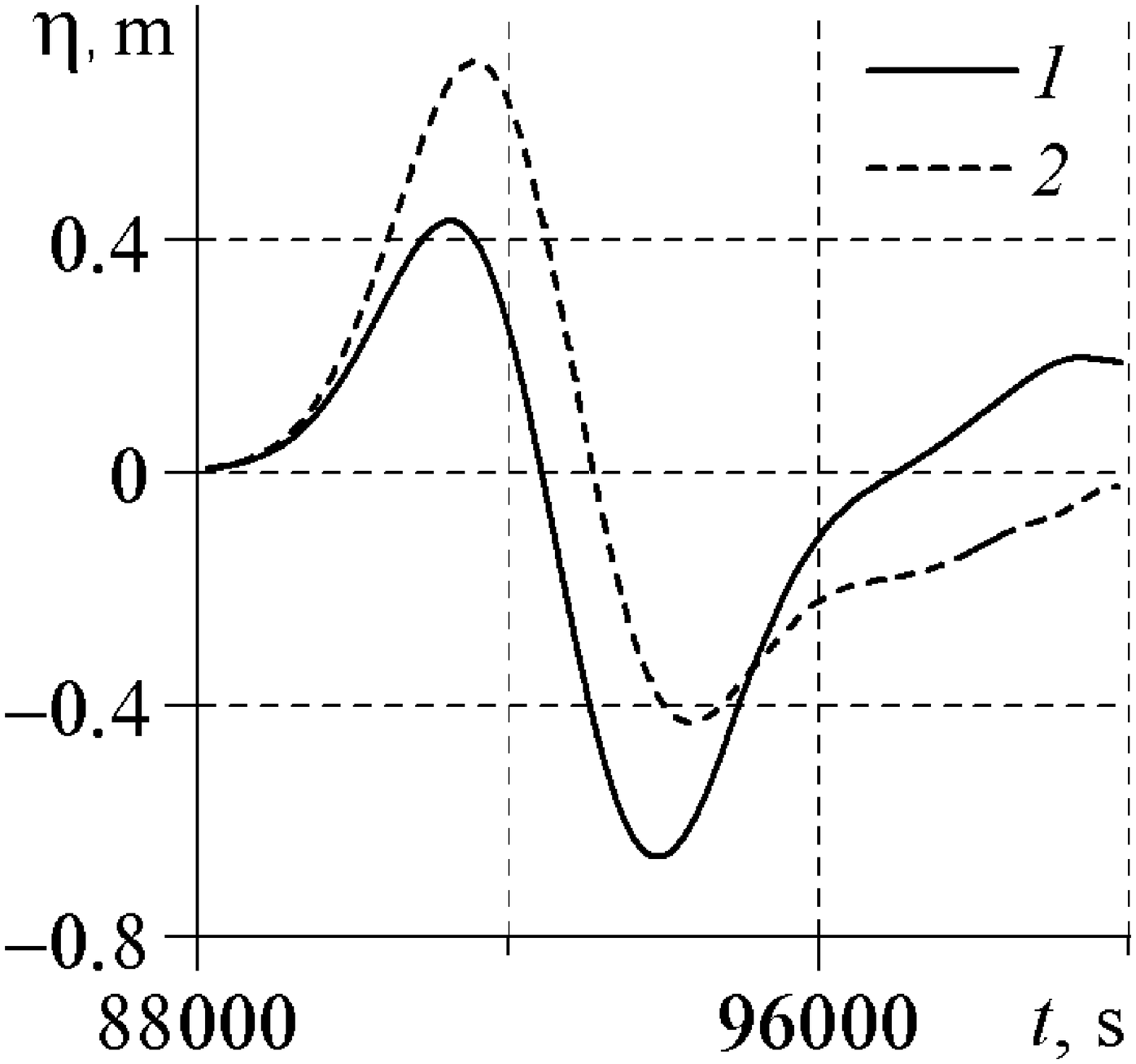}}
  \caption{\small\em Synthetic wave gauge records registered in points $\M_{\,1}\,$(a -- c) and $\M_{\,2}\,$(d -- f). FNWD predictions on a sphere with \textsc{Coriolis}'s force are shown with dashed lines (1) and without \textsc{Coriolis} is depicted with the solid line (2). The effective linear source sizes are $\W_{\,1}\,$(a, d), $\W_{\,2}\,$(b, e) and $\W_{\,3}\,$(c, f).}
  \label{fig:4}
\end{figure}

Let us estimate now the effect of \textsc{Earth}'s rotation on the wave propagation process. The main effect comes from the \textsc{Coriolis} force induced by the rotation of our planet. This force appears in the right hand sides of equations \eqref{eq:base2}, \eqref{eq:base3} and \eqref{eq:ell}. In Figure~\ref{fig:4} we show synthetic wave gauge records (for precisely the same test case and recorded in the same locations $\M_{\,1,\,2}$ as described in previous Section~\ref{sec:sphere}) with ($\Omega\ =\ 7.29\times 10^{\,-5}\;\s^{\,-1}$) and without ($\Omega\ =\ 0$) \textsc{Earth}'s rotation. We also considered initial conditions of different spatial extents $\W_{\,1,\,2,\,3}\,$. In particular, we can see that \textsc{Coriolis}'s force effect also increases with the source region size. However, \textsc{Earth}'s rotation seems to reduce somehow wave amplitudes (\ie elevation waves are decreased and depression waves are increased in absolute value). We conclude also that \textsc{Coriolis}'s force can alter significantly only the waves travelling at large distances.

Numerical simulations on a sphere, which rotates faster than \textsc{Earth}, show that \textsc{Coriolis} and centrifugal forces produce also much more visible effects on the wave propagation. In particular, important residual vortices remain in the generation region and the wave propagation speed is also reduced.

Concerning earlier investigations conducted in the framework of WNWD models, it was reported in \cite{Tkalich2007} that \textsc{Coriolis}'s force may change the maximal wave amplitude up to $15\%\,$, in \cite{Lovholt2008} --- $1.5$ -- $2.5$\% and in \cite{Kirby2013} --- up to $5\%\,$. Our results generally agree with these findings for corresponding source sizes. In the latter reference it is mentioned also that \textsc{Coriolis}'s force influence increases with source extent and it retains a portion of the initial perturbation in the source region, thus contributing to the formation of the residual wave field \cite{Nosov2014}.


\subsubsection{Dispersive effects}

\begin{figure}
  \centering
  \subfigure[]{\includegraphics[width=0.48\textwidth]{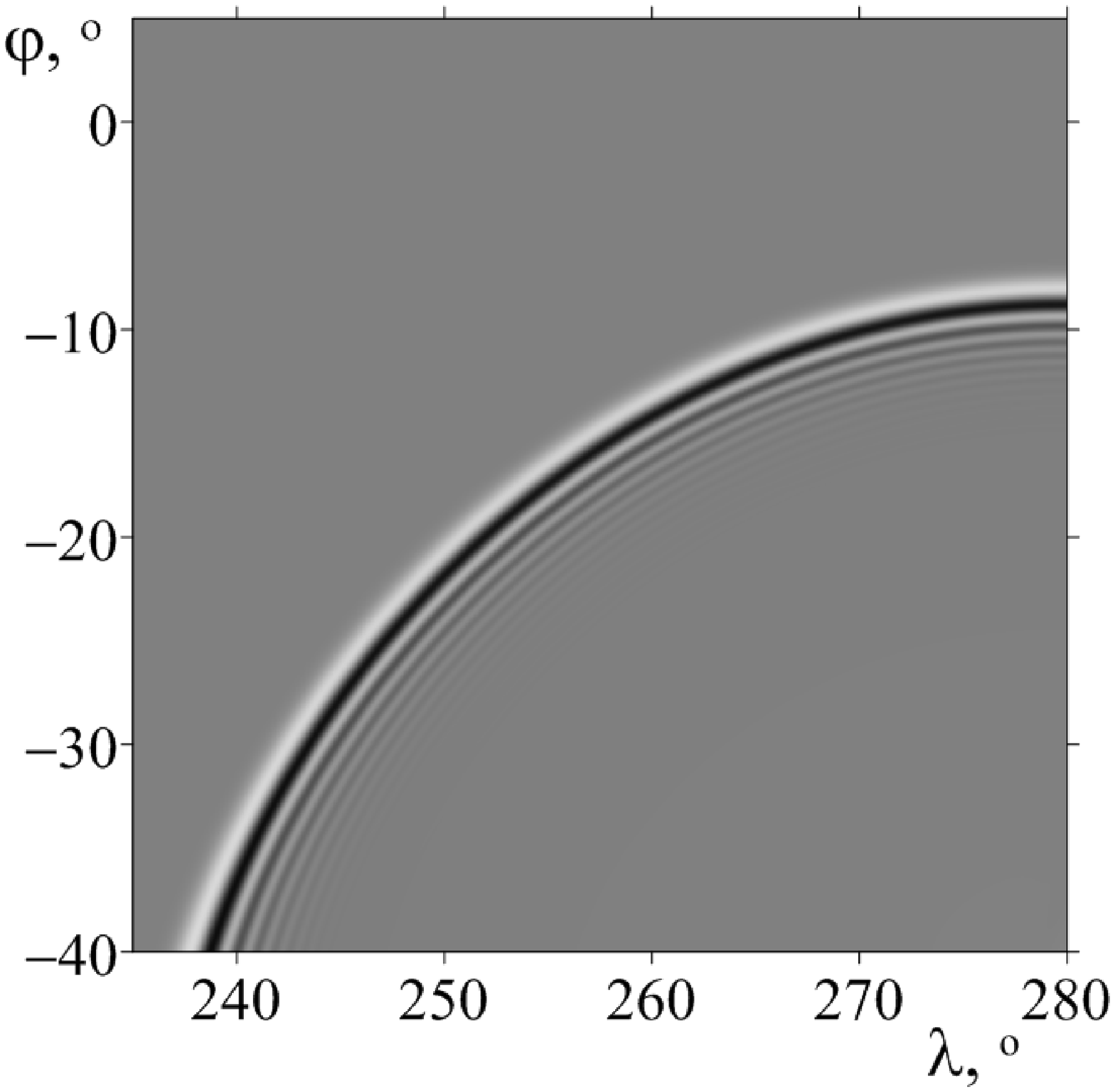}}
  \subfigure[]{\includegraphics[width=0.48\textwidth]{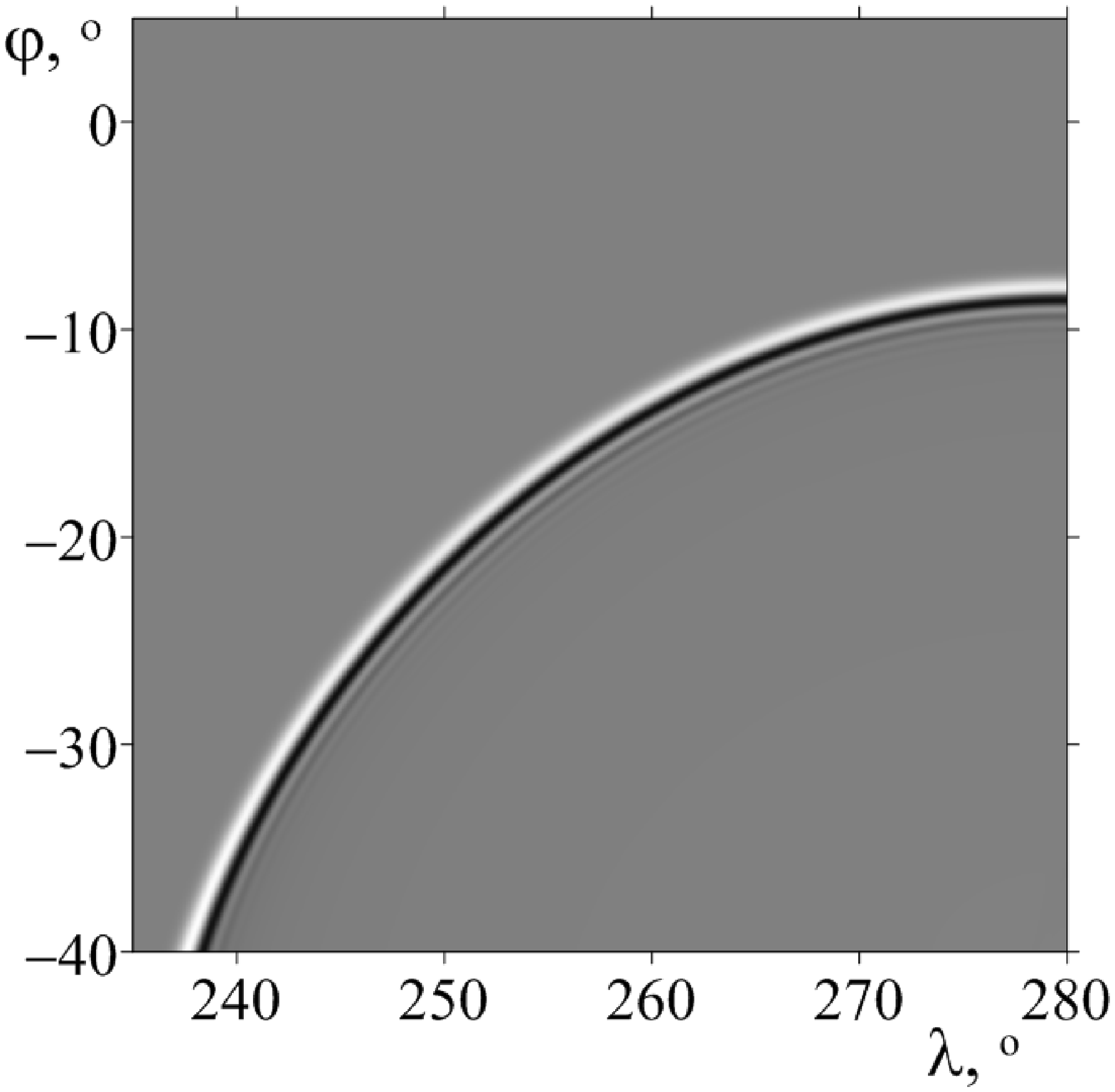}}
  \caption{\small\em Zoom on the wave field after $t\ =\ 5\,\h$ of free propagation over a sphere computed with FNWD (a) and NSWE (b) for the initial condition $\W_{\,1}\,$.}
  \label{fig:5}
\end{figure}

In order to estimate the contribution of the frequency dispersion effects on wave propagation, we are going to compare numerical predictions obtained with the FNWD and (hydrostatic non-dispersive) NSWE models on a \emph{rotating} sphere\footnote{We use the same angular velocity $\Omega\ =\ 7.29\times 10^{\,-5}\;\s^{\,-1}\,$.}. Both codes are fed with the same initial condition \eqref{eq:ic} as described above. In these simulations we use a fine grid with the angular resolution of $40^{\,\dprime}$ in order to resolve numerically shorter wave components. As it is shown in Figure~\ref{fig:5}(\textit{a}), FNWD model generates a dispersive tail behind the main wave front. The tail consists of shorter waves with smaller (decreasing) amplitudes. Obviously, the NSWE model does not reproduce this effect (see Figure~\ref{fig:5}(\textit{b})).

Computations based on the FNWD model show that dispersive effects are more pronounced for more compact initial perturbations. Signals in wave gauges (see Figure~\ref{fig:4}(\textit{a,d})) show that the initial condition of effective width $\W_{\,1}$ generates a dispersive tail quasi-absent in $\W_{\,2}$ and inexistent in $\W_{\,3}\,$. Another interesting particularity of dispersive wave propagation is the fact that during long distances the maximal amplitude may move into the dispersive tail. In other words, the amplitude of the first wave may becomes smaller than amplitude of waves from the dispersive tail. Moreover, the number of the highest wave may increase with the propagation distance \cite{Pelinovsky1996a}. A similar effect was observed in the framework of WNWD models in \cite{Lovholt2008, Glimsdal2013, Lovholt2010}.

\begin{figure}
  \centering
  \subfigure[]{\includegraphics[width=0.32\textwidth]{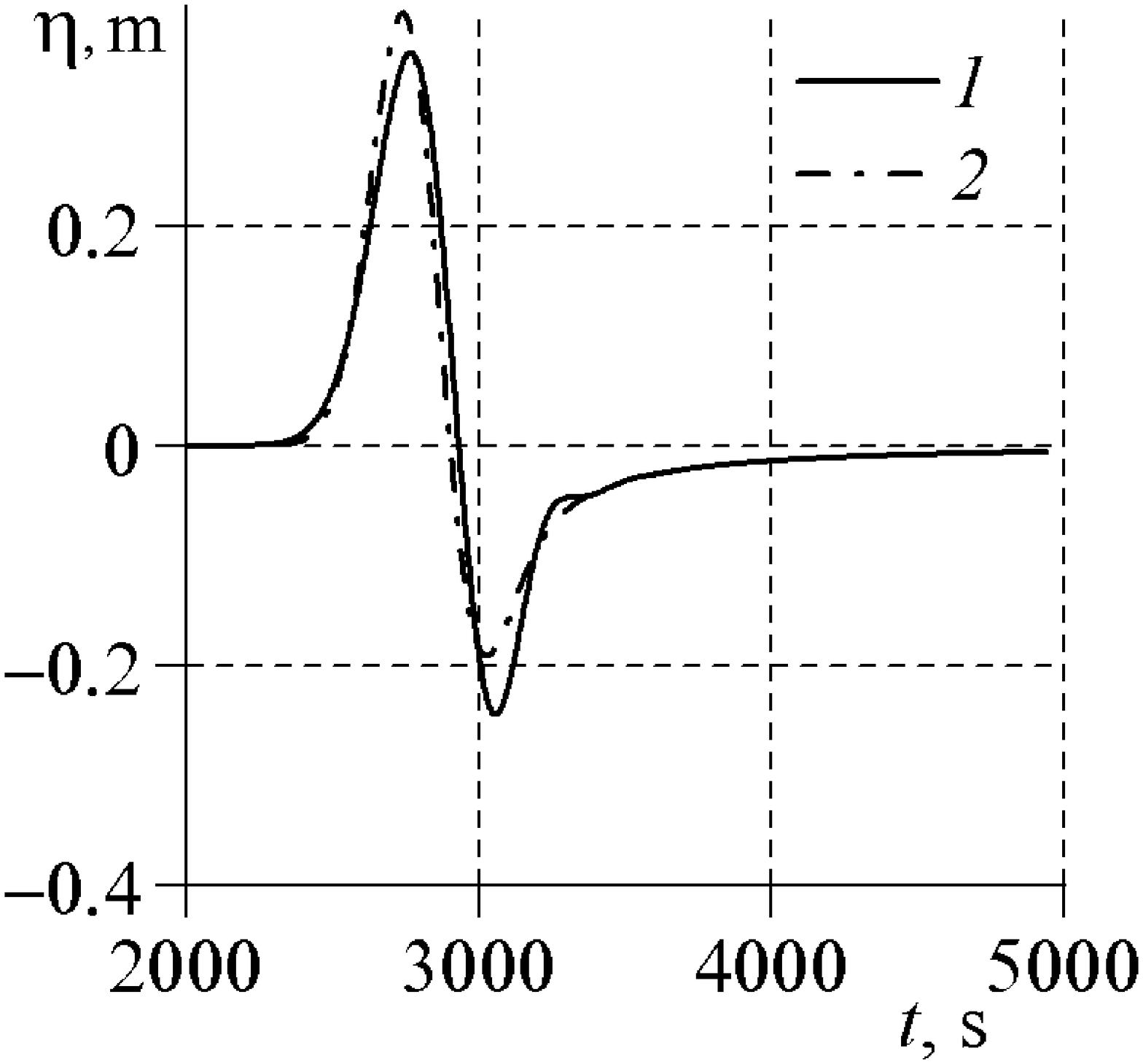}}
  \subfigure[]{\includegraphics[width=0.32\textwidth]{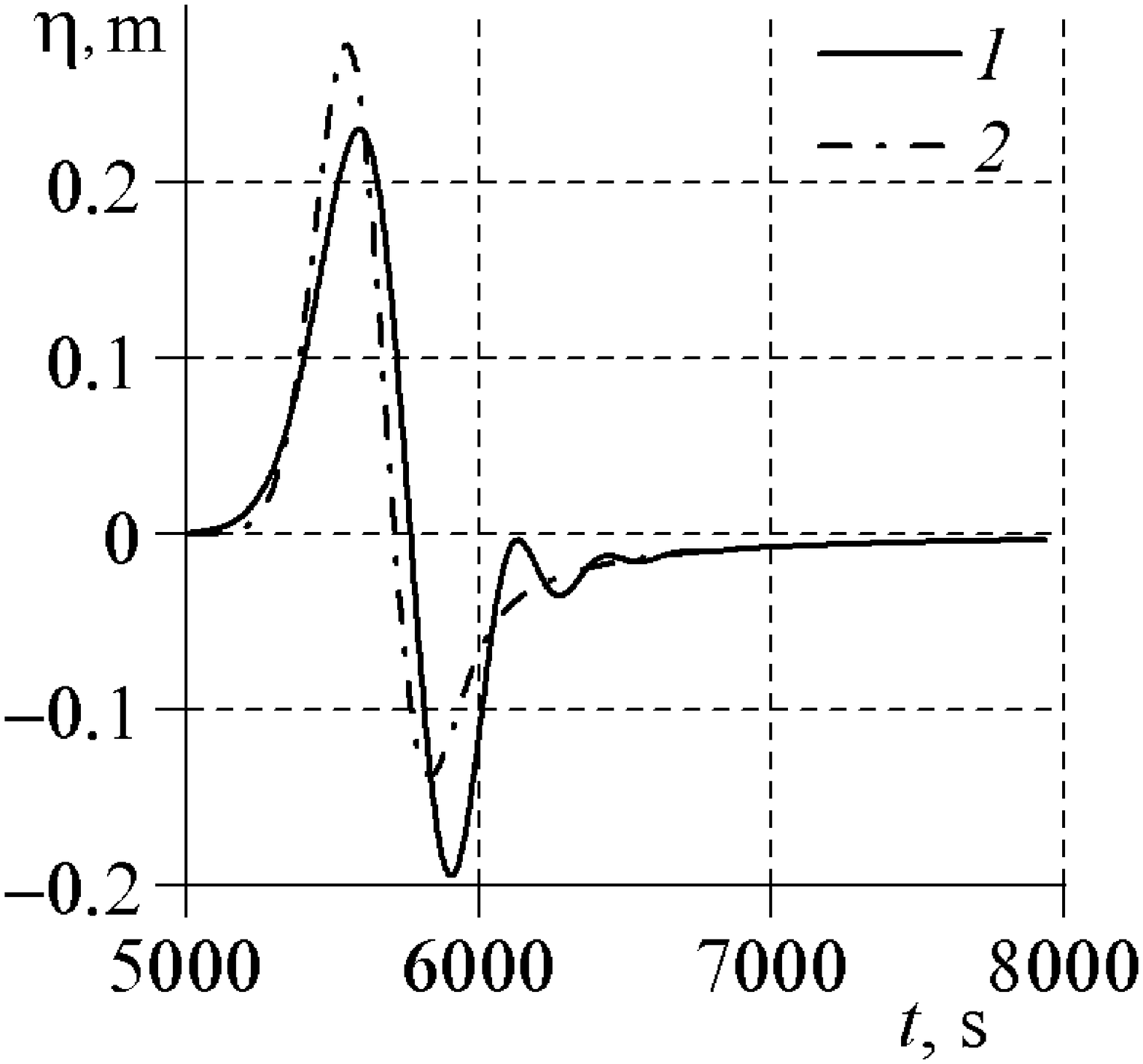}}
  \subfigure[]{\includegraphics[width=0.32\textwidth]{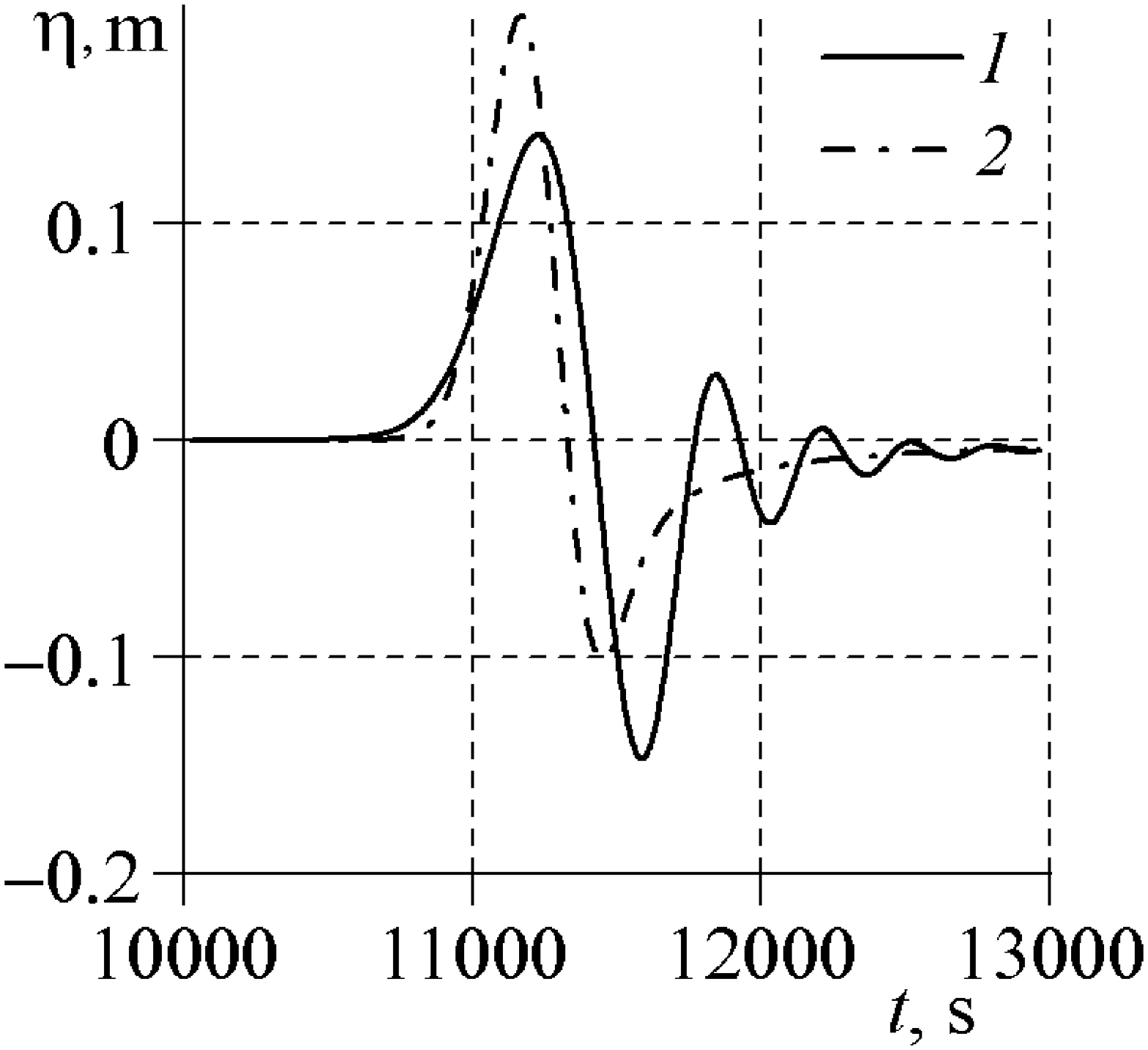}}
  \subfigure[]{\includegraphics[width=0.32\textwidth]{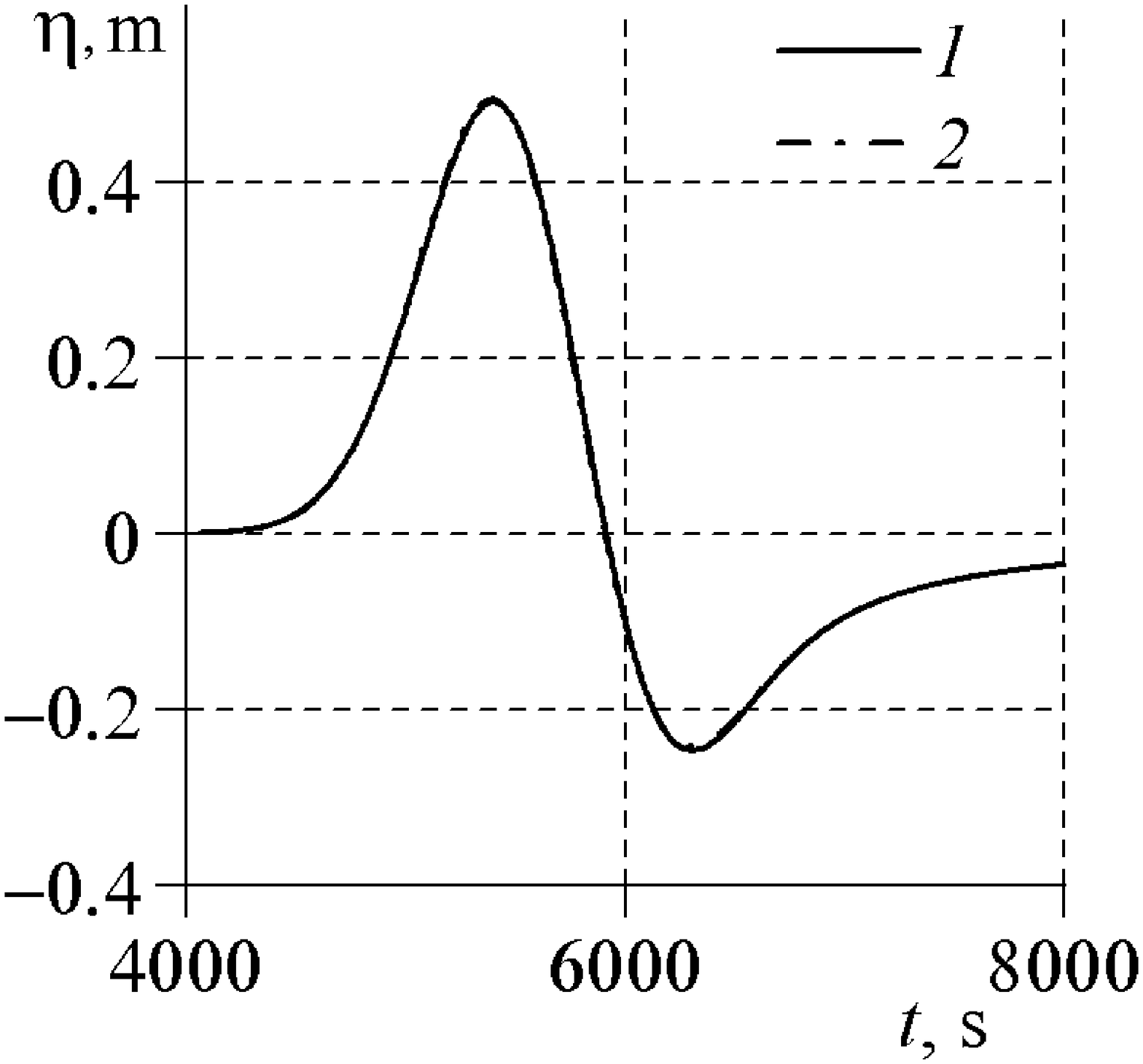}}
  \subfigure[]{\includegraphics[width=0.32\textwidth]{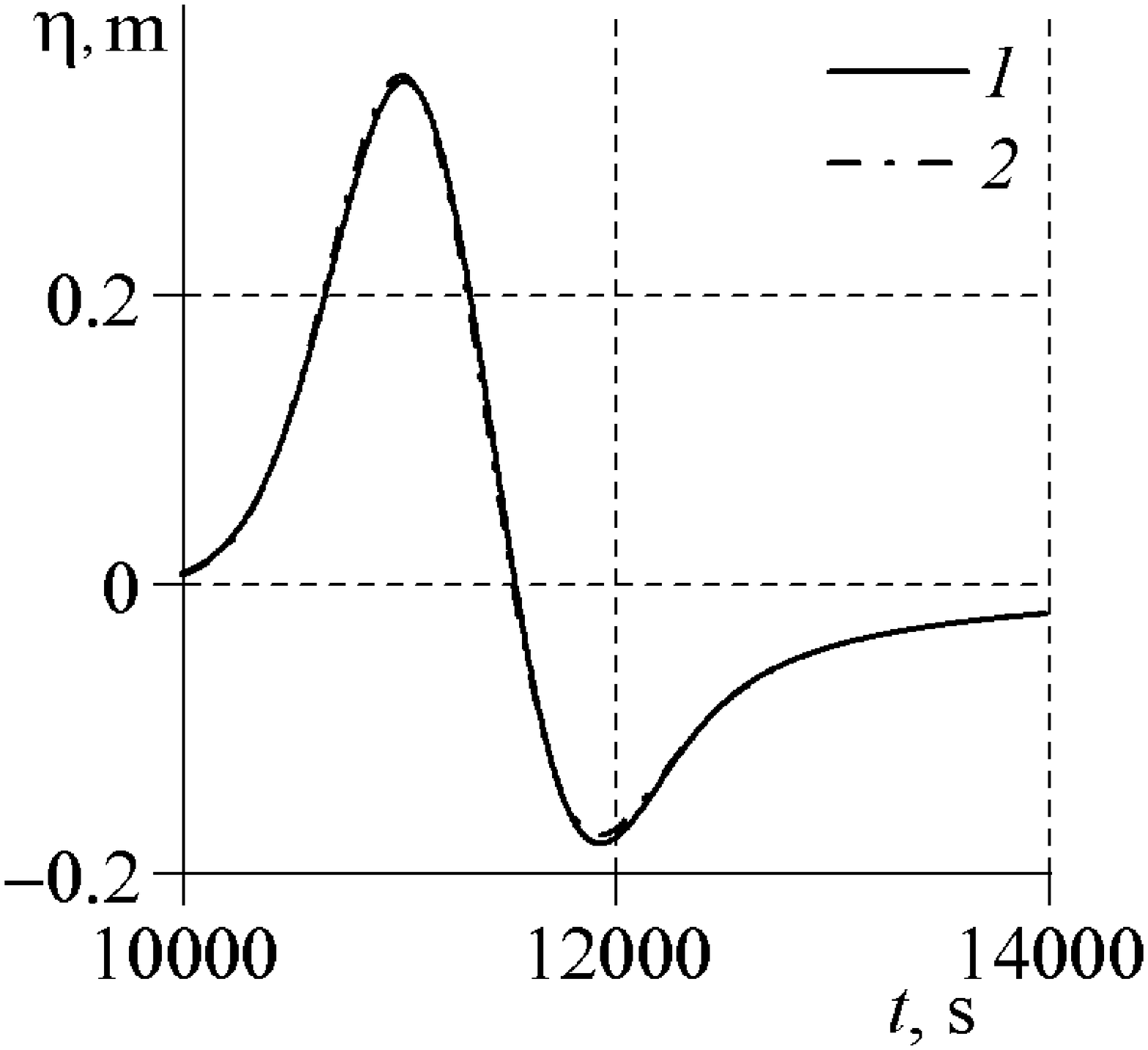}}
  \subfigure[]{\includegraphics[width=0.32\textwidth]{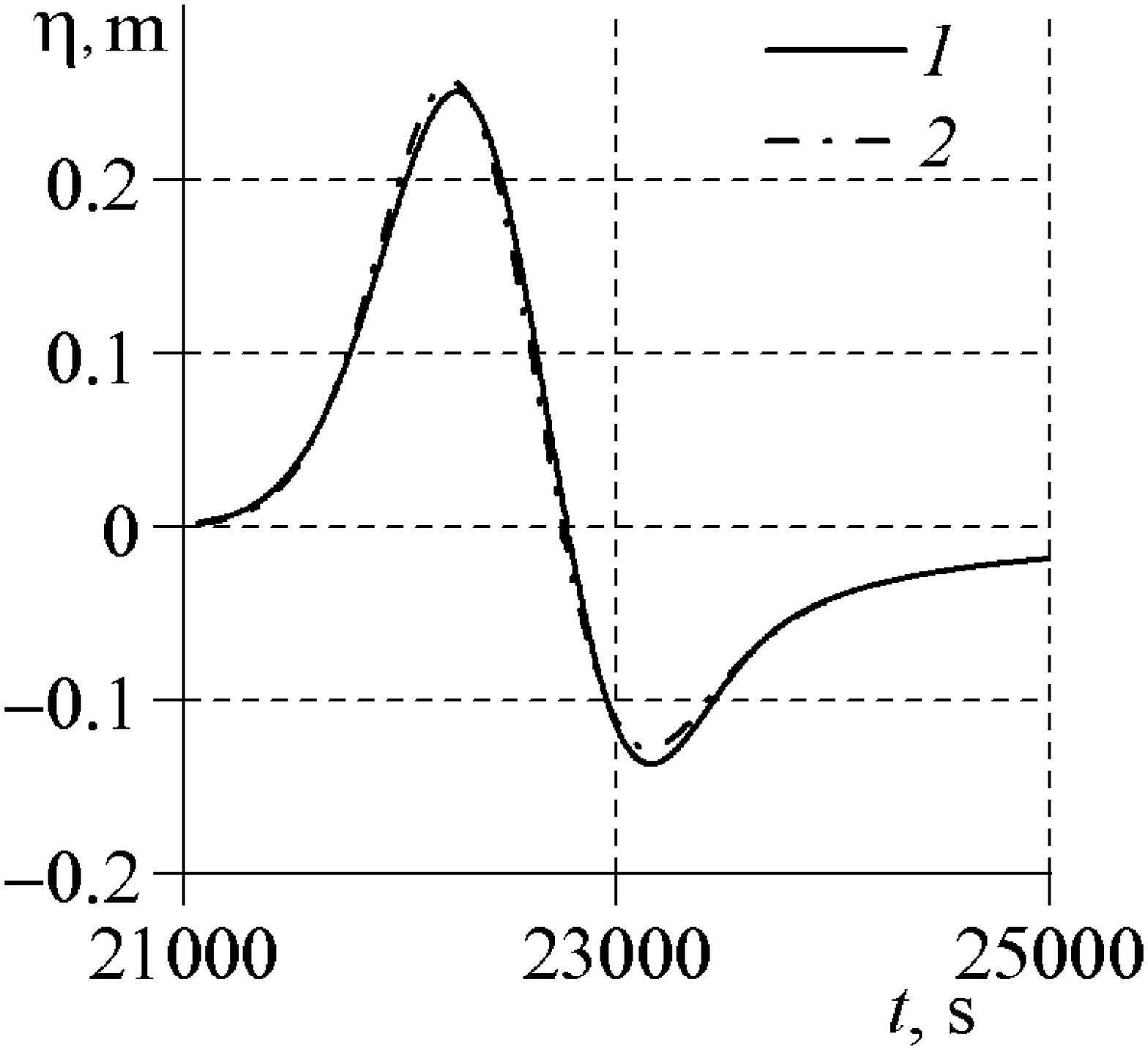}}
  \caption{\small\em Synthetic wave gauge records predicted by FNWD (1) and NSWE (2) spherical models for initial sources of sizes $\W_{\,1}\,$(a -- c) and $\W_{\,2}\,$(d -- f). Wave gauges are located in $\M_{\,3}\,$(a), $\M_{\,4}\,$(b, d), $\M_{\,5}\,$(c, e) and $\M_{\,6}\,$(f).}
  \label{fig:6}
\end{figure}

Moreover, the influence of dispersion is known to grow with the traveled distance \cite{Segur2007, Arcas2012}. It can be also observed in Figure~\ref{fig:4}. For example, for the initial condition of the size $\W_{\,2}\,$, the frequency dispersion effect is not apparent yet on wave gauge $\M_{\,1}$ (see Figure~\ref{fig:4}(\textit{b}), line 1), but it starts to emerge in gauge $\M_{\,2}$ (see Figure~\ref{fig:4}(\textit{e}), line 1). Concerning the source of size $\W_{\,1}\,$, here the dispersion becomes to play its r\^ole even before the point $\M_{\,1}\,$. In order to estimate better the distance on which dispersive effects may become apparent, we create additional synthetic wave gauges in points $\bigl\{\M_{\,i}\bigr\}_{\,i\,=\,3}^{\,6}$ with coordinates $\bigl\{\bigl(\lambda_{\,i},\,\phi_{\,i}\bigr)\bigr\}_{\,i\,=\,3}^{\,6}\,$. The longitude is taken to be that of the source center, \ie $\lambda_{\,i}\ \equiv\ \lambda_{\,0}\ =\ 280^{\circ}\,$, $\forall i\ \in\ 3,\,\ldots,\,6\,$, while the latitude takes the following values: $\phi_{\,3}\ =\ -35^{\circ}\,$, $\phi_{\,4}\ =\ -30^{\circ}\,$, $\phi_{\,5}\ =\ -20^{\circ}\,$, $\phi_{\,6}\ =\ 0^{\circ}\,$. Thus, these wave gauges are located much closer to the source $\bigl(\lambda_{\,0},\,\phi_{\,0}\bigr)$ than $\M_{\,1}\,$. The recorded data is represented in Figure~\ref{fig:6}. In panels \ref{fig:6}(\textit{a -- c}) one can see that for smallest source $\W_{\,1}$ the dispersion is already fully developed in gauge $\M_{\,5}$ (and becomes apparent in gauge $\M_{\,4}$), \ie at the distance of $\approx 2200\;\km$ from the source. Indeed, in panel \ref{fig:6}(\textit{c}) the second wave of depression has the amplitude larger than the first wave. One can notice in general that the dispersion always yields a slight reduction of the leading wave (compare with NSWE curves in dash dotted lines (2)). To our knowledge this fact was first reported in numerical simulations in \cite{Chubarov1987}. Concerning the initial condition of size $\W_{\,2}\,$, the dispersion does not seem to appear even at the point $\M_{\,6}\,$, located $\approx 4400\;\km$ from the source (see Figure~\ref{fig:6}(\textit{d -- f})). As a result, we conclude that, at least for tsunami applications, the frequency dispersion effect is mainly determined by the size of the generation region and thus, by the wavelength of initially excited waves. The dispersion effect is decreasing when the source extent increases. A common point between FNWD and NSWE models is that on relatively short distances the period of the main wave increases and the wave amplitude decreases.


\subsubsection{Some rationale on the dispersion}

In order to estimate approximatively the distance $\ld$ on which dispersive effects become significant for a given wavelength $\lambda$ we consider a model situation. For simplicity we assume the bottom to be even with constant depth $d\,$. Let us assume also that wave dynamics is described by the so-called linearized \textsc{Benjamin}--\textsc{Bona}--\textsc{Mahony} (BBM) equation \cite{Peregrine1966, bona}:
\begin{equation*}
  \eta_{\,t}\ +\ \ups_{\,0}\,\eta_{\,x}\ =\ \nu\,\eta_{\,x\,x\,t}\,,
\end{equation*}
which can be obtained from linearized FNWD plane model reported in \cite{Khakimzyanov2016}. Here $\ups_{\,0}\ \eqdef\ \sqrt{g\,d}$ is the linear long wave speed and $\nu\ \eqdef\ \frac{d^{\,2}}{6}\,$. For this equation, the dispersion relation $\omega(\k)\,$, the phase speeds $\nuc$ and $\nuc_{\,\k}\,$, corresponding to the given wavelength $\lambda$ and wavenumber $\k$ respectively, are given by the following formulas:
\begin{equation*}
  \omega(\k)\ =\ \frac{\ups_{\,0}\,\k}{1\ +\ \nu\,\k^{\,2}}\,, \qquad
  \nuc\ =\ \frac{\ups_{\,0}}{1\ +\ \nu\,\Bigl(\,\dfrac{2\,\pi}{\lambda}\,\Bigr)^{\,2}}\,, \qquad
  \nuc_{\,\k}\ =\ \frac{\ups_{\,0}}{1\ +\ \nu\,\Bigl(\,\dfrac{2\,\pi}{\lambda_{\,\k}}\,\Bigr)^{\,2}}\,,
\end{equation*}
where $\lambda_{\,\k}\ \eqdef\ \dfrac{2\,\pi}{\k}\,$.

\begin{figure}
  \centering
  \includegraphics[width=0.69\textwidth]{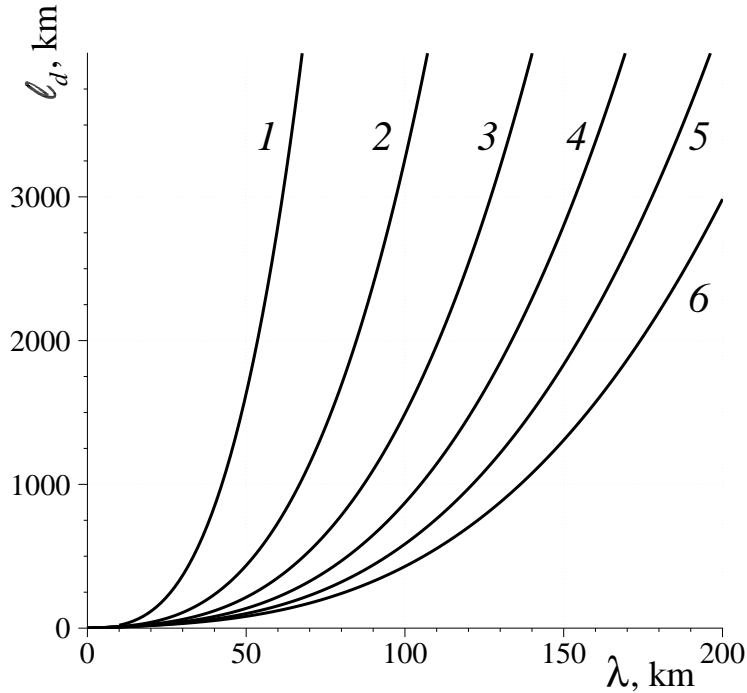}
  \caption{\small\em Dependence of the dispersion distance $\ld$ on the wavelength for parameter $\updelta\ =\ 0.33$ and for water depth $d\ =\ 1\;\km$ (1), $d\ =\ 2\;\km$ (2), $d\ =\ 3\;\km$ (3), $d\ =\ 4\;\km$ (4), $d\ =\ 5\;\km$ (5) and $d\ =\ 6\;\km$ (6).}
  \label{fig:7}
\end{figure}

During the time $t$ the generated wave travels the distance $\ld\ =\ \nuc\cdot t\,$, while a shorter wave of length $\lambda_{\,\k}\ <\ \lambda$ will travel a shorter distance
\begin{equation*}
  \ell_{\,\k}\ =\ \nuc_{\,\k}\cdot t\ =\ \ld\;\frac{\nuc_{\,\k}}{\nuc}\ <\ \ld\,.
\end{equation*}
The difference in traveled distances is a manifestation of the frequency dispersion. Now, we assume that the wavelength of a separated wave is related to $\lambda$ as
\begin{equation*}
  \lambda_{\,\k}\ =\ \updelta\,\lambda\,, \qquad \updelta\ \in\ (\,0,\,1\,)\,,
\end{equation*}
and we assume that by time $t$ the dispersion had enough time to act, \ie
\begin{equation*}
  \ell_{\,\k}\ =\ \ld\ -\ (\,1\ +\ \updelta\,)\;\frac{\lambda}{2}\,.
\end{equation*}
From this equality we can determine $\ld\,$:
\begin{equation*}
  \ld\,\Bigl(1\ -\ \frac{\nuc_{\,\k}}{\nuc}\Bigr)\ =\ \frac{1\ +\ \updelta}{2}\;\lambda\,,
\end{equation*}
and if we substitute the expressions of phase speeds $\nuc$ and $\nuc_{\,\k}$ we obtain:
\begin{equation*}
  \ld\;\Biggl\{\,1\ -\ \dfrac{\dfrac{\ups_{\,0}}{1\ +\ \nu\,\bigl(\frac{2\,\pi}{\lambda_{\,\k}}\bigr)^{\,2}}}{\dfrac{\ups_{\,0}}{1\ +\ \nu\,\bigl(\frac{2\,\pi}{\lambda}\bigr)^{\,2}}}\,\Biggr\}\ =\ \frac{1\ +\ \updelta}{2}\;\lambda\,.
\end{equation*}
By using the fact that $\lambda_{\,\k}\ =\ \updelta\,\lambda$ we obtain the final expression for the \emph{dispersion distance}:
\begin{equation}\label{eq:ld}
  \ld\ =\ \frac{1}{2\,(\,1\ -\ \updelta\,)}\,\biggl[\,\lambda\ +\ \frac{3}{2\,\pi^{\,2}}\;\updelta^{\,2}\;\frac{\lambda^{\,3}}{d^{\,2}}\,\biggr]\ \approx\ \frac{1}{2\,(\,1\ -\ \updelta\,)}\,\biggl[\,\lambda\ +\ 0.152\,\updelta^{\,2}\;\frac{\lambda^{\,3}}{d^{\,2}}\,\biggr]\,.
\end{equation}
The dependence of the dispersive distance $\ld$ on the wavelength $\lambda$ is depicted in Figure~\ref{fig:7}.

Let us mention that an analogue of formula \eqref{eq:ld} was obtained earlier in \cite{Mirchina1982, Pelinovsky1996a} based on the (linearized) \textsc{Korteweg}--\textsc{de Vries} equation \cite{Boussinesq1877, KdV}.

\bigskip
\paragraph*{Numerical application.}
In order to apply formula \eqref{eq:ld} (and check its validity) to our results, we have to estimate first the generated wavelength $\lambda$ and parameter $\updelta\,$. We saw above that for the initial condition of extent $\W_{\,1}$ the wave gauge $\M_{\,5}$ registered a well-separated secondary wave with the period about three times smaller. Thus, we can set $\updelta\ =\ 0.33\,$. Concerning the wavelength of the generated wave, in such estimations it is roughly identified\footnote{However, in the general case this approximation might be rather poor \cite{Pelinovsky1996a}.} with the size of the generated region $\W_{\,1}\,$ \cite{Kirby2013}. Under these assumptions, equation \eqref{eq:ld} gives us the dispersive distance $\ld\ =\ 1058\;\km\,$. Thus, we can conclude that dispersive effects should be apparent in wave gauge $\M_{\,5}\,$, which is in the perfect agreement with our numerical simulations. Let us take a source region with a larger horizontal extent $\W_{\,2}\,$. If we take $\updelta\ =\ 0.33$ (as above) and $\lambda\ =\ 339\;\km\,$, formula \eqref{eq:ld} gives us $\ld\ =\ 30\,139\;\km\,$. Thus, we conclude that dispersive effects will not be seen even in the farthest wave gauge $\M_{\,2}\,$.

\bigskip
\paragraph*{Alternative approaches.}
There exist other criteria of the importance of frequency dispersion effects. One popular criterium is the so-called \textsc{Kajiura} number $\Ka$ employed, for example, in \cite{Kirby2013}:
\begin{equation*}
  \Ka\ \eqdef\ \Bigl(\,\frac{6\,d}{\ell}\,\Bigr)^{\frac{1}{3}}\cdot\frac{\W}{d}\,,
\end{equation*}
where $\W$ is the source region width and $\ell$ is the distance traveled by the wave. It is believed that dispersive effects manifest if $\Ka\ <\ 4\,$. Let us check it against our numerical simulations. In the case depicted in Figure~\ref{fig:6}(\textit{c}) we have $\W\ =\ 107.3\;\km$ and $\ell\ =\ 2\,200\;\km\,$. Thus, $\Ka\ \approx\ 6\ >\ 4\,$. According, to \textsc{Kajiura}'s criterium, the dispersion should not appear yet at the point $\M_{\,5}\,$. However, it contradicts clearly our direct numerical simulations shown in Figure~\ref{fig:6}(\textit{c}). It was already noticed in \cite{Glimsdal2013} that criterium $\Ka\ <\ 4$ is too stringent. It was proposed instead to use a more sophisticated criterium based on the \emph{normalized dispersion time}:
\begin{equation*}
  \upvartheta\ \eqdef\ 6\,\sqrt{g\,d}\;\frac{d^{\,2}}{\lambda^{\,3}}\;t\,.
\end{equation*}
If $\upvartheta\ >\ 0.1$ the wave propagation is considered to be dispersive. In our simulations $t\ =\ 12\,000\;\s\,$. It is the time needed for the wave to travel to the location $\M_{\,5}\,$. Thus, in our case $\upvartheta\ \approx\ 0.18\ >\ 0.1$ and the dispersion should be already visible in location $\M_{\,5}$ for the source size $\W_{\,1}\,$, which is in perfect agreement with our synthetic wave record (see again Figure~\ref{fig:6}(\textit{c})).

In our opinion, the use of the criterium \eqref{eq:ld} based on the dispersion distance $\ld$ is preferable, since it gives directly an estimation of the distance, where the dispersive effects will become non-negligible. Outside the circle of radius $\ld$ and centered at the source the use of hydrostatic non-dispersive models is not advised.

Let us mention also a few earlier works where the importance of dispersive effects was studied for real-world events using WNWD models. In \cite{Grilli2007, Horrillo2006} the \textsc{Sumatra} 2004 \textsc{Indian} ocean tsunami was studied and it was shown that in the deep West part of the \textsc{Indian} ocean the discrepancy in wave amplitudes reaches 20\% (between WNWD and NSWE models). For the same tsunami event the discrepancy of 60\% was reported in \cite{Tkalich2007}. Numerical simulations of the \textsc{Tohoku} 2011 tsunami event reported in \cite{Kirby2013} confirmed again the difference of 60\% between hydrostatic and non-hydrostatic model predictions.


\subsection{Bulgarian 2007 tsunami}
\label{sec:bulgary}

The basin of the \textsc{Black Sea} is subject to a relatively important seismic activity and there exists a potential hazard of tsunami wave generation, which may be caused not only by earthquakes, but also by underwater landslides, which can be triggered even by weak seismic events. Geophysical surveys show that large portions of the \textsc{Black Sea} continental shelf contain unstable masses \cite{Evsyukov2009}, which have to be taken into account while assessing tsunami hazard for the population or and underwater infrastructure. Moreover, some past events are well documented \cite{Kazantsev1998}.

Some anomalous oscillations of the sea level were registered on the 7\up{th} of May 2007 at Bulgarian coasts. In \cite{Ranguelov2008} it was conjectured that a landslide may have provoked these oscillations and the authors considered four possible locations of this hypothetical landslide. In all these cases the landslide started at the water depth about $100\;\m$ and at the distances of $30$ -- $50\;\km$ from the coast. The suggested volume of the landslide is between $30$ and $60$ millions of $\m^{\,3}\,$. Landslide thickness is about $20$ -- $40\;\m\,$. In \cite{Ranguelov2008} the authors represented the landslide as a system of interconnected solid blocks which can move along the slope under the force of gravity whose action is compensated by frictional effects. The hydrodynamics was described by NSWE incorporating some viscous effects solved with the Finite Element Method (FEM). For all four considered cases simulations show that the landslide achieves the speed of about $20\;\m/\s$ in $200$ -- $300\;\s$ after the beginning of the motion. Landslide's motion stops at the depth around $1000\;\m$ after running out about $20\;\km$ from its initial position.

The hypothesis of landslide mechanism is studied by confronting numerical predictions with eyewitness reports and coastal wave gauge data. In particular, we know the values of lowest (\ie negative) and highest amplitudes for seven locations along the coast respectively:
\begin{description}
  \item[Shabla] $-1.5\;\m\,$, $0.9\;\m\,$
  \item[Bolata] $-1.3\;\m\,$, $0.9\;\m\,$
  \item[Dalboka] $-2.0\;\m\,$, $1.2\;\m\,$
  \item[Kavarna] $-1.8\;\m\,$, $0.9\;\m\,$
  \item[Balchik] $-1.5\;\m\,$, $1.2\;\m\,$
  \item[Varna] $-0.7\;\m\,$, $0.4\;\m\,$
  \item[Galata] $-0.2\;\m\,$, $0.1\;\m\,$.
\end{description}
The comparisons from \cite{Ranguelov2008} show that one can seemingly adopt the landslide mechanism hypothesis. However, even the most plausible landslide scenario (among four considered) does not give a satisfactory agreement with \emph{all} available field data. Moreover, the maximal synthetic amplitudes are shifted to the South (towards \textsc{Emine}), which was not observed during the real event.

In a companion study \cite{Vilibic2010} the authors investigated also the hypothesis of a meteo-tsunami responsible of anomalous waves recorded on the 7\up{th} of May 2007 at Bulgarian coasts. Their analysis showed that the weather conditions could provoke anomalous waves near \textsc{Bulgarian} coasts. The numerical simulations show again a good agreement of maximal amplitudes at some locations, even if we cannot speak yet about a good general agreement. In particular, numerical predictions seriously underestimate wave amplitudes in Northern parts of the coast such as \textsc{Shabla} and overestimate them in Southern ones (\eg large waves were not observed in \textsc{Burgas} Bay, but they existed in numerical predictions).

In the present study we continue to develop the landslide-generated hypothesis of anomalous waves. In contrast to the previous study \cite{Ranguelov2008}, we employ the FNWD model presented above and the landslide will be modeled using the quasi-deformable body paradigm \cite{Beisel2012}. The driving force was taken as the sum of the gravity, buoyancy, friction and water drag forces acting on elementary volumes. Quasi-deformability property means that the landslide can deform in order to follow complex bathymetry profiles by preserving the general shape (see \cite{Dutykh2011d, Beisel2012, Dutykh2012} for more details). However, the deformation process is such that the horizontal components of the velocity vector are the same throughout the sliding body (as in the absolutely rigid case). This model has been validated against experimental data \cite{Gusev2014} and direct numerical simulations of the free surface hydrodynamics \cite{Gusev2013}. Moreover, it was already successfully applied to study numerically a real world tsunami which occurred in \textsc{Papua New Guinea} on the 17\up{th} of July 1998 \cite{Khakimzyanov2015c}. Important parameters, which enter in our landslide model and, thus, that have to be prescribed are:
\begin{description}
  \item[$V$] landslide volume
  \item[$C_w$] added mass coefficient
  \item[$C_d$] drag coefficient
  \item[$C_f\ \equiv\ \tan\theta^{\,\ast}$] friction coefficient and $\theta^{\,\ast}$ is the friction angle
  \item[$\gamma\ \eqdef\ \frac{\rho_{s}}{\rho_{w}}\ >\ 1$] ratio between water $\rho_{w}$ and sliding mass $\rho_{s}$ densities.
\end{description}
The initial shape of the landslide is given by the following formula:
\begin{equation*}
  h_{\,s}^{\,0}\,(x,\,y)\; =\; \begin{dcases}
    \;\frac{T}{4}\;\biggl[\,1\; +\; \cos\Bigl(\,\frac{2\,\pi(x\; -\; x_{\,c}^{\,0})}{B_{\,x}}\,\Bigr)\,\biggr]\cdot\biggl[\,1\; +\; \cos\Bigl(\,\frac{2\,\pi(y\; -\; y_{\,c}^{\,0})}{B_{\,y}}\,\Bigr)\,\biggr]\,,\ &(x,\,y)\; \in\; \D_{\,0}\,, \\
    \;0\,,\ &(x,\,y)\; \notin\; \D_{\,0}\,,
  \end{dcases}
\end{equation*}
where $\D_{\,0}\ =\ \Bigl[\,x_{\,c}^{\,0}\ -\ \dfrac{B_{\,x}}{2},\, x_{\,c}^{\,0}\ +\ \dfrac{B_{\,x}}{2}\,\Bigr]\times\Bigl[\,y_{\,c}^{\,0}\ -\ \dfrac{B_{\,y}}{2}\,, y_{\,c}^{\,0}\ +\ \dfrac{B_{\,y}}{2}\,\Bigr]$ is the domain occupied by the sliding mass, $B_{\,x,\,y}$ are horizontal extensions of the landslide along the axes $O\,x$ and $O\,y$ respectively, $\bigl(\,x_{\,c}^{\,0},\,y_{\,c}^{\,0}\,\bigr)$ is the position of its barycenter and $T$ is its thickness.

\begin{remark}
For the sake of notation compactness, in the landslide description above we used \textsc{Cartesian} coordinates. This approximation is valid since landslide size is small enough to `feel' \textsc{Earth}'s sphericity. We place the origin in the left side center of the spherical rectangular computational domain, \ie in the point $\bigl(27^{\circ},\,43^{\circ}\bigr)\,$. Then, the local \textsc{Cartesian} coordinates are introduced in the following way:
\begin{equation*}
  x\ =\ R\;\frac{\pi}{180}\;(\lambda\ -\ 27)\,\cos\Bigl(\frac{\pi}{180}\;43\Bigr)\,, \qquad
  y\ =\ R\;\frac{\pi}{180}\;(\phi\ -\ 43)\,.
\end{equation*}
If $\bigl(\,\lambda_{\,c}^{\,0},\,\phi_{\,c}^{\,0}\,\bigr)$ are spherical coordinates of the landslide barycenter, then its local \textsc{Cartesian} coordinates $\bigl(\,x_{\,c}^{\,0},\,y_{\,c}^{\,0}\,\bigr)$ are computed accordingly:
\begin{equation*}
  x_{\,c}^{\,0}\ =\ R\;\frac{\pi}{180}\;(\lambda_{\,c}^{\,0}\ -\ 27)\,\cos\Bigl(\frac{\pi}{180}\;43\Bigr)\,, \qquad
  y_{\,c}^{\,0}\ =\ R\;\frac{\pi}{180}\;(\phi_{\,c}^{\,0}\ -\ 43)\,.
\end{equation*}
\end{remark}

\begin{figure}
  \centering
  \includegraphics[width=0.99\textwidth]{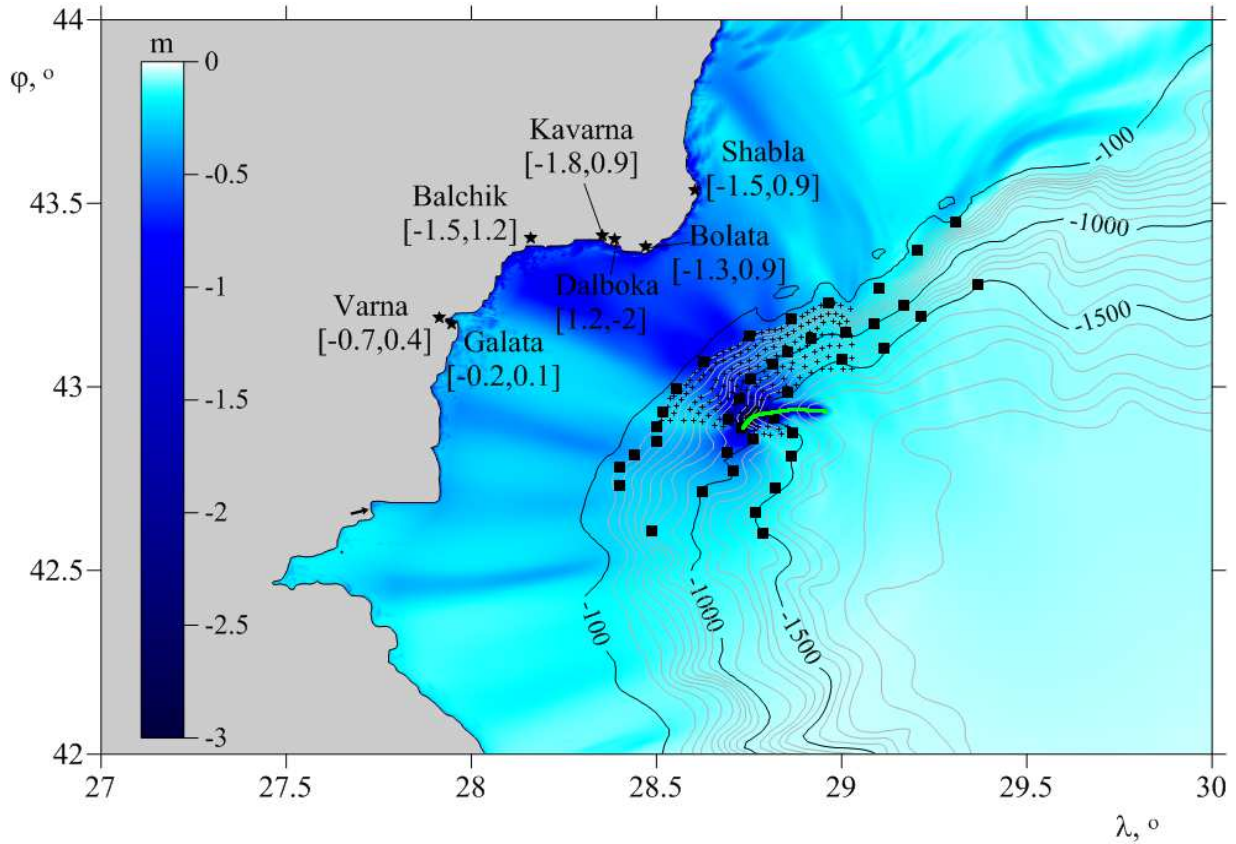}
  \caption{\small\em Computational domain and the distribution of minimal amplitudes of the waves generated by the \emph{optimal} landslide \eqref{eq:opt} computed during the first $3\;\h$ of the physical propagation time. With black lines we represent iso-values of the bathymetry. With symbols $\star$ we denote coastal towns where we know the maximal and minimal wave amplitudes. Little black rectangles and pluses denote various initial positions of the landslide considered in our study. The green line shows the most probably landslide trajectory. Finally, the little black arrow shows the starting point and direction along the coastline where we record maximal and minimal wave amplitudes.}
  \label{fig:8}
\end{figure}

During the modelling of landslide events in the \textsc{Black Sea} we used the parameters of some historical events \cite{Kazantsev1998}. We noticed that the most sensitive parameter is the initial location of the landslide. That is why in the present study we focus specifically on this aspect in order to shed some light on this unknown parameter. In this perspective we chose $40$ different initial locations along the \textsc{Bulgarian} coastline which were located mainly at the depth of $200\;\m\,$, $1000\;\m$ and $1\,500\;\m\,$. These locations are depicted with black rectangles in Figure~\ref{fig:8}. Other parameters are given in Table~\ref{tab:slide}. The volume $V$ of the landslide in our simulations is equal to $62.5\times 10^{\,6}\;\m^{\,3}\,$, which is close to the value used in \cite{Ranguelov2008}. We notice that there are two competing effects in our problem. If we increase the initial landslide depth, the amplitude of generated waves will be seriously reduced. However, this effect can be compensated by increasing the landslide thickness $T\,$. In general, the amplitude of waves is proportional to $T\,$.

\begin{table}
  \centering
  \begin{tabular}{l|r}
    \hline\hline
    \textit{Parameter} & \textit{Value} \\
    \hline\hline
    Added mass coefficient, $C_w$ & $1.0$ \\
    Drag coefficient, $C_d$ & $1.0$ \\
    Densities ratio, $\gamma$ & $2.0$ \\
    Friction angle, $\theta^{\,\ast}$ & $1^{\circ}$ \\
    Landslide thickness, $T$ & $40\;\m$ \\
    Landslide length, $B_{\,x}$ & $2\,500\;\m$ \\
    Landslide width, $B_{\,y}$ & $2\,500\;\m$ \\
    Landslide volume, $V$ & $62.5\times 10^{\,6}\;\m^{\,3}$ \\
    \hline\hline
  \end{tabular}
  \bigskip
  \caption{\small\em Physical parameters used to simulate the hypothetical landslide motion during Bulgarian tsunami of the 7\up{\,th} of May $2007\,$.}
  \label{tab:slide}
\end{table}

In these numerical simulations we use the finest angular resolution (in this section) of $3.75^{\,\dprime}\,$, since the domain is relatively compact. It corresponds to the grid of $2\,881\times 1\,921$ nodes. The bathymetry data was obtained by applying bilinear interpolation to data retrieved from ``GEBCO One Minute Grid -- 2008''. The computational (CPU) time of each run was about $35\;\h$ for this resolution. Some information on the wave propagation can be obtained using the so-called radiation diagrams, which represents the spatial distribution of maximal and minimal wave amplitudes\footnote{Here we mean maximal positive and minimal negative waves with respect to the still water level.} during the whole simulation time. After computing the first $40$ scenarii (marked with little black rectangles in Figure~\ref{fig:8}), we could delimit\footnote{Here we stress again that the computational domain was the same, but we delimit the search area for the landslide initial position.} the area where the hypothetical landslide could take place. Then, in the second time, this area was refined with additional $171$ initial landslide locations marked with pluses in Figure~\ref{fig:8}. In this way, by comparing simulation results with available field data, we could choose the most probable scenario of the initial location of the landslide. Finally, we performed the third optimisation cycle in order to determine the most likely landslide thickness $T\,$. The optimal parameters are given here:
\begin{equation}\label{eq:opt}
  \lambda_{\,c}^{\,0}\ =\ 28.7341^{\circ}\,, \qquad
  \phi_{\,c}^{\,0}\ =\ 42.8871^{\circ}\,, \qquad
  T\ =\ 320\;\m\,, \qquad
  B_{\,x}\ =\ B_{\,y}\ =\ 2\,500\;\m\,.
\end{equation}
The radiation diagram for this tsunami event is shown in Figure~\ref{fig:8}. One can see, in particular, that the radiation of minimal (negative) amplitudes is directed towards the coastal towns where the most significant oscillations of the sea level were observed. We notice also that in our simulations we do not obtain a two-tongue structure predicted in \cite[Figure~2]{Ranguelov2008}. However, such radiation diagrams are very sensitive to the initial location of the landslide. For some starting points we observed (as in an earlier work \cite{Khakimzyanov2015c}), for example, an abrupt termination of landslide motion, which affected quite a lot the radiation diagram.

\begin{figure}
  \centering
  \includegraphics[width=0.99\textwidth]{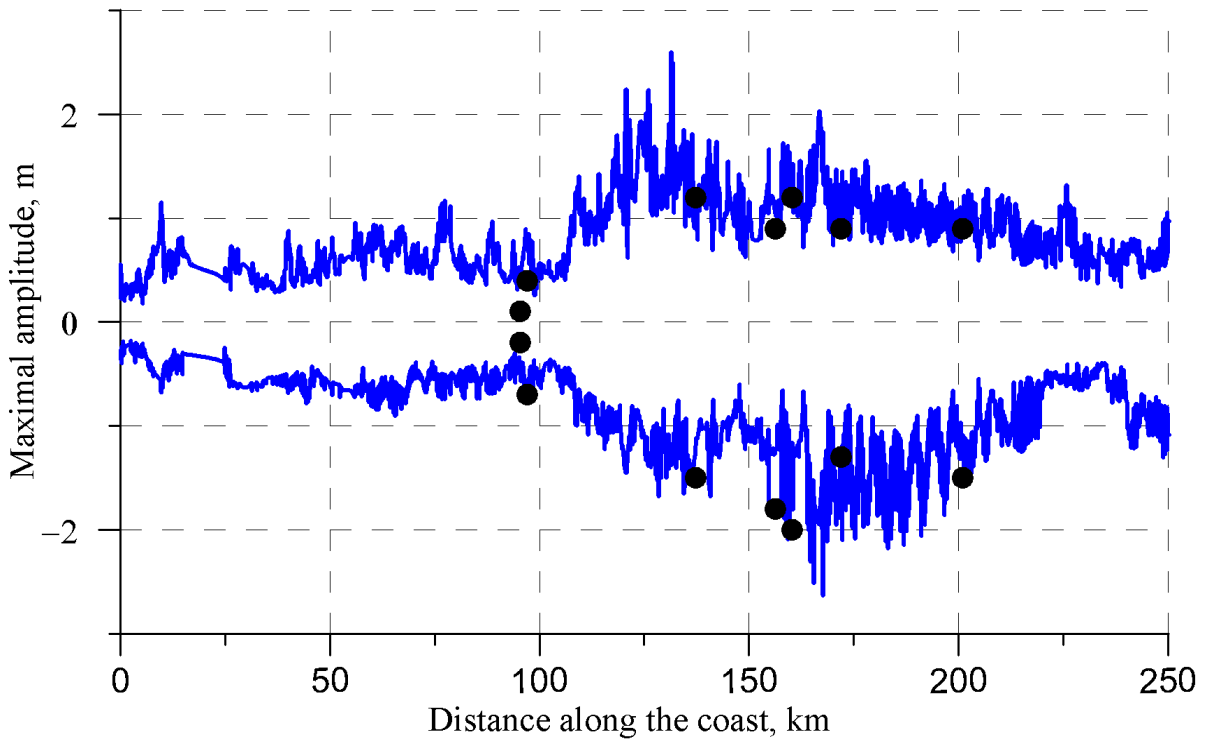}
  \caption{\small\em Minimal and maximal oscillations of the free surface along the \textsc{Bulgarian} coastline recorded during the whole numerical simulation of the most likely landslide event. Blue solid lines are numerical predictions and black circles represent field data.}
  \label{fig:9}
\end{figure}

Minimal and maximal wave amplitudes recorded along the \textsc{Bulgarian} coast are shown in Figure~\ref{fig:9}. The starting point of the path along the coastline is shown with a little black arrow in Figure~\ref{fig:8}. We mention also that the distance is computed on our grid (thus, the coastline is approximated \textit{in fine} by a polygon). Hence, there might be little discrepancies (in the sense of overestimations) with `real-world' distances. Figure~\ref{fig:9} shows an overall good agreement with field data. Moreover, in contrast to the previous study \cite{Ranguelov2008}, our scenario does not trigger large wave amplitudes in Southern parts of the \textsc{Bulgarian} coast.

We have to mention that a few other scenarii gave comparable agreement with field data. However, if we look carefully at corresponding landslide trajectories, they all pass through the termination point of the optimal landslide\footnote{We have to say that even the optimal landslide may move farther if we slightly decrease the friction angle $\theta^{\,\star}\,$.} \eqref{eq:opt}. Consequently, we can only suggest to study this area of the \textsc{Black Sea} in the perspective to discover eventually the deposits of this past landslide event.

We underline also that the geometrical extensions of the `optimal' landslide \eqref{eq:opt} can be changed without loosing too much the good agreement with observations demonstrated in Figure~\ref{fig:9}. For example, we used the following parameters to trigger a tsunami wave with similar amplitudes in observation points:
\begin{equation}\label{eq:opt2}
  \lambda_{\,c}^{\,0}\ =\ 28.7341^{\circ}\,, \qquad
  \phi_{\,c}^{\,0}\ =\ 42.8871^{\circ}\,, \qquad
  T\ =\ 110\;\m\,, \qquad
  B_{\,x}\ =\ B_{\,y}\ =\ 5\,000\;\m\,.
\end{equation}
In general, such modifications may alter seriously the landslide trajectory and velocity. However, in this particular case the new trajectory followed closely the green line depicted in Figure~\ref{fig:8}.


\subsubsection{Dispersive effects}
\label{sec:dispB}

\begin{figure}
  \centering
  \subfigure[]{\includegraphics[width=0.48\textwidth]{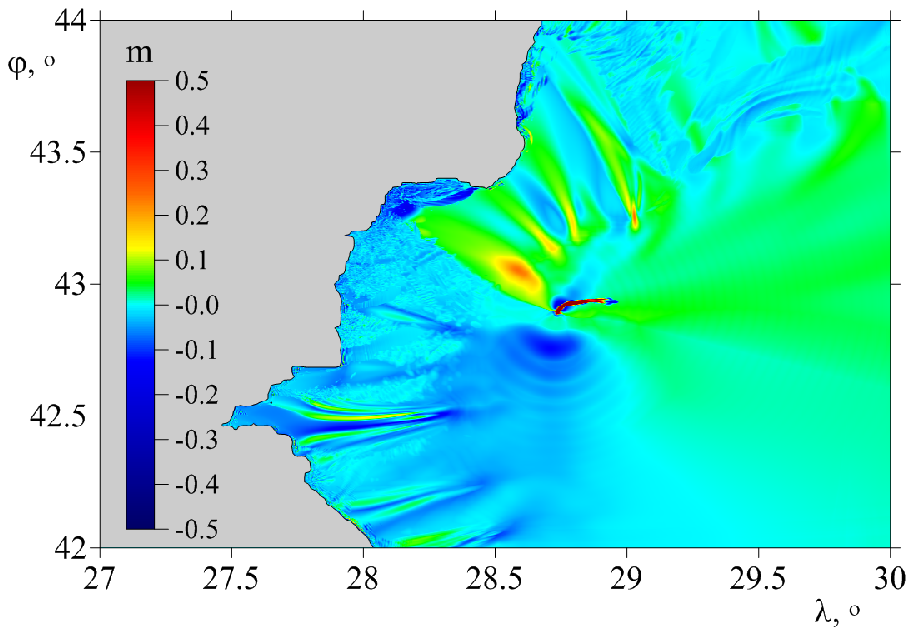}}
  \subfigure[]{\includegraphics[width=0.48\textwidth]{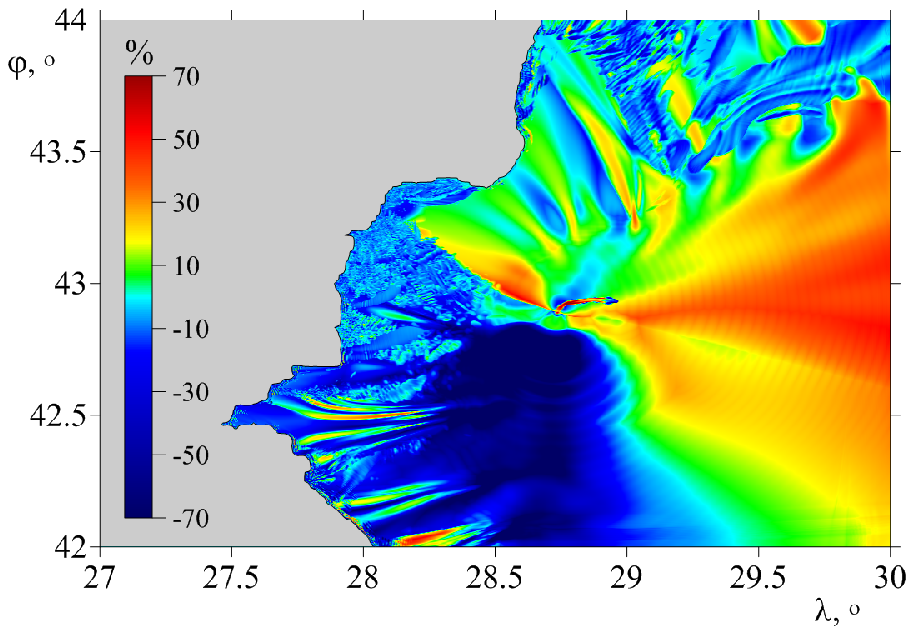}}
  \caption{\small\em The spatial distribution of absolute (a) and relative (b) differences in the maximal positive wave amplitudes computed according to FNWD and NSWE models for the optimal landslide \eqref{eq:opt}.}
  \label{fig:10}
\end{figure}

The influence of the frequency dispersion in this particular tsunami event will be estimated by computing the absolute and relative\footnote{In the relative difference we divide by the magnitude of the NSWE prediction.} differences between radiation diagrams (\ie maximal amplitudes) computed with FNWD and NSWE models. In this way, in Figure~\ref{fig:10} we show how the incorporation of non-hydrostatic effects modifies extreme (positive) wave amplitudes. In particular, one can see that relative differences can reach up to $70$\% in deep parts of the \textsc{Black Sea}. We computed differences of radiative diagrams for the nearly-optimal landslide \eqref{eq:opt2} (with smaller thickness $T$) and it seems that the dispersion plays smaller effect in that case. In general, when there is an abrupt termination of landslide motion, the differences between dispersive and non-dispersive predictions increase. Our computations show also that the period of principal wave components is between $250$ -- $400\;\s$ and waves of maximal (absolute) amplitude do not always come first (in a good agreement with observations). To save the space we do not provide these numerical results here. The main goal of this Section is to demonstrate that FNWD models can be successfully used to study real world events on all scales from regional to global ones. Even better, today one can successfully perform parametric studies with spherical FNWD models: in order to determine the optimal landslide parameters given in \eqref{eq:opt}, we had to run $211$ scenarii in total.

\subsection{Chilean 2010 tsunami}
\label{sec:chile}

In order to illustrate the application of our spherical FNWD model to a real-world large scale seismically generated tsunami event we consider the \textsc{Chilean} tsunami which took place on the 27\up{th} of February 2010. Earthquake epicenter was located under the Ocean $117\;\km$ to the North from \textsc{Concepci\'on} at the depth of about $35\;\km$ below the bottom. This event was estimated to have the seismic moment magnitude $\mathrm{M}_{\mathrm{w}}\ =\ 8.8\,$. This earthquake generated a tsunami wave, which was observed in the whole \textsc{Pacific Ocean}. The most important aspect for us is that this wave was registered at DART buoys. Many scientific works are devoted to the investigation of this particular event. Here we mention a few numerical studies which go along the lines of our own work \cite{Paranas-Carayannis2010, Abraimi2014, Wen2011}. Contrary to the catastrophic \textsc{Tohoku} 2011 event, where researchers had to introduce local landslide hypothesis in order to explain some extreme run-up values \cite{Tappin2014}, \textsc{Chilean} 2010 event seems to be purely seismic since the available data on this tsunami can be reproduced fairly well starting from the initial water column disturbances caused by the earthquake solely.

In order to reconstruct the displacement field of the \textsc{Earth} surface, some authors use GPS data \cite{Delouis2010, Moreno2010, Vigny2011}. Later, these seismic scenarii were tested in \cite{Abraimi2014} to confront them with available tsunami field data. The final agreement quality dependent on the chosen scenario. In the present work we consider the fourth alternative proposed by USGS. Earth surface displacements were reconstructed using the celebrated \textsc{Okada} solution \cite{okada92, Okada85}. Then, this displacement was transferred to the free surface as the initial condition for our hydrodynamic computations. This tsunami generation procedure is known as the `passive generation' approach \cite{ddk, Kervella2007}. We would like to mention that there are noticeable discrepancies among all these seismic inversions. Thus, there is an uncertainty in the initial condition for tsunami wave propagation \cite{Dias2006, Dutykh2006, Dias2014}. Our choice for the USGS inversion can be explained essentially by the immediate availability of their data through their web site.

\begin{figure}
  \centering
  \includegraphics[width=0.99\textwidth]{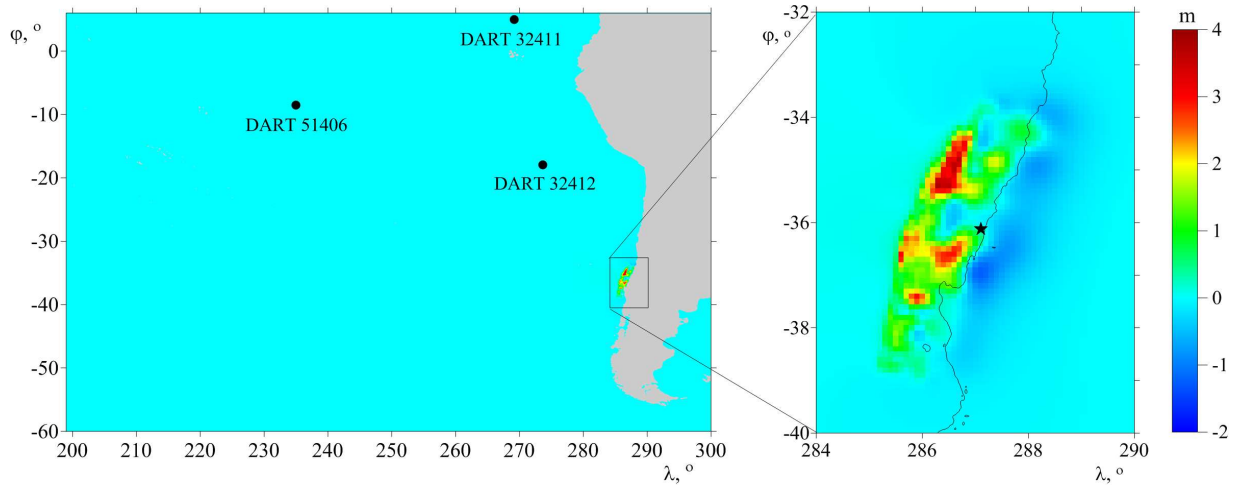}
  \caption{\small\em The computational domain (a) and the initial wave elevation (b) computed according to USGS inversion of the Chilean 2010 earthquake. Symbol $\star$ on the right panel shows the earthquake epicenter.}
  \label{fig:11}
\end{figure}

\begin{figure}
  \centering
  \includegraphics[width=0.99\textwidth]{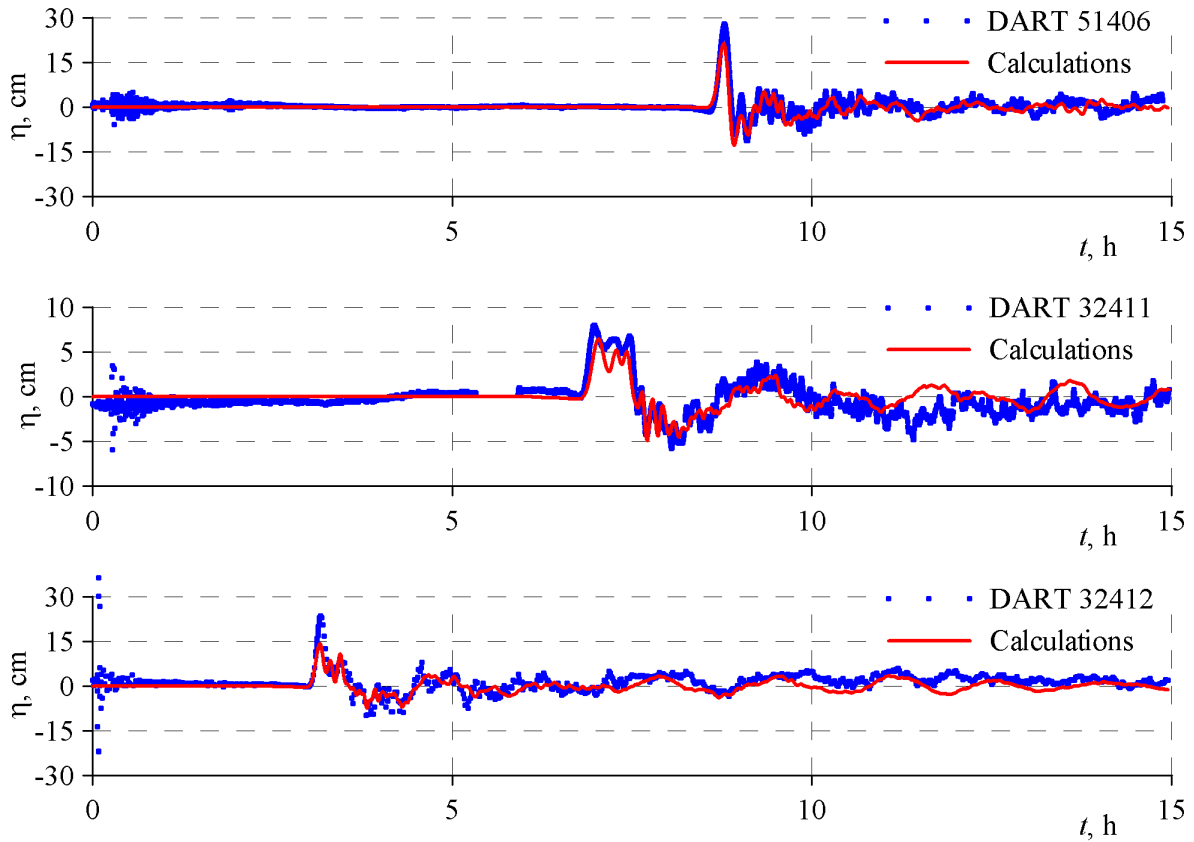}
  \caption{\small\em Comparison of our numerical predictions with the spherical FNWD model with DART data.}
  \label{fig:12}
\end{figure}

The computational domain used in our simulations covers a significant portion of the \textsc{Pacific} ocean --- $\bigl[\,199^{\circ},\,300^{\circ}\,\bigr]\times\bigl[\,-60^{\circ},\,5^{\circ}\,\bigr]\,$. We used a $1^{\prime}$ grid and the bathymetry data was taken from ``The GEBCO One Minute Grid --- 2008''. The use of finer grids or computation in significantly larger domains does not seem to be feasible with serial codes (see Section~\ref{sec:persp}). In order to validate our code we present the comparisons of predicted tsunami waves against three DART buoys: DART--32411, DART--32412 and DART--51406. The locations of these buoys are shown in Figure~\ref{fig:11}(\textit{a}) and the initial condition is represented in Figure~\ref{fig:11}(\textit{b}). Comparisons of FNWD predictions against aforementioned DART data is shown in Figure~\ref{fig:12}. Buoys data was downloaded from the National Oceanic and Atmospheric Administration (NOAA) web site. In the case of DART--32411 and DART--51406 buoys a vertical translation of data was needed to adjust the still water level. One can see an overall good agreement in Figure~\ref{fig:12} between our simulation and real-world data. The first oscillations present in DART data for $t\ <\ 1\;\h$ do not seem to be related to the studied tsunami event, since the wave did not have enough time to travel from the source region to the observation point. One can see also that synthetic records have somehow smaller amplitudes. It can be related to our choice of the initial condition (USGS) which did not take into account tsunami-related constraints during the inversion process \cite{Abraimi2014}.

\subsubsection{Dispersive effects}

\begin{figure}
  \centering
  \includegraphics[width=0.99\textwidth]{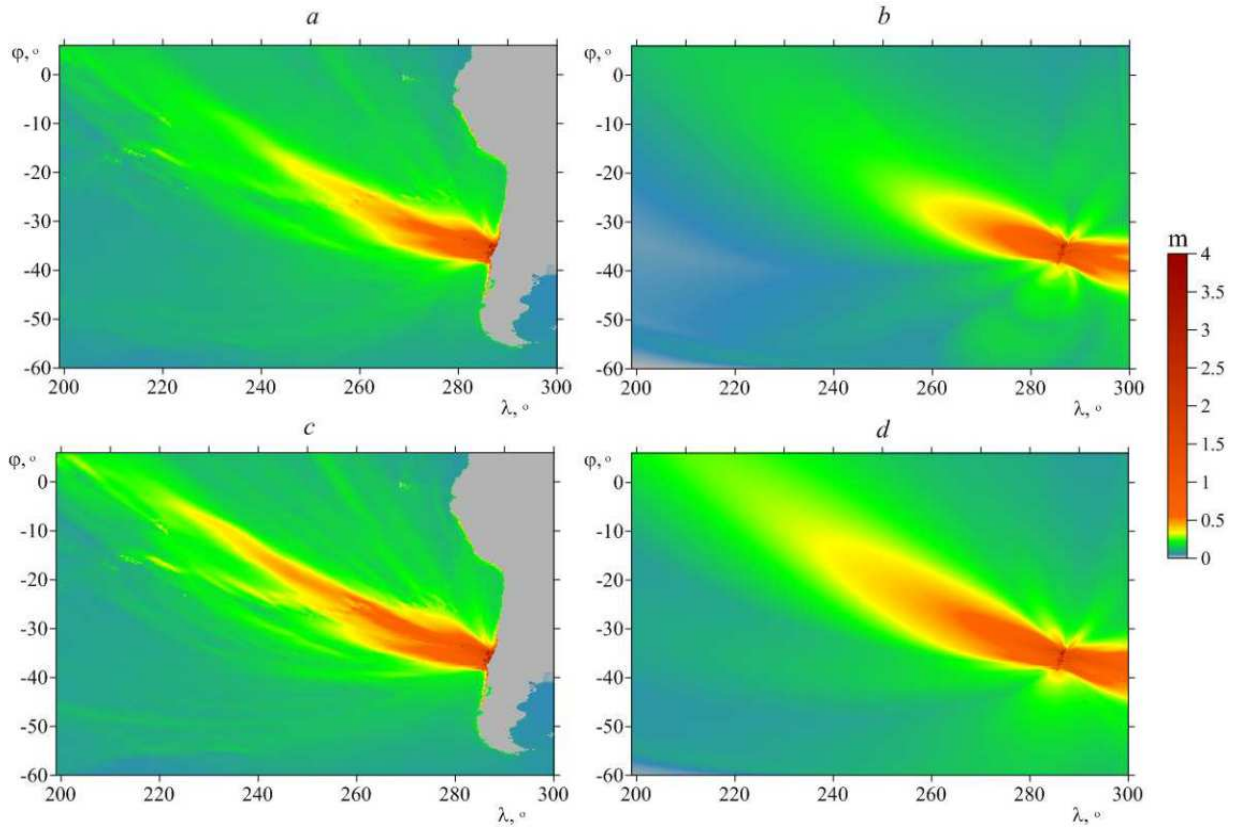}
  \caption{\small\em The distribution of maximal positive wave amplitudes predicted with FNWD model (a, b) and NSWE (c, d). The real bathymetry data is used in computations (a, c), while the even bottom of constant depth $d\ =\ 4\;\km$ is used in (b, d).}
  \label{fig:13}
\end{figure}

In order to estimate the influence of dispersive effects in this particular tsunami event, we perform two simulations --- with FNWD and NSWE models. Moreover, we consider one case with the real bathymetry data and another one with an even bottom of constant $d\ =\ 4\;\km$ depth. Radiation diagrams for all these four cases are shown in Figure~\ref{fig:13}. One can see that FNWD and NSWE predict significantly different radiation diagrams\footnote{Compare the top panels \ref{fig:13}(\textit{a, b}), which show FNWD result with lower panels \ref{fig:13}(\textit{c, d}) representing NSWE predictions.}. The difference becomes even more flagrant in the idealized constant depth case\footnote{Compare panels \ref{fig:13}(\textit{b}) and \ref{fig:13}(\textit{d}).}. Bottom irregularities contribute equally to radiation diagrams even if they fail to alter the maximal radiation direction (at least in this particular tsunami event). It seems that here the initial condition has the dominant r\^ole in shaping the triggered tsunami wave.

\begin{figure}
  \centering
  \includegraphics[width=0.99\textwidth]{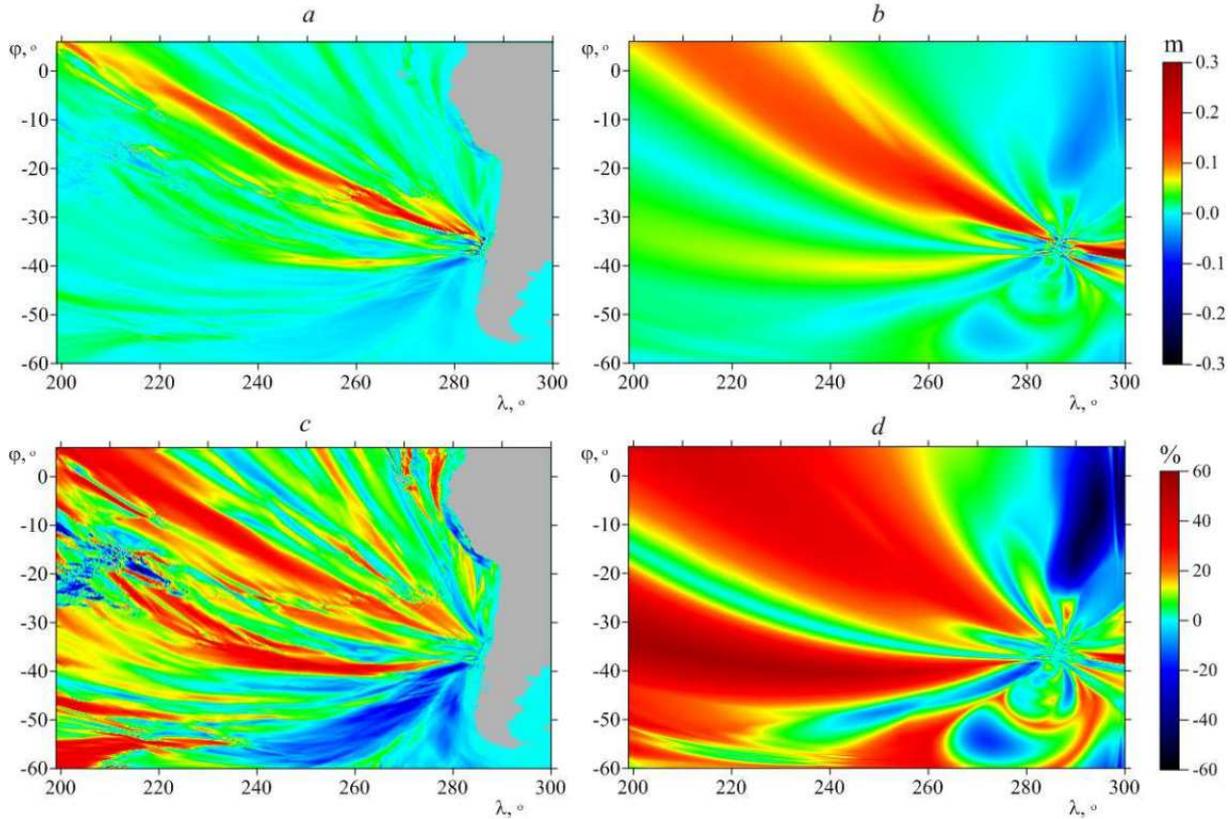}
  \caption{\small\em The absolute (a, b) and relative (c, d) differences among radiation diagrams computed according to FNWD and NSWE models on a real bathymetry (a, c) and on an even constant bottom (b, d).}
  \label{fig:14}
\end{figure}

In order to highlight the differences between NSWE and FNWD models predictions, we present in Figure~\ref{fig:14} the absolute and relative\footnote{While computing the relative difference, we divide by the magnitude of the NSWE prediction as we did it above in Section~\ref{sec:dispB}.} differences among the corresponding radiation diagrams. The biggest absolute differences are concentrated along the main radiation direction and NSWE model seems to overestimate substantially the wave amplitude. The picture of relative differences has a much more complex structure even in the idealized case. The largest relative differences attain easily $60$\% not only along the main radiation direction, but also to the South from the epicenter. We can only conclude that the frequency dispersion has to be taken into account in this event. However, the dispersion effect may vary with the tsunami initial condition \cite{Gusev2016}. Consequently, it is not excluded that for other seismic inversions the initial free surface shape may change such that the dispersion will play a more modest r\^ole.

\section{Discussion}
\label{sec:disc}

After the numerical developments and illustrations presented above, we finish this manuscript by outlining the main conclusions and perspectives of the present study and this series of papers in general.


\subsection{Conclusions}

In this work we presented a numerical algorithm to simulate the generation and propagation of long surface waves in the framework of a fully nonlinear weakly dispersive model on a globally spherical domain. Our model includes the rotation effects (\textsc{Coriolis} and centrifugal forces) of the \textsc{Earth}. Using a judicious choice of the still water level, we could `hide' the terms corresponding to the centrifugal force. However, if the still water level is assumed to be spherical, then these terms have to appear in governing equations.

Our numerical method is based on the numerical solution of the extended system, which consists of an elliptic equation to determine the dispersive component $\Pnh$ of the depth-integrated pressure and of a quasi-linear first order hyperbolic part common with Nonlinear Shallow Water Equations (NSWE). Both parts are coupled via source terms in the right hand of hyperbolic equations. These source terms come from non-hydrostatic effects. Thus, one can use the favourite numerical method for elliptic and hyperbolic problems. For the elliptic part we employed the finite differences, while hyperbolic equations were discretized with a two-step predictor--corrector scheme. On every stage of this scheme we solve both sub-problems.

The performance of the proposed algorithm is illustrated on several test cases. First, we consider an idealized situation of wave propagation over an even bottom. However, in this ideal setting we study the importance of sphericity, rotation, \textsc{Coriolis} and dispersion effects. For instance, we showed that the force of \textsc{Coriolis} becomes important only on large propagation distances (unless the angular velocity $\Omega$ is increased) and the rotation reduces somehow the maximal positive wave amplitudes (but only positive). The frequency dispersion also appears at rather large distances, but it depends greatly on the size of the initial free surface elevation, \ie more compact sources generate more dispersive waves. Contrary to the dispersion, the \textsc{Coriolis} effect becomes more important when we increase the generation area.

The question of dispersive effects importance is even more complex, since it can vary with the source size, but also with the source shape (\ie the initial condition). Consequently, for real-world problems a special investigation is needed in each particular case. It does not seem realistic that using simple criteria today we can select the most pertinent hydrodynamic model. In practical problems the bathymetry profile may have an important effect as well.

In the present study we did not rise the question of the numerical dispersion at all. Here we can mention that for grid resolutions of $40^{\,\dprime}$ and smaller, the numerical dispersion seems to be completely negligible comparing to the physical one (for the initial condition of the horizontal extension $\W_{\,1}\,$, for shorter waves the grid spacing has to be further reduced, of course). However, we are not sure that a weakly dispersive model is able to reproduce such waves with acceptable accuracy, without even speaking of the inherent computational complexity on such fine grids.

\subsubsection{General conclusions}

In this series of papers our main goal was to present a unified framework to modelling and numerical simulations of nonlinear dispersive waves. Indeed, in Part~I \cite{Khakimzyanov2016c} we presented a generalized derivation of dispersive wave models on a plane and the base model contained a free modeling variable --- the dispersive component of the horizontal velocity vector. By making special choices of this free variable, we could recover some known and some new models. Then, in Part~III \cite{Khakimzyanov2016a} the same approach was presented for globally spherical geometries with a similar degree of freedom at our disposal. The derivations on a sphere are more technical but in Part~III we really follow the main lines of Part~I. Finally, there is a similar interplay between numerical Parts~II \cite{Khakimzyanov2016} and IV \cite{Khakimzyanov2016b} on the globally flat and globally spherical geometries respectively. The key ingredient in both cases consists in deriving an elliptic equation to determine the dispersive component $\Pnh$ of the depth-integrated pressure. Then, the governing equations are decoupled into an elliptic and hyperbolic parts. For each part we apply the most suitable numerical method. Even if the details are quite different in flat and spherical cases, the philosophy remains invariant. As we hope, the numerical tests presented in Parts with even numbers, are convincing enough to show the operational qualities of the proposed algorithm.


\subsection{Perspectives}
\label{sec:persp}

One of the drawbacks of non-hydrostatic FNWD models is the inherent computational complexity. At every time step we have to solve an elliptic equation at least once. Most of runs presented in this study took about a week of CPU time to be completed in serial implementations. Consequently, in the future, we plan to develop a parallel version of the \textsc{NLDSW\_sphere} code in order to be able to handle much faster even higher grid resolutions. After this step we could try to implement fully nonlinear models with improved linear dispersion relation properties derived in \cite{Khakimzyanov2016a}. Another improvement would consist in a better representation of the shoreline. In the current implementation the shoreline is approximated by a polygon with sides parallel to (spherical) coordinate axes. Perhaps, a local mesh refinement could improve this point but it would create other numerical difficulties. Finally, the main point which remains to be solved is the development of a genuine run-up algorithm in the spirit of local analytical in time solutions \cite{Khakimzyanov2016d}, which takes into account the non-hydrostatic nature of FNWD governing equations.


\subsection*{Acknowledgments}
\addcontentsline{toc}{subsection}{Acknowledgments}

This research was supported by RSCF project No 14--17--00219. \\

We would like to thank also Professor Emmanuel \textsc{Audusse} (Universit\'e Paris 13, France) for rising the important question of boundary conditions in the \textsc{Serre}--\textsc{Green}--\textsc{Naghdi} equations. We did our best to explain our approach to this problem in the \emph{spherical} case, which is described in Section~\ref{sec:bc}.
\bigskip


\appendix
\section{Derivation of the equation for non-hydrostatic pressure component}
\label{app:der}

In this Appendix we provide the complete derivation of the elliptic equation \eqref{eq:ell} for the dispersive component $\Pnh$ of the depth-integrated pressure $p\,$. It is convenient to start with the base model \eqref{eq:base1} -- \eqref{eq:base3} written in the following equivalent form (see \cite{Khakimzyanov2016a} for more details):
\begin{equation}\label{eq:J1}
  (\,\J\,\H)_{\,t}\ +\ \bigl[\,\J\,\H\,u^{\,1}\,\bigr]_{\,\lambda}\ +\ \bigl[\,\J\,\H\,u^{\,2}\,\bigr]_{\,\theta}\ =\ 0\,,
\end{equation}
\begin{multline}\label{eq:J2}
  (\,\J\,\H\,\v)_{\,t}\ +\ \bigl[\,\J\,\H\,\v\,u^{\,1}\,\bigr]_{\,\lambda}\ +\ \bigl[\,\J\,\H\,\v\,u^{\,2}\,\bigr]_{\,\theta}\ +\ g\,\grad\Bigl(\,\J\;\frac{\H^{\,2}}{2}\Bigr)\ =\\ g\;\frac{\H^{\,2}}{2}\;\grad\J\ +\ \Bigl[\,g\,\H\,\grad h\ +\ \H\,\Sr\ +\ \grad\Pnh\ -\ \pb\,\grad h\,\Bigr]\,\J\,,
\end{multline}
where $\v\ =\ \bigl(\,v_{\,1},\,v_{\,2}\bigr)$ is the covariant velocity vector:
\begin{equation}\label{eq:0.2}
  v_{\,1}\ =\ (\Omega\ +\ u^{\,1})\,R^{\,2}\,\sin^{\,2}\theta\,, \qquad
  v_{\,2}\ =\ R^{\,2}\,u^{\,2}\,.
\end{equation}
Finally, the vector $\Sr\ =\ \bigl(0,\,s_{\,2}\bigr)$ with
\begin{equation*}
  s_{\,2}\ =\ \bigl[\,2\,\Omega\,u^{\,1}\ +\ (u^{\,1})^{\,2}\,\bigr]\,R^{\,2}\,\sin\theta\,\cos\theta\,.
\end{equation*}
Equation \eqref{eq:J2} can be rewritten in a non-conservative form:
\begin{equation}\label{eq:nc}
  \Dd\v\ =\ -g\,\grad\eta\ +\ \frac{\Pnh\ -\ \pb\,\grad h}{\H}\ +\ \Sr\,,
\end{equation}
where $\Dd\v\ \equiv\ \v_{\,t}\ +\ \u\scal\grad\v\,$.

As in the globally flat case, we shall express first the dispersive pressure component on the bottom $\pb$ as a function of $\Pnh$ and other variables. From definitions \eqref{eq:Pdef} we have
\begin{equation}\label{eq:pb}
  \pb\ =\ \frac{3\,\Pnh}{2\,\H}\ +\ \frac{\H}{4}\;\Rr_{\,2}\,,
\end{equation}
where the term $\Rr_{\,2}$ can be fully expanded:
\begin{multline*}
  \Rr_{\,2}\ \eqdef\ \Dd^{\,2}h\ \equiv\ \Dd\,(\Dd h)\ =\ (\Dd h)_{\,t}\ +\ \u\scal\grad(\Dd h)\ =\\
  \bigl[\,h_{\,t}\ +\ \u\scal\grad h\,\bigr]_{\,t}\ +\ \u\scal\grad\bigl[\,h_{\,t}\ +\ \u\scal\grad h\,\bigr]\ = \
  \underbrace{h_{\,t\,t}\ +\ 2\,\u\scal\grad h_{\,t}}_{\displaystyle{\defeq\ \B}}\ +\ \u_{\,t}\scal\grad h\ +\ \u\scal\grad(\u\scal\grad h)\,.
\end{multline*}
The term $\B$ contains all the terms involving the bottom motion. The last term in $\Rr_{\,2}$ can be further transformed\footnote{Indeed, \begin{multline*}\u\scal\grad(\u\scal\grad h)\ =\ u^{\,1}\,\bigl[\,u^{\,1}\,h_{\,\lambda}\ +\ u^{\,2}\,h_{\,\theta}\,\bigr]_{\,\lambda}\ +\ u^{\,2}\,\bigl[\,u^{\,1}\,h_{\,\lambda}\ +\ u^{\,2}\,h_{\,\theta}\,\bigr]_{\,\theta}\ =\\ u^{\,1}\,u_{\,\lambda}^{\,1}\,h_{\,\lambda}\ +\ (u^{\,1})^{\,2}\,h_{\,\lambda\,\lambda}\ +\ u^{\,1}\,u_{\,\lambda}^{\,2}\,h_{\,\theta}\ +\ u^{\,1}\,u^{\,2}\,h_{\,\lambda\,\theta}\ +\ u^{\,2}\,u_{\,\theta}^{\,1}\,h_{\,\lambda}\ +\ u^{\,2}\,u^{\,1}\,h_{\,\lambda\,\theta}\ +\ u^{\,2}\,u_{\,\theta}^{\,2}\,h_{\,\theta}\ +\ (u^{\,2})^{\,2}\,h_{\,\theta\,\theta}\ =\\ \bigl(u^{\,1}\,u^{\,1}_{\,\lambda}\ +\ u^{\,2}\,u^{\,1}_{\,\theta}\bigr)\,h_{\,\lambda}\ +\ \bigl(u^{\,1}\,u^{\,2}_{\,\lambda}\ +\ u^{\,2}\,u^{\,2}_{\,\theta}\bigr)\,h_{\,\theta}\ +\ u^{\,1}\,\bigl(u^{\,1}\,(h_{\,\lambda})_{\,\lambda}\ +\ u^{\,2}\,(h_{\,\lambda})_{\,\theta}\bigr)\ +\ u^{\,2}\,\bigl(u^{\,1}\,(h_{\,\theta})_{\,\lambda}\ +\ u^{\,2}\,(h_{\,\theta})_{\,\theta}\bigr)\ =\\ \bigl[\,(\u\scal\grad)\,\u\,\bigr]\scal\grad h\ +\ \u\scal\bigl[\,(\u\scal\grad)\,\grad h\,\bigr]\,.\end{multline*} $\qed$} equivalently as
\begin{equation*}
  \u\scal\grad(\u\scal\grad h)\ =\ \bigl[\,(\u\scal\grad)\,\u\,\bigr]\scal\grad h\ +\ \u\scal\bigl[\,(\u\scal\grad)\,\grad h\,\bigr]\,.
\end{equation*}
Consequently,
\begin{multline}\label{eq:r2}
  \Rr_{\,2}\ =\ \B\ +\ \u_{\,t}\scal\grad h\ +\ \bigl[\,(\u\scal\grad)\,\u\,\bigr]\scal\grad h\ +\ \u\scal\bigl[\,(\u\scal\grad)\,\grad h\,\bigr]\ \equiv \\
  \B\ +\ (\Dd\u)\scal\grad h\ +\ \u\scal\bigl[\,(\u\scal\grad)\,\grad h\,\bigr]\,,
\end{multline}
where in component-wise form we have
\begin{align*}
  \Dd u^{\,1}\ &=\ u_{\,t}^{\,1}\ +\ u^{\,1}\,u_{\,\lambda}^{\,1}\ +\ u^{\,2}\,u_{\,\theta}^{\,1}\,, \\
  \Dd u^{\,2}\ &=\ u_{\,t}^{\,2}\ +\ u^{\,1}\,u_{\,\lambda}^{\,2}\ +\ u^{\,2}\,u_{\,\theta}^{\,2}\,, \\
  (\Dd\u)\scal\grad h\ &=\ \bigl(u_{\,t}^{\,1}\ +\ u^{\,1}\,u_{\,\lambda}^{\,1}\ +\ u^{\,2}\,u_{\,\theta}^{\,1}\bigr)\,h_{\,\lambda}\ +\ \bigl(u_{\,t}^{\,2}\ +\ u^{\,1}\,u_{\,\lambda}^{\,2}\ +\ u^{\,2}\,u_{\,\theta}^{\,2}\bigr)\,h_{\,\theta}\,, \\
  \u\scal\bigl[\,(\u\scal\grad)\,\grad h\,\bigr]\ &=\ (u^{\,1})^{\,2}\,h_{\,\lambda\,\lambda}\ +\ 2\,u^{\,1}\,u^{\,2}\,h_{\,\lambda\,\theta}\ +\ (u^{\,2})^{\,2}\,h_{\,\theta\,\theta}\,.
\end{align*}
By substituting the last expression in \eqref{eq:r2} into equation \eqref{eq:pb} we obtain
\begin{equation}\label{eq:0.7}
  \pb\ =\ \frac{3\,\Pnh}{2\,\H}\ +\ \frac{\H}{4}\;\Bigl\{\,\B\ +\ (\Dd\u)\scal\grad h\ +\ \u\scal\bigl[\,(\u\scal\grad)\,\grad h\,\bigr]\,\Bigr\}\,.
\end{equation}
Now let us express the convective term $\Dd\,\u$ in terms of $\Dd\,\v\,$. According to formulas \eqref{eq:0.2}, the covariant vector of the velocity $\v$ can be rewritten as
\begin{equation*}
  \v\ =\ \Omega\;\Gg\ +\ \Gm\cdot\u\,,
\end{equation*}
where
\begin{equation*}
  \Gg\ \eqdef\ \begin{pmatrix}
    g_{\,1\,1} \\
    0
  \end{pmatrix}\,,
  \qquad
  \Gm\ \eqdef\ \begin{pmatrix}
    g_{\,1\,1} & 0 \\
    0 & g_{\,2\,2}
  \end{pmatrix}\,,
\end{equation*}
with $g_{\,1\,1}\,$, $g_{\,2\,2}$ being covariant components of the metric tensor on a sphere \cite{Khakimzyanov2016a}, \ie
\begin{equation*}
  g_{\,1\,1}\ =\ R^{\,2}\,\sin^{\,2}\theta\,, \qquad
  g_{\,2\,2}\ =\ R^{\,2}\,.
\end{equation*}
Then, we have 
\begin{equation}\label{eq:0.8}
  \Dd\,\v\ =\ \Omega\,\Dd\,\Gg\ +\ \Dd\,(\Gm\cdot\u)\ =\ u^{\,2}\,\Omega\,\Gg_{\,\theta}\ +\ \Gm\cdot\Dd\,\u\ +\ u^{\,1}\,u^{\,2}\,\Gg_{\,\theta}\ =\ \Gm\cdot\Dd\,\u\ +\ u^{\,2}\,\bigl(\Omega\ +\ u^{\,1}\bigr)\,\Gg_{\,\theta}\,,
\end{equation}
where
\begin{equation*}
  \Gg_{\,\theta}\ =\ \begin{pmatrix}
    2\,R^{\,2}\,\sin\theta\,\cos\theta \\
    0
  \end{pmatrix}\,.
\end{equation*}
By inverting equation \eqref{eq:0.8}, we obtain:
\begin{equation*}
  \Dd\,\u\ =\ \Gm^{\,-1}\cdot\Dd\,\v\ -\ u^{\,2}\,\bigl(\Omega\ +\ u^{\,1})\,\Gm^{\,-1}\cdot\Gg_{\,\theta}\,,
\end{equation*}
and using the non-conservative equation \eqref{eq:nc} we arrive to the required representation:
\begin{equation}\label{eq:Du}
  \Dd\,\u\ =\ \Gm^{\,-1}\cdot\Bigl\{\,-g\,\grad\eta\ +\ \frac{\Pnh\ -\ \pb\,\grad h}{\H}\,\Bigr\}\ +\ \Gm^{\,-1}\cdot\Lab\,,
\end{equation}
with
\begin{equation*}
  \Gm^{\,-1}\ =\ \begin{pmatrix}
    g^{\,1\,1} & 0 \\
    0 & g^{\,2\,2}
  \end{pmatrix}\,, \qquad
  g^{\,1\,1}\ \equiv\ \frac{1}{g_{\,1\,1}}\ =\ \frac{1}{R^{\,2}\,\sin^{\,2}\theta}\,, \qquad
  g^{\,2\,2}\ \equiv\ \frac{1}{g_{\,2\,2}}\ =\ \frac{1}{R^{\,2}}\,,
\end{equation*}
\begin{equation*}
  \Lab\ =\ \begin{pmatrix}
    \La_{\,1} \\
    \La_{\,2}
  \end{pmatrix}\ =\ \Sr\ -\ u^{\,2}\,\bigl(\Omega\ +\ u^{\,1}\bigr)\,\Gg_{\,\theta}\ =\ R^{\,2}\,\sin\theta\,\cos\theta\begin{pmatrix}
    -2\,u^{\,2}\,\bigl(\Omega\ +\ u^{\,1}\bigr) \\
    u^{\,1}\,\bigl(\,2\,\Omega\ +\ u^{\,1}\bigr)
  \end{pmatrix}\,.
\end{equation*}
Finally, by substituting expression \eqref{eq:Du} into equation \eqref{eq:0.7} we obtain:
\begin{equation*}
  \pb\ =\ \frac{3\,\Pnh}{2\,\H}\ +\ \frac{\H}{4}\;\Bigl\{\,\frac{\grad\Pnh\scal\grad h\ -\ \pb\,\abs{\grad h}^{\,2}}{\H}\ +\ \Qq\,\Bigr\}\,,
\end{equation*}
where
\begin{align*}
  \grad\Pnh\scal\grad h\ &\equiv\ g^{\,1\,1}\,\Pnh_{\,\lambda}\,h_{\,\lambda}\ +\ g^{\,2\,2}\,\Pnh_{\,\theta}\,h_{\,\theta}\ =\ \frac{1}{R^{\,2}}\;\biggl[\,\frac{\Pnh_{\,\lambda}\,h_{\,\lambda}}{\sin^{\,2}\theta}\ +\ \Pnh_{\,\theta}\,h_{\,\theta}\,\biggr]\,, \\
  \abs{\grad h}^{\,2}\ &\equiv\ g^{\,1\,1}\,h_{\,\lambda}^{\,2}\ +\ g^{\,2\,2}\,h_{\,\theta}^{\,2}\ =\ \frac{1}{R^{\,2}}\;\biggl[\,\frac{h_{\,\lambda}^{\,2}}{\sin^{\,2}\theta}\ +\ h_{\,\theta}^{\,2}\,\biggr]\,, \\
  \Qq\ &\eqdef\ \bigl[\,\Lab\ -\ g\,\grad\eta\,\bigr]\scal\grad h\ +\ \B\ +\ \u\scal\bigl((\u\scal\grad)\,\grad h\bigr)\,, \\
  \grad\eta\scal\grad h\ &\equiv\ g^{\,1\,1}\,\eta_{\,\lambda}\,h_{\,\lambda}\ +\ g^{\,2\,2}\,\eta_{\,\theta}\,h_{\,\theta}\ =\ \frac{1}{R^{\,2}}\;\biggl[\,\frac{\eta_{\,\lambda}\,h_{\,\lambda}}{\sin^{\,2}\theta}\ +\ \eta_{\,\theta}\,h_{\,\theta}\,\biggr]\,, \\
  \Lab\scal\grad h\ &\equiv\ g^{\,1\,1}\,\La_{\,1}\,h_{\,\lambda}\ +\ g^{\,2\,2}\,\La_{\,2}\,h_{\,\theta}\\
  &=\ \cot\theta\;\Bigl[\,-2\,u^{\,2}\bigl(\Omega\ +\ u^{\,1}\bigr)\,h_{\,\lambda}\ +\ u^{\,1}\,\bigl(\,2\,\Omega\ +\ u^{\,1}\bigr)\,h_{\,\theta}\,\sin^{\,2}\theta\,\Bigr]\,.
\end{align*}
Consequently, the dispersive part of the fluid pressure at the bottom can be expressed in terms of other variables as
\begin{equation}\label{eq:0.9}
  \pb\ =\ \frac{1}{\r}\;\biggl\{\,\frac{6\,\Pnh}{\H}\ +\ \H\,\Qq\ +\ \grad\Pnh\scal\grad h\,\biggr\}\,,
\end{equation}
where we introduced a new dependent variable
\begin{equation*}
  \r\ \eqdef\ 4\ +\ \abs{\grad h}^{\,2}\,.
\end{equation*}

Now we can proceed to the derivation of an equation for $\Pnh\,$. From definitions \eqref{eq:Pdef} it follows that
\begin{equation}\label{eq:0.11}
  \Pnh\ =\ \frac{\H^{\,3}}{12}\,\Rr_{\,1}\ +\ \frac{\H}{2}\;\pb\,,
\end{equation}
where $\Rr_{\,1}$ was defined as
\begin{equation*}
  \Rr_{\,1}\ \eqdef\ \Dd\,(\div\u)\ -\ (\div\u)^2\,.
\end{equation*}
Let us transform the last expression for $\Rr_{\,1}$ using the definition of the divergence $\div(\cdot)$ and material derivative $\Dd$ operators
\begin{multline*}
  \Dd\,(\div\u)\ =\ (\div\u)_{\,t}\ +\ u^{\,1}\,(\div\u)_{\,\lambda}\ +\ u^{\,2}\,(\div\u)_{\,\theta}\\
  =\ \div\u_{\,t}\ +\ u^{\,1}\,\Bigl\{\,u_{\,\lambda}^{\,1}\ +\ \frac{\bigl(\,\J\,u^{\,2}\bigr)_{\,\theta}}{\J}\,\Bigr\}_{\,\lambda}\ +\ u^{\,2}\,\Bigl\{\,u_{\,\lambda}^{\,1}\ +\ \frac{\bigl(\,\J\,u^{\,2}\bigr)_{\,\theta}}{\J}\,\Bigr\}_{\,\theta}\\
  =\ \div\u_{\,t}\ +\ \bigl(u^{\,1}\,u_{\,\lambda}^{\,1}\bigr)_{\,\lambda}\ -\ (u_{\,\lambda}^{\,1})^{\,2}\ +\ \bigl(u^{\,2}\,u_{\,\theta}^{\,1}\bigr)_{\,\lambda}\ -\ u_{\,\lambda}^{\,2}\,u_{\,\theta}^{\,1}\ +\ u^{\,1}\;\frac{\bigl(\,\J\,u^{\,2}\bigr)_{\,\lambda\,\theta}}{\J}\ +\ u^{\,2}\;\Bigl[\,\frac{\bigl(\,\J\,u^{\,2}\bigr)_{\,\theta}}{\J}\,\Bigr]_{\,\theta}\\
  =\ \underbrace{(u_{\,t}^{\,1})_{\,\lambda}}_{\displaystyle\openbigstar_{\,1}}\ +\ \frac{\bigl(\,\J\,u_{\,t}^{\,2}\bigr)_{\,\theta}}{\J}\ +\ \underbrace{\bigl[\,u^{\,1}\,u_{\,\lambda}^{\,1}\ +\ u^{\,2}\,u_{\,\theta}^{\,1}\,\bigr]_{\,\lambda}}_{\displaystyle\openbigstar_{\,1}}\ +\ \frac{\bigl[\,u^{\,1}\,\bigl(\,\J\,u^{\,2}\bigr)_{\,\lambda}\,\bigr]_{\,\theta}}{\J}\ -\ \frac{u_{\,\theta}^{\,1}\,\bigl(\,\J\,u^{\,2}\bigr)_{\,\lambda}}{\J}\\
  +\ \biggl[\,u^{\,2}\;\frac{\bigl(\,\J\,u^{\,2}\bigr)_{\,\theta}}{\J}\,\biggr]_{\,\theta}\ -\ u_{\,\theta}^{\,2}\;\frac{\bigl(\,\J\,u^{\,2}\bigr)_{\,\theta}}{\J}\ -\ \bigl(u_{\,\lambda}^{\,1}\bigr)^{\,2}\ -\ u_{\,\lambda}^{\,2}\,u_{\,\theta}^{\,1}\,.
\end{multline*}
It is not difficult to see that terms marked with $(\openbigstar_{\,1})$ can be aggregated into $(\Dd\,u^{\,1})_{\,\lambda}\,$. Consequently, we have
\begin{multline*}
  \Dd\,(\div\u)\ =\ (\Dd\,u^{\,1})_{\,\lambda}\ +\ \underbrace{\frac{\bigl(\,\J\,u_{\,t}^{\,2}\bigr)_{\,\theta}\ +\ \bigl(\,\J\,u^{\,2}\,u_{\,\theta}^{\,2}\bigr)_{\,\theta}}{\J}}_{\displaystyle\openbigstar_{\,2}}\ -\ u_{\,\theta}^{\,1}\,u_{\,\lambda}^{\,2}\ +\ \underbrace{\frac{\bigl[\,\J\,u^{\,1}\,u_{\,\lambda}^{\,2}\,\bigr]_{\,\theta}}{\J}}_{\displaystyle\openbigstar_{\,2}}\\
  +\ \frac{\bigl[\,\J_{\,\theta}\,(u^{\,2}\,)^{\,2}\,\bigr]_{\,\theta}}{\J}\ -\ \underbrace{\biggl[\,\frac{u^{\,2}\,\bigl(\,\J\,u^{\,2}\bigr)_{\,\theta}}{\J^{\,2}}\;\J_{\,\theta}\ +\ \frac{u_{\,\theta}^{\,2}\,\bigl(\,\J\,u^{\,2}\bigr)_{\,\theta}}{\J}\ +\ (u_{\,\lambda}^{\,1})^{\,2}\,\biggr]}_{\displaystyle\bigstar}\ -\ u_{\,\lambda}^{\,2}\,u_{\,\theta}^{\,1}\,.
\end{multline*}
The terms marked above with $(\openbigstar_{\,2})$ give another interesting combination:
\begin{equation*}
  \frac{\bigl[\,\J\,\bigl(u_{\,t}^{\,2}\ +\ u^{\,1}\,u_{\,\lambda}^{\,2}\ +\ u^{\,2}\,u_{\,\theta}^{\,2}\bigr)\,\bigr]_{\,\theta}}{\J}\ \equiv\ \frac{\bigl(\,\J\,\Dd\,u^{\,2}\bigr)_{\,\theta}}{\J}\,.
\end{equation*}
According to the definition of the divergence $\div\u\,$, the term $(\bigstar)$ can be transformed as
\begin{equation*}
  (u_{\,\lambda}^{\,1}\,)^{\,2}\ +\ \biggl[\,\frac{\bigl(\,\J\,u^{\,2}\bigr)_{\,\theta}}{\J}\,\biggr]^{\,2}\ \equiv\ \bigl(\div\u\bigr)^{\,2}\ -\ 2\,u_{\,\lambda}^{\,1}\;\frac{\bigl(\,\J\,u^{\,2}\bigr)_{\,\theta}}{\J}\,.
\end{equation*}
Consequently, we can rewrite $\Dd\,(\div\u)$ using mentioned above simplifications:
\begin{multline*}
  \Dd\,(\div\u)\ =\ \bigl(\Dd\,u^{\,1}\bigr)_{\,\lambda}\ +\ \frac{\bigl(\,\J\,\Dd\,u^{\,2}\bigr)_{\,\theta}}{\J}\ -\ (\div\u)^{\,2}\\
  +\ 2\,u_{\,\lambda}^{\,1}\,u_{\,\theta}^{\,2}\ +\ 2\,u_{\,\lambda}^{\,1}\,u^{\,2}\;\frac{\J_{\,\theta}}{\J}\ +\ 2\,u^{\,2}\,u_{\,\theta}^{\,2}\;\frac{\J_{\,\theta}}{\J}\ +\ (u^{\,2}\,)^{\,2}\;\frac{\J_{\,\theta\,\theta}}{\J}\ -\ 2\,u_{\,\theta}^{\,1}\,u_{\,\lambda}^{\,2}\,.
\end{multline*}
Finally, taking into account the fact that $\J_{\,\theta\,\theta}\ \equiv\ -\J\,$, we obtain:
\begin{equation*}
  \Dd\,(\div\u)\ =\ \div(\Dd\,\u)\ -\ (\div\u)^{\,2}\ +\ 2\,\bigl(u_{\,\lambda}^{\,1}\,u_{\,\theta}^{\,2}\ -\ u_{\,\theta}^{\,1}\,u_{\lambda}^{\,2}\,\bigr)\ +\ 2\,u^{\,2}\,\bigl(u_{\,\lambda}^{\,1}\ +\ u_{\,\theta}^{\,2}\bigr)\,\cot\theta\ -\ (u^{\,2}\,)^{\,2}\,.
\end{equation*}
By assembling all the ingredients together, we obtain the following expression for $\Rr_{\,1}\,$:
\begin{equation*}
  \Rr_{\,1}\ =\ \div(\Dd\,\u)\ -\ 2\,(\div\u)^{\,2}\ +\ 2\,\bigl(u_{\,\lambda}^{\,1}\,u_{\,\theta}^{\,2}\ -\ u_{\,\theta}^{\,1}\,u_{\lambda}^{\,2}\,\bigr)\ +\ 2\,u^{\,2}\,\bigl(u_{\,\lambda}^{\,1}\ +\ u_{\,\theta}^{\,2}\bigr)\,\cot\theta\ -\ (u^{\,2}\,)^{\,2}\,.
\end{equation*}
By substituting the last expression for $\Rr_{\,1}$ into equation \eqref{eq:0.11} and using formulas \eqref{eq:Du}, \eqref{eq:0.9} for $\Dd\,\u$ and $\pb$ correspondingly, we obtain the first version of the required equation for $\Pnh\,$:
\begin{multline}\label{eq:0.14}
  \Pnh\ =\ \frac{\H^{\,3}}{12}\;\biggl[\,\div\Bigl\{\,\Gm^{\,-1}\cdot\Bigl(\Lab\ -\ g\,\grad\eta\ +\ \frac{\grad\Pnh}{\H}\ -\ \frac{6\,\grad h}{\H^{\,2}\,\r}\;\Pnh\ -\ \grad h\;\frac{\Qq}{\r}\ -\ \frac{\bigl(\grad\Pnh\scal\grad h\bigr)\,\grad h}{\H\,\r}\Bigr)\,\Bigr\} \\
  -\ 2\,(\div\u)^{\,2}\ +\ 2\,\bigl(u_{\,\lambda}^{\,1}\,u_{\,\theta}^{\,2}\ -\ u_{\,\theta}^{\,1}\,u_{\lambda}^{\,2}\,\bigr)\ +\ 2\,u^{\,2}\,\bigl(u_{\,\lambda}^{\,1}\ +\ u_{\,\theta}^{\,2}\bigr)\,\cot\theta\ -\ (u^{\,2}\,)^{\,2}\,\biggr] \\
  +\ \frac{\H}{2\,\r}\;\Bigl\{\,\frac{6\,\Pnh}{\H}\ +\ \H\,\Qq\ +\ \grad\Pnh\scal\grad h\,\Bigr\}\,.
\end{multline}
We can notice that the multiplication of covariant vectors $\Lab\,$, $\grad\eta\,$, $\grad\Pnh$ and $\grad h$ by matrix $\Gm^{\,-1}$ transforms them into contravariant vectors which enter into the definition of the divergence operator, for example
\begin{align*}
  \div\grad\eta\ &\eqdef\ \frac{\bigl(\,\J\,g^{\,1\,1}\,\eta_{\,\lambda}\bigr)_{\,\lambda}\ +\ \bigl(\,\J\,g^{\,2\,2}\,\eta_{\,\theta}\bigr)_{\,\theta}}{\J}\ =\ \frac{1}{R^{\,2}\,\sin\theta}\;\Bigl[\,\frac{\eta_{\,\lambda\,\lambda}}{\sin\theta}\ +\ \bigl(\eta_{\,\theta}\,\sin\theta\bigr)_{\,\theta}\,\Bigr]\ \equiv\ \DDelta\eta\,, \\
  \div\Lab\ &\eqdef\ \frac{\bigl(\,\J\,g^{\,1\,1}\,\La_{\,1}\bigr)_{\,\lambda}\ +\ \bigl(\,\J\,g^{\,2\,2}\,\La_{\,2}\bigr)_{\,\theta}}{\J}\ =\ \frac{1}{R^{\,2}\,\sin\theta}\;\Bigl[\,\frac{\La_{\,1,\,\lambda}}{\sin\theta}\ +\ \bigl(\La_{\,2}\,\sin\theta\bigr)_{\,\theta}\,\Bigr]\,.
\end{align*}
We can notice also that the following identity holds:
\begin{equation*}
  \div\biggl[\,\frac{6\,\grad h}{\H^{\,2}\,\r}\;\Pnh\,\biggr]\ =\ \frac{6}{\H^{\,2}\,\r}\;\grad\Pnh\scal\grad h\ +\ 6\,\Pnh\,\div\biggl[\,\frac{\grad h}{\H^{\,2}\,\r}\,\biggr]\,.
\end{equation*}
Taking into account the last two remarks, we can rewrite equation \eqref{eq:0.14} in a more compact form:
\begin{equation}\label{eq:0.16}
  \Ll\Pnh\ =\ \F^{\,\star}\,,
\end{equation}
where the linear operator $\Ll$ and the right hand side $\F^{\,\star}$ are defined as
\begin{equation*}
  \Ll\Pnh\ \eqdef\ \div\biggl\{\,\frac{\grad\Pnh}{\H}\ -\ \frac{\bigl(\grad\Pnh\scal\grad h\bigr)\,\grad h}{\H\,\r}\,\biggr\}\ -\ 6\,\Pnh\;\biggl\{\,\frac{2}{\H^{\,3}}\;\frac{\r\ -\ 3}{\r}\ +\ \div\Bigl[\,\frac{\grad h}{\H^{\,2}\,\r}\,\Bigr]\,\biggr\}\,,
\end{equation*}
\begin{multline*}
  \F^{\,\star}\ \eqdef\ g\,\DDelta\eta\ +\ \div\biggl\{\,\frac{\Qq\,\grad h}{\r}\ -\ \Lab\,\biggr\}\ -\ \frac{6\,\Qq}{\H\,\r}\\
  +\ 2\,(\div\u)^{\,2}\ -\ 2\,\bigl(u_{\,\lambda}^{\,1}\,u_{\,\theta}^{\,2}\ -\ u_{\,\theta}^{\,1}\,u_{\lambda}^{\,2}\,\bigr)\ -\ 2\,u^{\,2}\,\bigl(u_{\,\lambda}^{\,1}\ +\ u_{\,\theta}^{\,2}\bigr)\,\cot\theta\ +\ (u^{\,2}\,)^{\,2}\,.
\end{multline*}
By expanding divergences of covariant vectors, we can express the operator $\Ll(\cdot)$ through partial derivatives:
\begin{multline*}
  \Ll\Pnh\ \equiv\ \frac{1}{R^{\,2}\,\sin\theta}\;\biggl\{\,\frac{1}{\sin\theta}\;\biggl[\,\frac{\Pnh_{\,\lambda}}{\H}\ -\ \frac{\grad\Pnh\scal\grad h}{\H\,\r}\;h_{\,\lambda}\,\biggr]_{\,\lambda}\ +\ \biggl[\,\Bigl(\frac{\Pnh_{\,\theta}}{\H}\ -\ \frac{\grad\Pnh\scal\grad h}{\H\,\r}\;h_{\,\theta}\Bigr)\,\sin\theta\,\biggr]_{\,\theta}\,\biggr\} \\
  -\ 6\,\Pnh\;\biggl\{\,\frac{2}{\H^{\,3}}\;\frac{\r\ -\ 3}{\r}\ +\ \frac{1}{R^{\,2}\,\sin\theta}\;\biggl[\,\frac{1}{\sin\theta}\Bigl(\frac{h_{\,\lambda}}{\H^{\,2}\,\r}\Bigr)_{\,\lambda}\ +\ \Bigl(\frac{h_{\,\theta}}{\H^{\,2}\,\r}\;\sin\theta\Bigr)_{\,\theta}\,\biggr]\,\biggr\}\,.
\end{multline*}
After multiplying both sides of equation \eqref{eq:0.16} by $R^{\,2}\,\sin\theta$ and switching to linear components of the velocity $u$ and $v\,$, we obtain the following equation for $\Pnh\,$:
\begin{multline}\label{eq:0.17}
  \biggl[\,\frac{1}{\sin\theta}\;\Bigl\{\,\frac{\Pnh_{\,\lambda}}{\H}\ -\ \frac{\grad\Pnh\scal\grad h}{\H\,\r}\;h_{\,\lambda}\,\Bigr\}\,\biggr]_{\,\lambda}\ +\ \biggl[\,\Bigl\{\,\frac{\Pnh_{\,\theta}}{\H}\ -\ \frac{\grad\Pnh\scal\grad h}{\H\,\r}\;h_{\,\theta}\,\Bigr\}\,\sin\theta\,\biggr]_{\,\theta}\\
  -\ 6\,\Pnh\;\biggl[\,\frac{2}{\H^{\,3}}\;\frac{\r\ -\ 3}{\r}\;R^{\,2}\,\sin\theta\ +\ \Bigl(\frac{h_{\,\lambda}}{\H^{\,2}\,\r\,\sin\theta}\Bigr)_{\,\lambda}\ +\ \Bigl(\frac{h_{\,\theta}}{\H^{\,2}\,\r}\;\sin\theta\Bigr)_{\,\theta}\,\biggr]\ =\ \F\,,
\end{multline}
where the right hand side $\F$ is defined as
\begin{multline*}
  \F\ \eqdef\ \biggl[\,\frac{1}{\sin\theta}\;\Bigl\{\,g\,\eta_{\,\lambda}\ +\ \frac{\Qq}{\r}\;h_{\,\lambda}\ -\ \La_{\,1}\,\Bigr\}\,\biggr]_{\,\lambda}\ +\ \biggl[\,\Bigl\{\,g\,\eta_{\,\theta}\ +\ \frac{\Qq}{\r}\;h_{\,\theta}\ -\ \La_{\,2}\,\Bigr\}\,\sin\theta\,\biggr]_{\,\theta}\\ 
  -\ R^{\,2}\;\frac{6\,\Qq}{\H\,\r}\;\sin\theta\ + \ \frac{2}{\sin\theta}\;\Bigl\{\,u_{\,\lambda}\ +\ (\,v\,\sin\theta)_{\,\theta}\,\Bigr\}^{\,2}\\ 
  -\ 2\,\bigl(u_{\,\lambda}\,v_{\,\theta}\ -\ v_{\,\lambda}\,u_{\,\theta}\bigr)\ -\ 2\,(u\,v)_{\,\lambda}\cot\theta\ -\ \bigl(\,v^{\,2}\,\cos\theta\bigr)_{\,\theta}\,.
\end{multline*}
The quantities $\Qq\,$, $\B$ and $\Lab$ are expressed through linear velocity components as
\begin{equation*}
  \Qq\ \equiv\ \bigl(\Lab\ -\ g\,\grad\eta\bigr)\scal\grad h\ +\ \frac{1}{R^{\,2}\,\sin\theta}\;\biggl\{\,\frac{u^{\,2}}{\sin\theta}\;h_{\,\lambda\,\lambda}\ +\ 2\,u\,v\,h_{\,\lambda\,\theta}\ +\ v^{\,2}\,h_{\,\theta\,\theta}\,\sin\theta\,\biggr\}\ +\ \B\,,
\end{equation*}
\begin{equation*}
  \B\ \equiv\ h_{\,t\,t}\ +\ 2\,\Bigl\{\,\frac{u}{R\,\sin\theta}\;h_{\,\lambda\,t}\ +\ \frac{v}{R}\;h_{\,\theta\,t}\,\Bigr\}\,,
\end{equation*}
\begin{equation*}
  \La_{\,1}\ \eqdef\ -\bigl(\,2\,u\,v\,\cot\theta\ +\ \digamma\,v\,R\bigr)\,\sin\theta\,, \qquad
  \La_{\,2}\ \eqdef\ u^{\,2}\,\cot\theta\ +\ \digamma\,u\,R\,,
\end{equation*}
\begin{equation*}
  \Lab\scal\grad h\ \equiv\ \frac{\La_{\,1}\,h_{\,\lambda}}{R^{\,2}\,\sin^{\,2}\theta}\ +\ \frac{\La_{\,2}\,h_{\,\theta}}{R^{\,2}}\,.
\end{equation*}
This concludes naturally the derivation of the elliptic equation \eqref{eq:0.17} for the dispersive part of the depth-integrated pressure $\Pnh\,$.

\begin{remark}
Let us assume that the angular velocity vanishes, \ie $\Omega\ \equiv\ 0\,$. In a vicinity of a fixed admissible point $\bigl(\lambda^{\,\star},\,\theta^{\,\star}\bigr)$ we introduce a non-degenerate transformation of coordinates:
\begin{equation*}
  x^{\,\checkmark}\ \eqdef\ R\,(\lambda\ -\ \lambda^{\,\star})\,\sin\theta^{\,\star}\,, \qquad
  y^{\,\checkmark}\ \eqdef\ -R\,(\theta\ -\ \theta^{\,\star})\,,
\end{equation*}
along with corresponding velocity components:
\begin{equation*}
  u^{\,\checkmark}\ \equiv\ \dot{x^{\,\checkmark}}\ =\ R\,\dot{\lambda}\,\sin\theta\ \equiv\ \varsigma\,u\,, \qquad
  v^{\,\checkmark}\ \equiv\ \dot{y^{\,\checkmark}}\ =\ -R\,\dot{\theta}\ =\ -v\,,
\end{equation*}
where $\varsigma\ \eqdef\ \dfrac{\sin\theta^{\,\star}}{\sin\theta}\,$. By assuming that the considered neighbourhood is small in the latitude, \ie the quantity $\upvarepsilon\ \eqdef\ \abs{\theta\ -\ \theta^{\,\star}}\ \ll\ 1$ is small. After transforming equation \eqref{eq:0.17} to new independent $\bigl(x^{\,\checkmark},\,y^{\,\checkmark}\bigr)$ and new dependent $\bigl(u^{\,\checkmark},\,v^{\,\checkmark}\bigr)$ variables and neglecting the terms of the order $\O\,(\upvarepsilon)$ and $\O\,(R^{\,-1})$ one can obtain the familiar to us elliptic equation for $\Pnh$ in the globally plane case \cite{Khakimzyanov2016}. This fact is a further confirmation of the correctness of our derivation procedure described above.
\end{remark}


\section{Acronyms}

In the text above the reader could encounter the following acronyms:

\begin{description}
  \item[BBM] \textsc{Benjamin}--\textsc{Bona}--\textsc{Mahony}
  \item[FEM] Finite Element Method
  \item[SOR] Successive Over--Relaxation
  \item[NOAA] National Oceanic and Atmospheric Administration
  \item[NSWE] Nonlinear Shallow Water Equations
  \item[FNWD] Fully Nonlinear Weakly Dispersive
  \item[WNWD] Weakly Nonlinear Weakly Dispersive
\end{description}


\addcontentsline{toc}{section}{References}
\bigskip
\bibliographystyle{abbrv}

\begin{thebibliography}{10}

\bibitem{Abraimi2014}
R.~Abraimi.
\newblock {\em {Modelling the 2010 Chilean Tsunami using the H2Ocean
  unstructured mesh model}}.
\newblock Master thesis, TU Delft, 2014.

\bibitem{Arcas2012}
D.~Arcas and H.~Segur.
\newblock {Seismically generated tsunamis}.
\newblock {\em Phil. Trans. R. Soc. A}, 370:1505--1542, 2012.

\bibitem{Beck2016}
J.~Beck and S.~Guillas.
\newblock {Sequential Design with Mutual Information for Computer Experiments
  (MICE): Emulation of a Tsunami Model}.
\newblock {\em SIAM/ASA J. Uncertainty Quantification}, 4(1):739--766, jan
  2016.

\bibitem{Beisel2012}
S.~A. Beisel, L.~B. Chubarov, D.~Dutykh, G.~S. Khakimzyanov, and N.~Y. Shokina.
\newblock {Simulation of surface waves generated by an underwater landslide in
  a bounded reservoir}.
\newblock {\em Russ. J. Numer. Anal. Math. Modelling}, 27(6):539--558, 2012.

\bibitem{bona}
T.~B. Benjamin, J.~L. Bona, and J.~J. Mahony.
\newblock {Model equations for long waves in nonlinear dispersive systems}.
\newblock {\em Philos. Trans. Royal Soc. London Ser. A}, 272:47--78, 1972.

\bibitem{BS}
J.~L. Bona and R.~Smith.
\newblock {A model for the two-way propagation of water waves in a channel}.
\newblock {\em Math. Proc. Camb. Phil. Soc.}, 79:167--182, 1976.

\bibitem{Boussinesq1877}
J.~V. Boussinesq.
\newblock {Essai sur la th{\'{e}}orie des eaux courantes}.
\newblock {\em M{\'{e}}moires pr{\'{e}}sent{\'{e}}s par divers savants {\`{a}}
  l'Acad. des Sci. Inst. Nat. France}, XXIII:1--680, 1877.

\bibitem{Cherevko2009a}
A.~A. Cherevko and A.~P. Chupakhin.
\newblock {Equations of the shallow water model on a rotating attracting
  sphere. 1. Derivation and general properties}.
\newblock {\em Journal of Applied Mechanics and Technical Physics},
  50(2):188--198, mar 2009.

\bibitem{Chubarov2005}
L.~B. Chubarov, S.~V. Eletsky, Z.~I. Fedotova, and G.~S. Khakimzyanov.
\newblock {Simulation of surface waves by an underwater landslide}.
\newblock {\em Russ. J. Numer. Anal. Math. Modelling}, 20(5):425--437, 2005.

\bibitem{Chubarov1987}
L.~B. Chubarov and Y.~I. Shokin.
\newblock {The numerical modelling of long wave propagation in the framework of
  non-linear dispersion models}.
\newblock {\em Comput. {\&} Fluids}, 15(3):229--249, jan 1987.

\bibitem{Clamond2015c}
D.~Clamond, D.~Dutykh, and D.~Mitsotakis.
\newblock {Conservative modified Serre--Green--Naghdi equations with improved
  dispersion characteristics}.
\newblock {\em Comm. Nonlin. Sci. Num. Sim.}, 45:245--257, 2017.

\bibitem{DGK}
R.~A. Dalrymple, S.~T. Grilli, and J.~T. Kirby.
\newblock {Tsunamis and challenges for accurate modeling}.
\newblock {\em Oceanography}, 19:142--151, 2006.

\bibitem{Tkalich2007}
M.~H. Dao and P.~Tkalich.
\newblock {Tsunami propagation modelling - a sensitivity study}.
\newblock {\em Nat. Hazards Earth Syst. Sci.}, 7:741--754, 2007.

\bibitem{Delouis2010}
B.~Delouis, J.-M. Nocquet, and M.~Vall{\'{e}}e.
\newblock {Slip distribution of the February 27, 2010 Mw = 8.8 Maule
  Earthquake, central Chile, from static and high-rate GPS, InSAR, and
  broadband teleseismic data}.
\newblock {\em Geophys. Res. Lett.}, 37(17), sep 2010.

\bibitem{Dias2006}
F.~Dias and D.~Dutykh.
\newblock {\em {Dynamics of tsunami waves}}, pages 35--60.
\newblock Springer Netherlands, 2007.

\bibitem{Dias2014}
F.~Dias, D.~Dutykh, L.~O'Brien, E.~Renzi, and T.~Stefanakis.
\newblock {On the Modelling of Tsunami Generation and Tsunami Inundation}.
\newblock {\em Procedia IUTAM}, 10:338--355, 2014.

\bibitem{Dutykh2006}
D.~Dutykh and F.~Dias.
\newblock {Water waves generated by a moving bottom}.
\newblock In A.~Kundu, editor, {\em Tsunami and Nonlinear waves}, pages 65--96.
  Springer Verlag (Geo Sc.), 2007.

\bibitem{Dutykh2007b}
D.~Dutykh and F.~Dias.
\newblock {Tsunami generation by dynamic displacement of sea bed due to
  dip-slip faulting}.
\newblock {\em Mathematics and Computers in Simulation}, 80(4):837--848, 2009.

\bibitem{ddk}
D.~Dutykh, F.~Dias, and Y.~Kervella.
\newblock {Linear theory of wave generation by a moving bottom}.
\newblock {\em C. R. Acad. Sci. Paris, Ser. I}, 343:499--504, 2006.

\bibitem{Dutykh2011d}
D.~Dutykh and H.~Kalisch.
\newblock {Boussinesq modeling of surface waves due to underwater landslides}.
\newblock {\em Nonlin. Processes Geophys.}, 20(3):267--285, may 2013.

\bibitem{Dutykh2012}
D.~Dutykh, D.~Mitsotakis, S.~A. Beisel, and N.~Y. Shokina.
\newblock {Dispersive waves generated by an underwater landslide}.
\newblock In E.~Vazquez-Cendon, A.~Hidalgo, P.~Garcia-Navarro, and L.~Cea,
  editors, {\em Numerical Methods for Hyperbolic Equations: Theory and
  Applications}, pages 245--250. CRC Press, Boca Raton, London, New York,
  Leiden, 2013.

\bibitem{Dutykh2009a}
D.~Dutykh, R.~Poncet, and F.~Dias.
\newblock {The VOLNA code for the numerical modeling of tsunami waves:
  Generation, propagation and inundation}.
\newblock {\em Eur. J. Mech. B/Fluids}, 30(6):598--615, 2011.

\bibitem{Evsyukov2009}
Y.~D. Evsyukov.
\newblock {Distribution of landslide bodies on the continental slope of the
  North-Eastern part of the Black Sea}.
\newblock {\em Bulletin of the North Caucasus Scientific Center of the Higher
  School. Natural Sciences}, 6:100--104, 2009.

\bibitem{Fedotova2006}
Z.~I. Fedotova.
\newblock {On application of the MacCormack difference scheme for problems of
  long-wave hydrodynamics}.
\newblock {\em Comput. Technologies}, 11(5):53--63, 2006.

\bibitem{Fedotova2010}
Z.~I. Fedotova and G.~S. Khakimzyanov.
\newblock {Nonlinear-dispersive shallow water equations on a rotating sphere}.
\newblock {\em Russian Journal of Numerical Analysis and Mathematical
  Modelling}, 25(1), jan 2010.

\bibitem{Fedotova2014a}
Z.~I. Fedotova and G.~S. Khakimzyanov.
\newblock {Nonlinear dispersive shallow water equations on a rotating sphere
  and conservation laws}.
\newblock {\em J. Appl. Mech. Tech. Phys.}, 55(3):404--416, 2014.

\bibitem{Glimsdal2006}
S.~Glimsdal, G.~K. Pedersen, K.~Atakan, C.~B. Harbitz, H.~P. Langtangen, and
  F.~Lovholt.
\newblock {Propagation of the Dec. 26, 2004, Indian Ocean Tsunami: Effects of
  Dispersion and Source Characteristics}.
\newblock {\em Int. J. Fluid Mech. Res.}, 33(1):15--43, 2006.

\bibitem{Glimsdal2013}
S.~Glimsdal, G.~K. Pedersen, C.~B. Harbitz, and F.~L{\o}vholt.
\newblock {Dispersion of tsunamis: does it really matter?}
\newblock {\em Natural Hazards and Earth System Science}, 13(6):1507--1526, jun
  2013.

\bibitem{Godlewski1990}
E.~Godlewski and P.-A. Raviart.
\newblock {\em {Hyperbolic systems of conservation laws}}.
\newblock Ellipses, Paris, 1990.

\bibitem{Godunov1987}
S.~K. Godunov and V.~S. Ryabenkii.
\newblock {\em {Difference Schemes}}.
\newblock North-Holland, Amsterdam, 1987.

\bibitem{Grilli2012}
S.~T. Grilli, J.~C. Harris, T.~S. {Tajalli Bakhsh}, T.~L. Masterlark,
  C.~Kyriakopoulos, J.~T. Kirby, and F.~Shi.
\newblock {Numerical Simulation of the 2011 Tohoku Tsunami Based on a New
  Transient FEM Co-seismic Source: Comparison to Far- and Near-Field
  Observations}.
\newblock {\em Pure Appl. Geophys.}, jul 2012.

\bibitem{Grilli2007}
S.~T. Grilli, M.~Ioualalen, J.~Asavanant, F.~Shi, J.~T. Kirby, and P.~Watts.
\newblock {Source Constraints and Model Simulation of the December 26, 2004,
  Indian Ocean Tsunami}.
\newblock {\em Journal of Waterway, Port, Coastal, and Ocean Engineering},
  133(6):414--428, nov 2007.

\bibitem{Grue2008}
J.~Grue, E.~N. Pelinovsky, D.~Fructus, T.~Talipova, and C.~Kharif.
\newblock {Formation of undular bores and solitary waves in the Strait of
  Malacca caused by the 26 December 2004 Indian Ocean tsunami}.
\newblock {\em J. Geophys. Res.}, 113(C5):C05008, may 2008.

\bibitem{Gusev2014}
O.~I. Gusev.
\newblock {Algorithm for surface waves calculation above a movable bottom
  within the frame of plane nonlinear dispersive wave model}.
\newblock {\em Comput. Technologies}, 19(6):19--40, 2014.

\bibitem{Gusev2016}
O.~I. Gusev and S.~A. Beisel.
\newblock {Tsunami dispersion sensitivity to seismic source parameters}.
\newblock {\em Science of Tsunami Hazards}, 35(2):84--105, 2016.

\bibitem{Gusev2013}
O.~I. Gusev, N.~Y. Shokina, V.~A. Kutergin, and G.~S. Khakimzyanov.
\newblock {Numerical modelling of surface waves generated by underwater
  landslide in a reservoir}.
\newblock {\em Comput. Technologies}, 18(5):74--90, 2013.

\bibitem{Gusyakov2008}
V.~K. Gusyakov, Z.~I. Fedotova, G.~S. Khakimzyanov, L.~B. Chubarov, and Y.~I.
  Shokin.
\newblock {Some approaches to local modelling of tsunami wave runup on a
  coast}.
\newblock {\em Russ. J. Numer. Anal. Math. Modelling}, 23(6), jan 2008.

\bibitem{Horrillo2015}
J.~Horrillo, S.~T. Grilli, D.~Nicolsky, V.~Roeber, and J.~Zhang.
\newblock {Performance Benchmarking Tsunami Models for NTHMP's Inundation
  Mapping Activities}.
\newblock {\em Pure Appl. Geophys.}, 172(3-4):869--884, mar 2015.

\bibitem{Horrillo2006}
J.~Horrillo, Z.~Kowalik, and Y.~Shigihara.
\newblock {Wave Dispersion Study in the Indian Ocean-Tsunami of December 26,
  2004}.
\newblock {\em Marine Geodesy}, 29(3):149--166, dec 2006.

\bibitem{Imamura1996a}
F.~Imamura.
\newblock {Simulation of wave-packet propagation along sloping beach by
  TUNAMI-code}.
\newblock In {\em Long-wave Runup Models}, pages 231--241. World Scientific,
  Singapore, 1996.

\bibitem{Kazantsev1998}
R.~A. Kazantsev and V.~V. Kruglyakov.
\newblock {Giant landslide on the Black Sea floor}.
\newblock {\em Priroda}, 10:86--87, 1998.

\bibitem{Kervella2007}
Y.~Kervella, D.~Dutykh, and F.~Dias.
\newblock {Comparison between three-dimensional linear and nonlinear tsunami
  generation models}.
\newblock {\em Theor. Comput. Fluid Dyn.}, 21:245--269, 2007.

\bibitem{Khakimzyanov2016a}
G.~S. Khakimzyanov, D.~Dutykh, and Z.~I. Fedotova.
\newblock {Dispersive shallow water wave modelling. Part III: Model derivation
  on a globally spherical geometry}.
\newblock {\em Commun. Comput. Phys.}, pages 1--40, 2017.

\bibitem{Khakimzyanov2016c}
G.~S. Khakimzyanov, D.~Dutykh, Z.~I. Fedotova, and D.~E. Mitsotakis.
\newblock {Dispersive shallow water wave modelling. Part I: Model derivation on
  a globally flat space}.
\newblock {\em Commun. Comput. Phys.}, pages 1--40, 2017.

\bibitem{Khakimzyanov2016b}
G.~S. Khakimzyanov, D.~Dutykh, and O.~Gusev.
\newblock {Dispersive shallow water wave modelling. Part IV: Numerical
  simulation on a globally spherical geometry}.
\newblock {\em Commun. Comput. Phys.}, pages 1--40, 2017.

\bibitem{Khakimzyanov2016}
G.~S. Khakimzyanov, D.~Dutykh, O.~Gusev, and N.~Y. Shokina.
\newblock {Dispersive shallow water wave modelling. Part II: Numerical
  modelling on a globally flat space}.
\newblock {\em Commun. Comput. Phys.}, pages 1--40, 2017.

\bibitem{Khakimzyanov2015a}
G.~S. Khakimzyanov, D.~Dutykh, D.~E. Mitsotakis, and N.~Y. Shokina.
\newblock {Numerical solution of conservation laws on moving grids}.
\newblock {\em Submitted}, pages 1--28, 2016.

\bibitem{Khakimzyanov2015c}
G.~S. Khakimzyanov, O.~I. Gusev, S.~A. Beizel, L.~B. Chubarov, and N.~Y.
  Shokina.
\newblock {Simulation of tsunami waves generated by submarine landslides in the
  Black Sea}.
\newblock {\em Russ. J. Numer. Anal. Math. Modelling}, 30(4):227--237, jan
  2015.

\bibitem{Khakimzyanov2001}
G.~S. Khakimzyanov, Y.~I. Shokin, V.~B. Barakhnin, and N.~Y. Shokina.
\newblock {\em {Numerical Simulation of Fluid Flows with Surface Waves}}.
\newblock Sib. Branch, Russ. Acad. Sci., Novosibirsk, 2001.

\bibitem{Khakimzyanov2016d}
G.~S. Khakimzyanov, N.~Y. Shokina, D.~Dutykh, and D.~Mitsotakis.
\newblock {A new run-up algorithm based on local high-order analytic
  expansions}.
\newblock {\em J. Comp. Appl. Math.}, 298:82--96, may 2016.

\bibitem{Kirby2013}
J.~T. Kirby, F.~Shi, B.~Tehranirad, J.~C. Harris, and S.~T. Grilli.
\newblock {Dispersive tsunami waves in the ocean: Model equations and
  sensitivity to dispersion and Coriolis effects}.
\newblock {\em Ocean Modelling}, 62:39--55, feb 2013.

\bibitem{Kolar1994}
R.~L. Kolar, W.~G. Gray, J.~J. Westerink, and R.~A. Luettich.
\newblock {Shallow water modeling in spherical coordinates: equation
  formulation, numerical implementation, and application}.
\newblock {\em J. Hydr. Res.}, 32(1):3--24, jan 1994.

\bibitem{KdV}
D.~J. Korteweg and G.~de~Vries.
\newblock {On the change of form of long waves advancing in a rectangular
  canal, and on a new type of long stationary waves}.
\newblock {\em Phil. Mag.}, 39(5):422--443, 1895.

\bibitem{Kulikov}
E.~A. Kulikov, P.~P. Medvedev, and S.~S. Lappo.
\newblock {Satellite recording of the Indian Ocean tsunami on December 26,
  2004}.
\newblock {\em Doklady Earth Sciences A}, 401:444--448, 2005.

\bibitem{Ladyzhenskaya1973}
O.~A. Ladyzhenskaya and N.~N. Uraltseva.
\newblock {\em {Linear and quasilinear elliptic equations}}.
\newblock Nauka, Moscow, 1973.

\bibitem{Lovholt2009}
F.~L{\o}vholt and G.~Pedersen.
\newblock {Instabilities of Boussinesq models in non-uniform depth}.
\newblock {\em Int. J. Num. Meth. Fluids}, 61(6):606--637, oct 2009.

\bibitem{Lovholt2008}
F.~L{\o}vholt, G.~Pedersen, and G.~Gisler.
\newblock {Oceanic propagation of a potential tsunami from the La Palma
  Island}.
\newblock {\em J. Geophys. Res.}, 113(C9):C09026, sep 2008.

\bibitem{Lovholt2010}
F.~Lovholt, G.~Pedersen, and S.~Glimsdal.
\newblock {Coupling of Dispersive Tsunami Propagation and Shallow Water Coastal
  Response}.
\newblock {\em The Open Oceanography Journal}, 4(1):71--82, may 2010.

\bibitem{Lynett2002}
P.~Lynett and P.~L.~F. Liu.
\newblock {A numerical study of submarine-landslide-generated waves and
  run-up}.
\newblock {\em Proc. R. Soc. A}, 458(2028):2885--2910, dec 2002.

\bibitem{Mirchina1982}
N.~R. Mirchina and E.~N. Pelinovsky.
\newblock {Nonlinear and dispersive effects for tsunami waves in the open
  ocean}.
\newblock {\em Int. J. Tsunami Soc.}, 2(4):1073--1081, 1982.

\bibitem{Moreno2010}
M.~Moreno, M.~Rosenau, and O.~Oncken.
\newblock {2010 Maule earthquake slip correlates with pre-seismic locking of
  Andean subduction zone}.
\newblock {\em Nature}, 467(7312):198--202, sep 2010.

\bibitem{Murty2006}
T.~S. Murty, A.~D. Rao, N.~Nirupama, and I.~Nistor.
\newblock {Numerical modelling concepts for tsunami warning systems}.
\newblock {\em Current Science}, 90(8):1073--1081, 2006.

\bibitem{Nosov2014}
M.~A. Nosov, G.~N. Nurislamova, A.~V. Moshenceva, and S.~V. Kolesov.
\newblock {Residual hydrodynamic fields after tsunami generation by an
  earthquake}.
\newblock {\em Izvestiya, Atmospheric and Oceanic Physics}, 50(5):520--531, sep
  2014.

\bibitem{Nwogu1993}
O.~Nwogu.
\newblock {Alternative form of Boussinesq equations for nearshore wave
  propagation}.
\newblock {\em J. Waterway, Port, Coastal and Ocean Engineering}, 119:618--638,
  1993.

\bibitem{Okada85}
Y.~Okada.
\newblock {Surface deformation due to shear and tensile faults in a
  half-space}.
\newblock {\em Bull. Seism. Soc. Am.}, 75:1135--1154, 1985.

\bibitem{okada92}
Y.~Okada.
\newblock {Internal deformation due to shear and tensile faults in a
  half-space}.
\newblock {\em Bull. Seism. Soc. Am.}, 82:1018--1040, 1992.

\bibitem{Paranas-Carayannis2010}
G.~Paranas-Carayannis.
\newblock {The earthquake and tsunami of 27 February 2010 in Chile - evaluation
  of source mechanism and of near and far-field tsunami effects}.
\newblock {\em Science of Tsunami Hazards}, 29(2):96--126, 2010.

\bibitem{Pelinovsky1996a}
E.~N. Pelinovsky.
\newblock {\em {Tsunami wave hydrodynamics}}.
\newblock Institute of Applied Physics Press, Nizhny Novgorod, 1996.

\bibitem{Peregrine1966}
D.~H. Peregrine.
\newblock {Calculations of the development of an undular bore}.
\newblock {\em J. Fluid Mech.}, 25(02):321--330, mar 1966.

\bibitem{Ranguelov2008}
B.~Ranguelov, S.~Tinti, G.~Pagnoni, R.~Tonini, F.~Zaniboni, and A.~Armigliato.
\newblock {The nonseismic tsunami observed in the Bulgarian Black Sea on 7 May
  2007: Was it due to a submarine landslide?}
\newblock {\em Geophys. Res. Lett.}, 35(18):L18613, sep 2008.

\bibitem{Samarskii2001}
A.~A. Samarskii.
\newblock {\em {The Theory of Difference Schemes}}.
\newblock CRC Press, New York, 2001.

\bibitem{Segur2007}
H.~Segur.
\newblock {Waves in shallow water, with emphasis on the tsunami of 2004}.
\newblock In A.~Kundu, editor, {\em Tsunamis and Nonlinear waves}, pages 3--29.
  Springer, New York, 2007.

\bibitem{Shi2012}
F.~Shi, J.~T. Kirby, J.~C. Harris, J.~D. Geiman, and S.~T. Grilli.
\newblock {A high-order adaptive time-stepping TVD solver for Boussinesq
  modeling of breaking waves and coastal inundation}.
\newblock {\em Ocean Modelling}, 43-44:36--51, 2012.

\bibitem{Shi2012a}
F.~Shi, J.~T. Kirby, and B.~Tehranirad.
\newblock {Tsunami Benchmark Results for Spherical Coordinate Version of
  FUNWAVE-TVD (Version 2.0)}.
\newblock Technical report, University of Delaware, Newark, Delaware, USA,
  2012.

\bibitem{Shokin2008}
Y.~I. Shokin, V.~V. Babailov, S.~A. Beisel, L.~B. Chubarov, S.~V. Eletsky,
  Z.~I. Fedotova, and V.~K. Gusiakov.
\newblock {Mathematical Modeling in Application to Regional Tsunami Warning
  Systems Operations}.
\newblock In {\em Computational Science and High Performance Computing III},
  pages 52--68. 2008.

\bibitem{Shokin2015}
Y.~I. Shokin, Z.~I. Fedotova, and G.~S. Khakimzyanov.
\newblock {Hierarchy of nonlinear models of the hydrodynamics of long surface
  waves}.
\newblock {\em Doklady Physics}, 60(5):224--228, may 2015.

\bibitem{Shokin2007}
Y.~I. Shokin, Z.~I. Fedotova, G.~S. Khakimzyanov, L.~B. Chubarov, and S.~A.
  Beisel.
\newblock {Modelling surface waves generated by a moving landslide with
  allowance for vertical flow structure}.
\newblock {\em Russ. J. Numer. Anal. Math. Modelling}, 22(1):63--85, 2007.

\bibitem{Shokina2012}
N.~Y. Shokina.
\newblock {To the problem of construction of difference schemes on movable
  grids}.
\newblock {\em Russ. J. Numer. Anal. Math. Modelling}, 27(6), jan 2012.

\bibitem{Stoker1957}
J.~J. Stoker.
\newblock {\em {Water Waves: The mathematical theory with applications}}.
\newblock Interscience, New York, 1957.

\bibitem{Syno2006}
C.~E. Synolakis and E.~N. Bernard.
\newblock {Tsunami science before and beyond Boxing Day 2004}.
\newblock {\em Phil. Trans. R. Soc. A}, 364:2231--2265, 2006.

\bibitem{Synolakis2008}
C.~E. Synolakis, E.~N. Bernard, V.~V. Titov, U.~K{\^{a}}noglu, and F.~I.
  Gonz{\'{a}}lez.
\newblock {Validation and Verification of Tsunami Numerical Models}.
\newblock {\em Pure Appl. Geophys.}, 165:2197--2228, 2008.

\bibitem{Tang2012}
L.~Tang, V.~V. Titov, E.~N. Bernard, Y.~Wei, C.~D. Chamberlin, J.~C. Newman,
  H.~O. Mofjeld, D.~Arcas, M.~C. Eble, C.~Moore, B.~Uslu, C.~Pells,
  M.~Spillane, L.~Wright, and E.~Gica.
\newblock {Direct energy estimation of the 2011 Japan tsunami using deep-ocean
  pressure measurements}.
\newblock {\em J. Geophys. Res.: Oceans}, 117(C8), aug 2012.

\bibitem{Tappin2014}
D.~R. Tappin, S.~T. Grilli, J.~C. Harris, R.~J. Geller, T.~Masterlark, J.~T.
  Kirby, F.~Shi, G.~Ma, K.~Thingbaijam, and P.~M. Mai.
\newblock {Did a submarine landslide contribute to the 2011 Tohoku tsunami?}
\newblock {\em Marine Geology}, 357:344--361, nov 2014.

\bibitem{Titov1997}
V.~V. Titov and F.~I. Gonz{\'{a}}lez.
\newblock {Implementation and testing of the method of splitting tsunami (MOST)
  model}.
\newblock Technical Report ERL PMEL-112, Pacific Marine Environmental
  Laboratory, NOAA, 1997.

\bibitem{Titov1996}
V.~V. Titov and C.~E. Synolakis.
\newblock {Numerical modeling of 3-D long wave runup using VTCS-3}.
\newblock In H.~Yeh, P.~L.-F. Liu, and C.~E. Synolakis, editors, {\em Long wave
  runup models}, pages 242--248. World Scientific, Singapore, 1996.

\bibitem{Vigny2011}
C.~Vigny, A.~Socquet, S.~Peyrat, J.-C. Ruegg, M.~Metois, R.~Madariaga,
  S.~Morvan, M.~Lancieri, R.~Lacassin, J.~Campos, D.~Carrizo, M.~Bejar-Pizarro,
  S.~Barrientos, R.~Armijo, C.~Aranda, M.-C. Valderas-Bermejo, I.~Ortega,
  F.~Bondoux, S.~Baize, H.~Lyon-Caen, A.~Pavez, J.~P. Vilotte, M.~Bevis,
  B.~Brooks, R.~Smalley, H.~Parra, J.-C. Baez, M.~Blanco, S.~Cimbaro, and
  E.~Kendrick.
\newblock {The 2010 Mw 8.8 Maule Megathrust Earthquake of Central Chile,
  Monitored by GPS}.
\newblock {\em Science}, 332(6036):1417--1421, jun 2011.

\bibitem{Vilibic2010}
I.~Vilibic, J.~Sepic, B.~Ranguelov, N.~S. Mahovic, and S.~Tinti.
\newblock {Possible atmospheric origin of the 7 May 2007 western Black Sea
  shelf tsunami event}.
\newblock {\em J. Geophys. Res.}, 115(C7):C07006, jul 2010.

\bibitem{Wei2008}
Y.~Wei, E.~N. Bernard, L.~Tang, R.~Weiss, V.~V. Titov, C.~Moore, M.~Spillane,
  M.~Hopkins, and U.~Kanoglu.
\newblock {Real-time experimental forecast of the Peruvian tsunami of August
  2007 for U.S. coastlines}.
\newblock {\em Geophys. Res. Lett.}, 35(4):L04609, feb 2008.

\bibitem{Wen2011}
R.~Wen, Y.~Ren, X.~Li, and R.~Pan.
\newblock {Comparison of two great Chile tsunamis in 1960 and 2010 using
  numerical simulation}.
\newblock {\em Earthquake Science}, 24(5):475--483, oct 2011.

\bibitem{Young1950}
D.~M. Young.
\newblock {\em {Iterative Methods for Solving Partial Difference Equations of
  Elliptic Type}}.
\newblock Phd, Harvard University, 1950.

\bibitem{Zaitsev2005}
A.~I. Zaitsev, A.~A. Kurkin, B.~V. Levin, E.~N. Pelinovsky, A.~C. Yalciner,
  Y.~I. Troitskaya, and S.~A. Ermakov.
\newblock {Numerical simulation of carastrophic tsunami propagation in the
  Indian Ocean}.
\newblock {\em Doklady Earth Sciences}, 402(4):614--618, 2005.

\end{thebibliography}

\bigskip

\end{document}